\documentclass[a4paper,10pt]{article}
\pdfoutput=1

\usepackage{afterpage}
\usepackage{flafter}
\usepackage[export]{adjustbox}
\usepackage{jheppub} %
\usepackage{cmap}
\usepackage{bold-extra}
\usepackage{makecell}
\usepackage{cleveref}
\usepackage{booktabs}
\usepackage{tablefootnote}
\crefformat{footnote}{#2\footnotemark[#1]#3}

\usepackage{mathtools}
\usepackage{xspace}
\RequirePackage{snapshot}
\usepackage{algpseudocode,algorithm}
\usepackage{amsmath}
\usepackage{etoolbox}
\usepackage{enumitem,amssymb}
\usepackage{pdflscape}

\usepackage[margin=0pt,font+=smaller,labelformat=parens,labelsep=space,skip=6pt,list=false,hypcap=false]{subcaption}

\preprint{CERN-TH-2025-231/MCnet-25-26}

\title{KrkNLO matching and phenomenology for vector boson processes}

\author[a]{Pratixan Sarmah,}
\author[a,b]{Andrzej Siódmok,}
\author[a]{James Whitehead}

\affiliation[a]{Jagiellonian University, \\
ul.\ prof.\ Stanis\l{}awa \L{}ojasiewicza 11, 30-348 Krak\'{o}w, Poland}

\affiliation[b]{Theoretical Physics Department, CERN, \\
1211 Geneva 23, Switzerland}

\emailAdd{pratixan.sarmah@doctoral.uj.edu.pl}
\emailAdd{andrzej.siodmok@uj.edu.pl}
\emailAdd{james.whitehead@uj.edu.pl}

\abstract{
The combination of NLO matrix elements with parton showers is indispensable
for LHC physics.
Differences between matching methods introduce matching uncertainties, corresponding to formally higher-order terms.
We recently presented the process-independent generalisation of the \krknlo method for NLO matching,
which employs a modified PDF factorisation scheme
to achieve NLO accuracy.
With this factorisation scheme, the method can be used for colour-singlet final-states,
and was previously implemented in the \herwig Monte Carlo Event Generator
and applied to the diphoton-production process.

Here we present the extension of the implementation of the 
\krknlo method within \herwig
to support the full class of applicable processes,
using an external matrix-element library.
We re-validate the implementation, and use it to study the
NLO matching uncertainty for four vector-boson production processes at the
LHC: $W$, $Z\gamma$, $WW$ and $ZZ$.
We demonstrate that the KrkNLO method effectively eliminates the negative-weight problem in NLO event generation, across the four processes studied.
We provide detailed comparisons between \krknlo and variants of the \mcatnlo method with different shower starting-scale choices, across processes and throughout phase-space, including double-differential observables. 
For each process, we compare the predictions to LHC data from \atlas.
}


\DeclareRobustCommand{\ensuremathrm}[1]{\ensuremath{\mathrm{#1}}\xspace}

\newcommand{\nlo}{\ensuremath{\mathrm{NLO}}\xspace}
\newcommand{\nnlo}{\ensuremath{\mathrm{NNLO}}\xspace}
\newcommand{\nnnlo}{\ensuremath{\mathrm{N^3LO}}\xspace}

\newcommand{\cut}{\mathrm{cut}}

\newcommand{\dd}{\mathrm{d}}

\newcommand{\qbar}{\bar{q}}

\newcommand{\alphas}{\alpha_\mathrm{s}}

\newcommand{\msbar}{\ensuremath{{\overline{\mathrm{MS}}}}\xspace}

\newcommand{\rB}{\mathrm{B}}

\newcommand{\rF}{\mathrm{F}}
\newcommand{\rFS}{\mathrm{FS}}
\newcommand{\rI}{\mathrm{I}}

\newcommand{\rR}{\mathrm{R}}

\newcommand{\rT}{\mathrm{T}}
\newcommand{\rV}{\mathrm{V}}

\newcommand{\pt}{p_\rT}

\DeclareRobustCommand{\cut}{\ensuremathrm{cut}\xspace}

\newcommand{\muf}{\mu_\rF}
\newcommand{\mur}{\mu_\rR}

\newcommand{\cf}{C_\mathrm{F}}

\newcommand{\krk}{\ensuremath{\mathrm{Krk}}\xspace}
\newcommand{\krknlo}{\ensuremath{\mathrm{KrkNLO}}\xspace}
\newcommand{\mcatnlo}{\ensuremath{\mathrm{\textsc{Mc@Nlo}}}\xspace}
\newcommand{\powheg}{\ensuremath{\mathrm{\textsc{Powheg}}}\xspace}

\newcommand{\herwig}{\textsf{Herwig}\xspace}
\newcommand{\herwigseven}{\textsf{Herwig 7}\xspace}
\newcommand{\matchbox}{\textsf{Matchbox}\xspace}
\newcommand{\madgraph}{\textsf{MadGraph}\xspace}
\newcommand{\openloops}{\textsf{OpenLoops}\xspace}

\newcommand{\rivet}{\textsf{Rivet}\xspace}

\newcommand{\atlas}{\textsc{Atlas}\xspace}

\DeclareRobustCommand{\GeV}{\ensuremathrm{GeV}\xspace}
\DeclareRobustCommand{\TeV}{\ensuremathrm{TeV}\xspace}

\DeclareRobustCommand{\pt}{\ensuremath{p_\rT}\xspace}
\DeclareRobustCommand{\ptcut}{\ensuremath{p_\rT^\cut}\xspace}
\DeclareRobustCommand{\ptof}[1]{\ensuremath{p_{\rT,{#1}}}\xspace}
\DeclareRobustCommand{\ptofp}[1]{\ensuremath{p_{\rT}^{#1}}\xspace}

\DeclareRobustCommand{\kt}{\ensuremath{k_\rT}\xspace}

\DeclareRobustCommand{\ptg}[1]{\ensuremath{p_{\rT}^{\gamma_{#1}}}\xspace}

\DeclareRobustCommand{\ptell}[1]{\ensuremath{p_{\rT}^{\ell_{#1}}}\xspace}
\DeclareRobustCommand{\ptnu}[1]{\ensuremath{p_{\rT}^{\nu_{#1}}}\xspace}

\DeclareRobustCommand{\ptj}[1]{\ensuremath{p_{\rT}^{j_{#1}}}\xspace}

\DeclareRobustCommand{\mt}{\ensuremath{m_{\rT}}\xspace}

\DeclareRobustCommand{\Etiso}{\ensuremath{E_{\rT}^{\text{iso}}}\xspace}

\DeclareRobustCommand{\Etmiss}{\ensuremath{E_{\rT}^{\text{miss}}}\xspace}

\DeclareRobustCommand{\dRlg}{\ensuremath{\Delta R_{\ell\gamma}}\xspace}

\DeclareRobustCommand{\dRlj}{\ensuremath{\Delta R_{\ell j}}\xspace}

\DeclareRobustCommand{\absetaell}{\ensuremath{\left| \eta^{\ell} \right| }\xspace}

\DeclareRobustCommand{\absyell}{\ensuremath{\left| y^{\ell} \right| }\xspace}

\DeclareRobustCommand{\absyof}[1]{\ensuremath{\left| y^{#1} \right| }\xspace}

\DeclareRobustCommand{\Mll}{\ensuremath{M_{\ell\ell}}\xspace}
\DeclareRobustCommand{\Mllg}{\ensuremath{M_{\ell\ell\gamma}}\xspace}
\DeclareRobustCommand{\Mlnu}{\ensuremath{M_{\ell\nu}}\xspace}

\makeatletter
\newlength{\negph@wd}
\DeclareRobustCommand{\negphantom}[1]{%
  \ifmmode
    \mathpalette\negph@math{#1}%
  \else
    \negph@do{#1}%
  \fi
}
\newcommand{\negph@math}[2]{\negph@do{$\m@th#1#2$}}
\newcommand{\negph@do}[1]{%
  \settowidth{\negph@wd}{#1}%
  \hspace*{-\negph@wd}%
}

\def\negstrip#1 #2\relax{-#1}

\makeatother

\usepackage[switch]{lineno}

\begin{document}
\maketitle
\flushbottom

\section{Introduction}
\label{sec:intro}

`Matched' calculations
combining next-to-leading-order (NLO) perturbative accuracy with the logarithmic resummation
of parton shower algorithms remain indispensable for LHC phenomenology.
Two methods for matching at NLO, the \mcatnlo \cite{Frixione:2002ik}
and \powheg methods \cite{Nason:2004rx,Frixione:2007vw,Alioli:2010xd},
have been widely used for LHC physics.%
\footnote{A number of others, at NLO and beyond, have been proposed or are in development or use \cite{Giele:2007di,Nason:2021xke,vanBeekveld:2025lpz,Hamilton:2012np,Alioli:2012fc,Hoche:2014uhw,Monni:2019whf,Campbell:2021svd,Prestel:2021vww}.}
The matching uncertainties associated with these methods 
have been investigated for a 
number of processes and independent implementations in
\cite{Alioli:2008gx,Alioli:2008tz,Alioli:2009je,Hamilton:2010mb,Platzer:2011bc,Hoeche:2011fd,Nason:2012pr,Frederix:2012dh,Heinrich:2017kxx,Jones:2017giv,Cormier:2018tog,Jager:2020hkz,ATL-PHYS-PUB-2023-029}.

The differences between results generated by alternative matching schemes
may be considered `matching uncertainties', 
and correspond to
formally higher-order terms introduced beyond NLO by the choice of matching method.
The presence of logarithms of ratios of scales, generated by the evolution of the parton shower,
allows such uncertainties to be numerically larger than their formal perturbative suppression
would na\"ively suggest.

In \cite{Sarmah:2024hdk} we recapitulated the \krknlo method \cite{Jadach:2011cr,Jadach:2015mza,Jadach:2016qti},
generalised to the production of arbitrary colour-singlet final-states which proceed via $q\qbar$-annihilation at leading-order,
and outlined its implementation in \herwigseven \cite{Bellm:2015jjp,Bewick:2023tfi}.%
\footnote{For convenience, in this work we adopt the notation and conventions of \cite{Sarmah:2024hdk},
which may be considered a companion paper to this one.}
The \krknlo method uses the freedom, at NLO and beyond,
to choose a factorisation scheme for the parton distribution functions (PDFs).
The scheme used, the \krk scheme \cite{Jadach:2016acv}, simplifies the matching calculation by moving
collinear counterterms from the hard process into the PDFs.
We used diphoton production as a test process, with hard-coded real-emission and virtual matrix elements.
This was the first process calculated with the \krknlo method that was not used for the definition of the \krk factorisation scheme.

In this work, we report on the extension of the \krknlo code
to support arbitrary processes calculable by the \krknlo method,
using the \openloops library \cite{Buccioni:2019sur} to compute the matrix-elements.
This enables the \krknlo method to be applied to the full class of suitable processes.
Here, we focus on four processes involving the production of at least one massive vector-boson:
$pp \to \{ W, Z\gamma, WW, ZZ \}$.%
\footnote{In \cite{Jadach:2015mza} the \krknlo method was applied to neutral-current
Drell--Yan production in an approximation factorising $Z$-boson production from its decay.
In this work we use exact matrix-elements, unapproximated.}
With this we demonstrate the full breadth of applicability of the method,
and the computational readiness of the code.

For each included process, we follow the general strategy applied to diphoton-production in \cite{Sarmah:2024hdk}: we
(i) demonstrate the numerical validation of the \krknlo implementation, in \cref{sec:validation};
(ii) perform a theory comparison to fixed-order NLO and \mcatnlo, 
both with the parton shower truncated to a single emission, and running to completion, in \cref{sec:matching},
and (iii) examine the resulting phenomenology using experimental data from the LHC, in \cref{sec:LHCpheno}.
We find that the \krknlo method gives the expected level of agreement with the LHC
data for an NLO calculation, and that the \krknlo prediction
is not necessarily within the matching-uncertainty envelope obtained by 
shower starting-scale variation within the \mcatnlo method.

\section{The KrkNLO method}
\label{sec:nlomatching_krknlo}

The \krknlo method generates events from the Born $m$-particle phase-space $\Phi_m$,
and uses the shower algorithm itself to generate events with real-emission kinematics $\Phi_{m+1}$.
According to the decision tree implied by the shower algorithm, events are reweighted multiplicatively to
incorporate the real and virtual matrix-elements, $\rR$ and $\rV$,
in a way that leads to positive event-weights
subject to the positivity of the PDFs and the virtual reweight.
By avoiding subtraction,
the method eliminated a common source of negative weights in the \mcatnlo method.%
\footnote{Positivity is discussed further in \cref{sec:app_pos}.}

The method may be summarised succinctly as:
\begin{algorithm}
	\begin{algorithmic}
		\ForAll{Born events} {shower}
		\If {first emission generated, from kernel $(\alpha)$}
		\State $w\gets w \times \frac{\rR(\Phi_{m+1})}{P_m^{(\alpha)}(\Phi_{m+1}) \, \rB(\Phi_m)}$ 
		\EndIf
		\State $w \gets w \times
		\left[1 + \frac{\alphas(\mur)}{2\pi}\left(\frac{\rV(\Phi_m;\, \mur)}{\rB(\Phi_m)} + \frac{\rI(\Phi_m;\, {\mu}_\rR)}{\rB (\Phi_m)} + \Delta_0^\rFS \right)\right]$
		\EndFor
	\end{algorithmic}
\end{algorithm}

Here $P_m^{(\alpha)}$ denotes the splitting function that generated the selected splitting within
the shower algorithm;
$\rI$, the contribution from the Catani $\mathbf{I}$-operator,
corresponding here to integrated Catani--Seymour dipoles;
$\mur$, the renormalisation scale, and
$\Delta_0^{\rFS}$ a factorisation-scheme-dependent constant
which corrects for endpoint-contributions absorbed into the PDFs.

When combined with the use of PDFs in the \krk factorisation scheme \cite{Jadach:2016acv,Delorme:2025teo},
no further collinear convolutions are required, allowing 
the NLO matching condition to be satisfied through multiplicative reweighting alone.

A full exposition, including formulae for the resulting
contributions to a given IR-safe observable $\mathcal{O}$,
is given in \cite{Sarmah:2024hdk}.

\section{Validation}
\label{sec:validation}

In this section we apply the validation strategy introduced for the diphoton process
in \cite{Sarmah:2024hdk}, illustrating the validation of the \krknlo code
for each of the processes considered in this study.
This also documents the validation of the \openloops \krknlo implementation that will
be made publicly available as part of \herwig 7.4.
These tests are performed with generation parameters and analysis cuts
(via \rivet \cite{Bierlich:2019rhm}) as used in \cref{sec:matching}, except where stated otherwise.

\subsection{Real validation}
\label{sec:validation_real}

As in \cite{Sarmah:2024hdk}, we test the reweight implementing the real-emission correction
by truncating the parton shower after one-emission, and vetoing the no-emission contributions;
the Sudakov factor generated by the veto algorithm within the parton shower is calculated numerically,
and its reciprocal applied as a reweight.

Since the one-emission truncation of the shower algorithm changes the result only at $\mathcal{O}(\alphas^2)$,
and the Sudakov factor (and its reciprocal) are formally 1 to leading-order,
this provides numerical verification that the method achieves NLO-accuracy in this region of phase-space.

\begin{figure}[p]
	\centering
	\begin{subfigure}[t]{\textwidth}
		\centering
        \makebox[\textwidth][c]{
		\includegraphics[width=.32\textwidth]{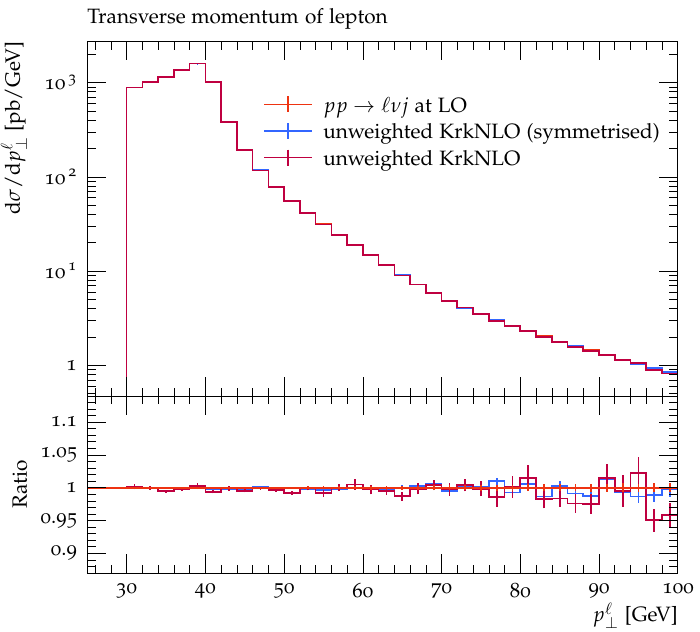}
		\includegraphics[width=.32\textwidth]{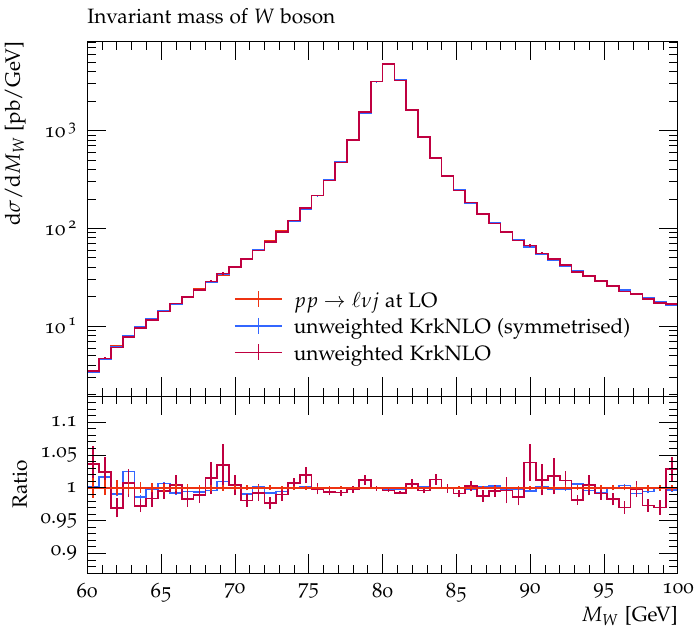}
        \includegraphics[width=.32\textwidth]{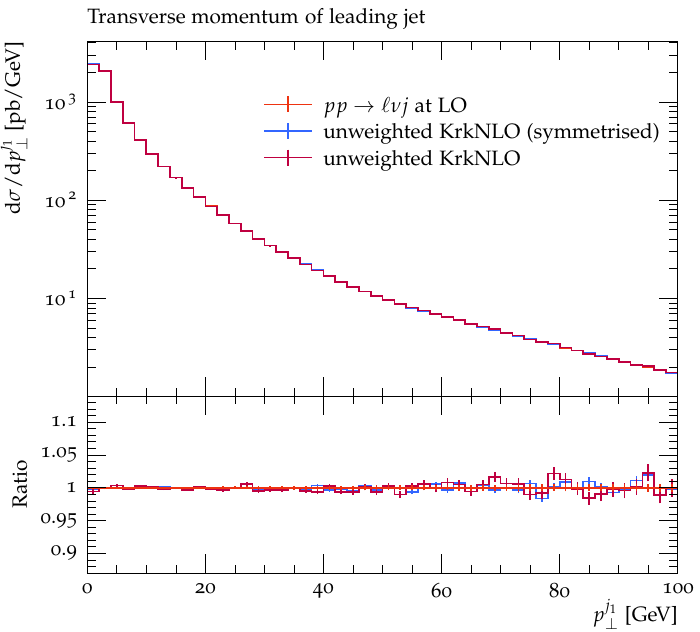}
		}
		\caption{$p p \to \ell \nu j$ \label{fig:validation_W_real_pp}}
    \end{subfigure}
        \begin{subfigure}[t]{\textwidth}
        \centering
        \makebox[\textwidth][c]{
        \includegraphics[width=.32\textwidth]{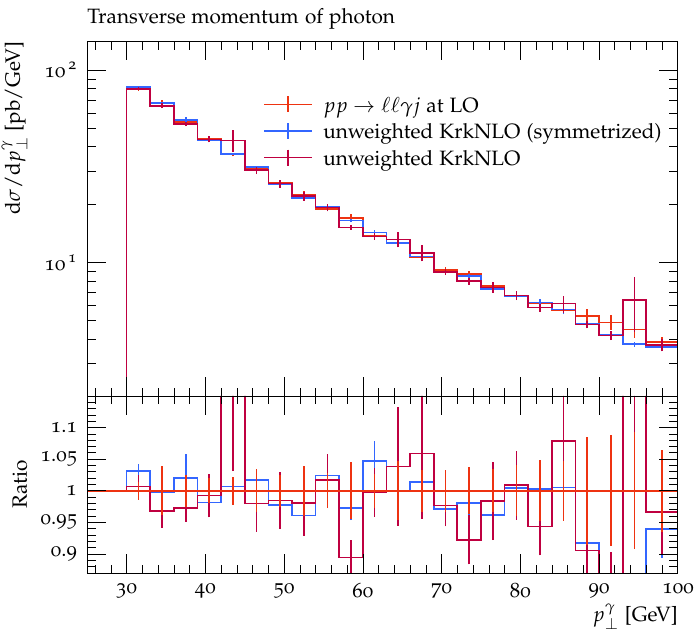}
        \includegraphics[width=.32\textwidth]{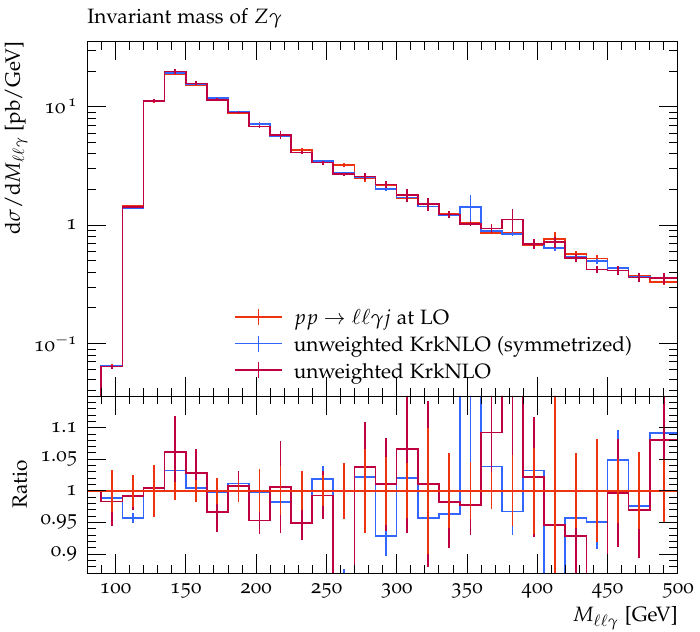}
        \includegraphics[width=.32\textwidth]{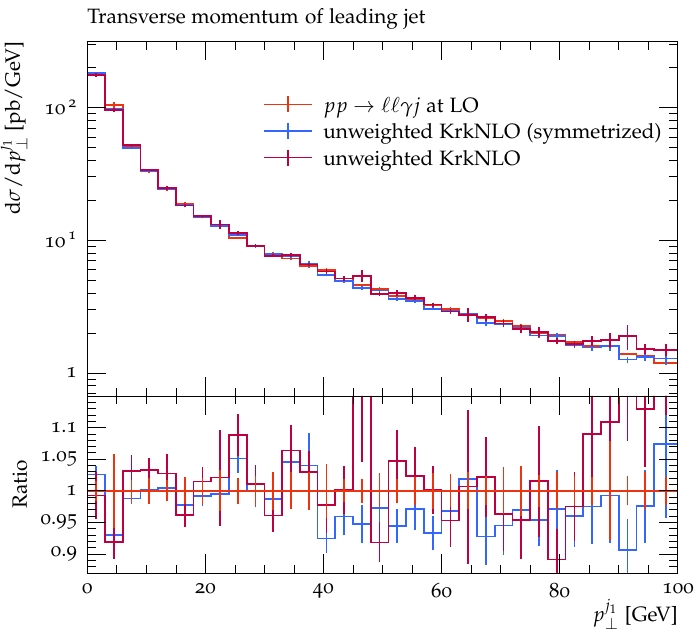}
        }
    \caption{$p p \to \ell \ell \gamma j$ \label{fig:validation_Zgam_real_pp}}
	\end{subfigure}
    \begin{subfigure}[t]{\textwidth}
        \centering
        \makebox[\textwidth][c]{
        \includegraphics[width=.32\textwidth]{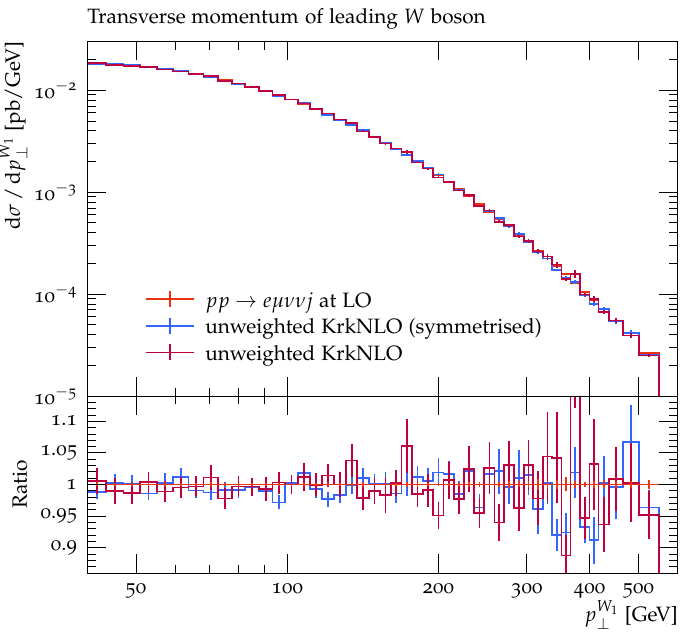}
        \includegraphics[width=.32\textwidth]{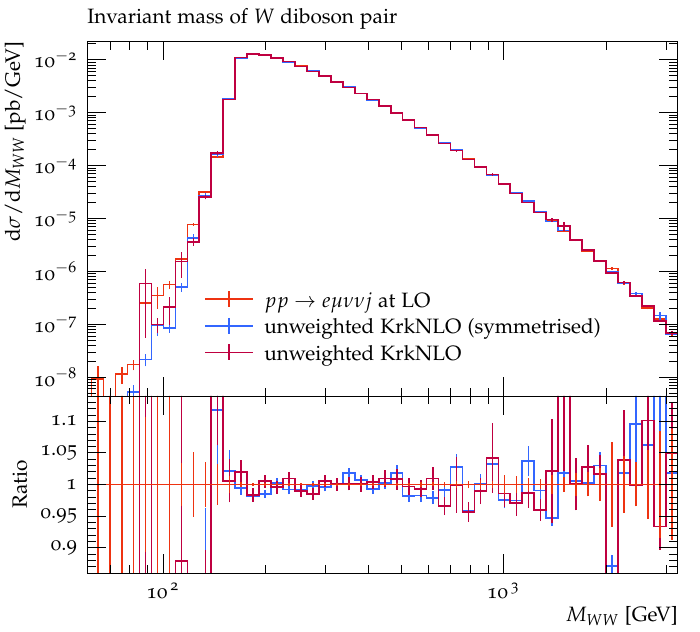}
        \includegraphics[width=.32\textwidth]{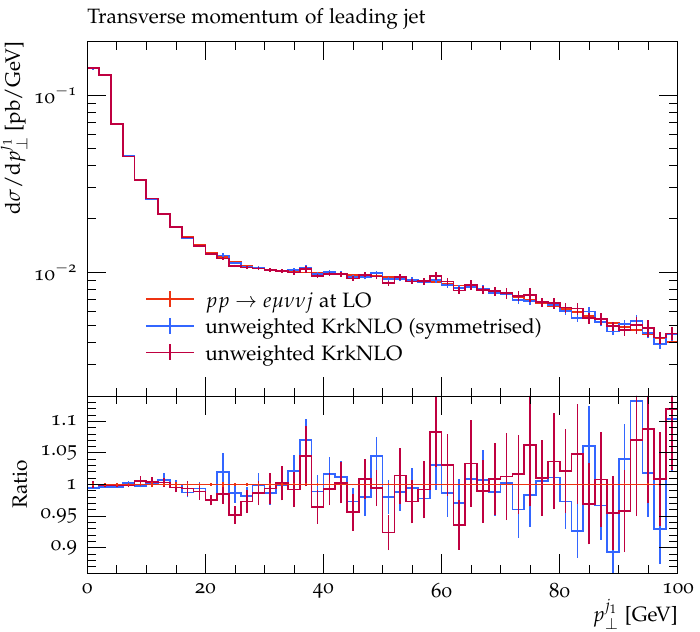}
        }
    \caption{$p p \to e \mu \nu \nu j$ \label{fig:validation_WW_real_pp}}
    \end{subfigure}
        \begin{subfigure}[t]{\textwidth}
        \centering
        \makebox[\textwidth][c]{
        \includegraphics[width=.32\textwidth]{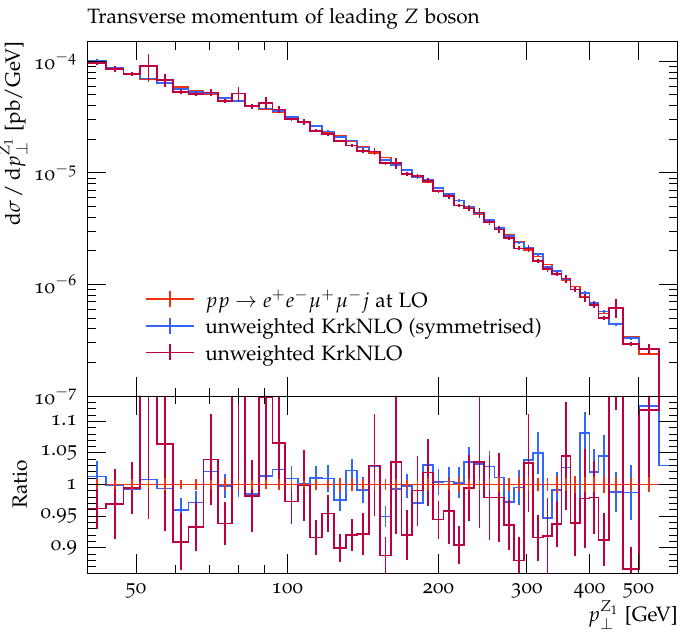}
        \includegraphics[width=.32\textwidth]{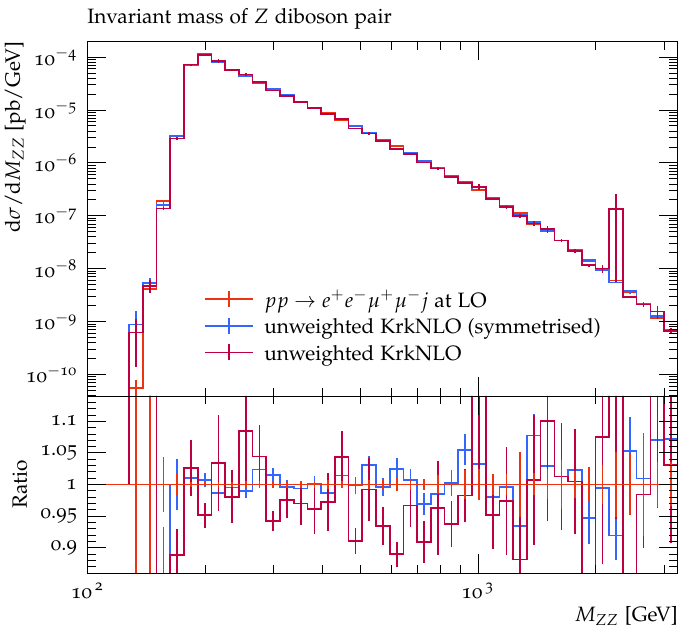}
        \includegraphics[width=.32\textwidth]{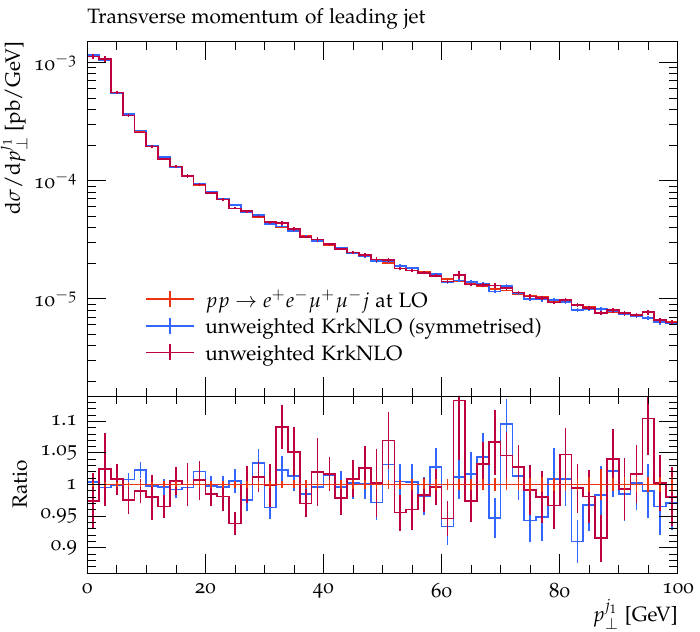}
        }
    \caption{$p p \to e^+ e^- \mu^+ \mu^- j$ \label{fig:validation_ZZ_real_pp}}
    \end{subfigure}
	\caption{Validation of the real weight, unweighting by the first-emission Sudakov factor 
	$\Delta \bigr\vert_{\ptof{1}}^{Q(\Phi_m)} (\Phi_m)$
	(independently calculated for each emission by numerical integration of the kernels over the radiative phase-space)
	to isolate the real matrix element within the \krknlo implementation.
    Plots shown for the $W$, $Z\gamma$, $WW$, and $ZZ$ processes respectively.
    \label{fig:validation_real}}
\end{figure}

The resulting differential cross-sections are shown in  \cref{fig:validation_real},
compared against those arising from the
real-emission matrix-element, i.e.\ the sole contribution to $\mathcal{O}(\Phi_{m+1})$
within an NLO calculation of $pp \to X$
(equivalently, within a leading-order calculation of $pp \to X+j$),
calculated with \matchbox \cite{Platzer:2011bc} within \herwigseven.

We show both the symmetrised and unsymmetrised versions of \krknlo as presented in \cite{Sarmah:2024hdk}.%
\footnote{The symmetrised real weight averages over the two possible underlying Born
phase-space configurations for each real-emission configuration, rather than
only using the Born phase-space point actually generated as the initial condition for the shower, in order to regulate the real-reweight distribution.
As outlined in \cite{Sarmah:2024hdk}, we consider the symmetrised version to be the default,
and use it for the remainder of this work.}
We see agreement to the percent level for the symmetrised version. For the unsymmetrised \krknlo variant, agreement is at the few-percent level and within statistical uncertainties, limited by slower statistical convergence (due to large weights, some visible) and fewer generated events.

\subsection{Virtual validation}
\label{sec:validation_virtual}

As in \cite{Sarmah:2024hdk},
we isolate the implementation of the virtual reweight within the \krknlo
code by setting the parton shower cutoff scale sufficiently high to prohibit any parton-shower radiation.
The radiative phase-space is then empty, and as a consequence the Sudakov factor
representing the no-emission probability is identically 1.
To isolate the one-loop matrix-element contribution $\rV$ and Catani--Seymour $\rI$-operator terms within
the virtual reweight, we disable the Born and $\Delta_0^\krk$ terms
(setting them to zero within the code).%

\begin{figure}[p]
    \centering
	\begin{subfigure}[t]{\textwidth}
    \centering
    \makebox[\textwidth][c]{
	\includegraphics[width=.32\textwidth]{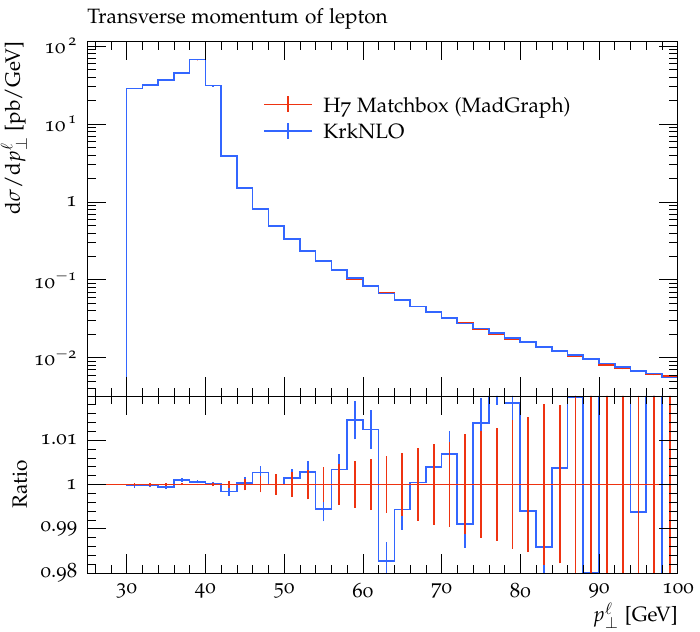}
	\includegraphics[width=.32\textwidth]{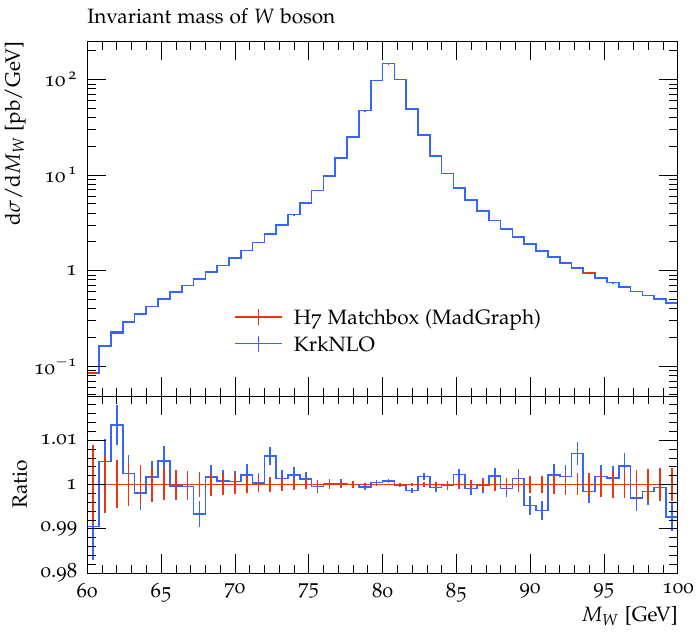}
	\includegraphics[width=.32\textwidth]{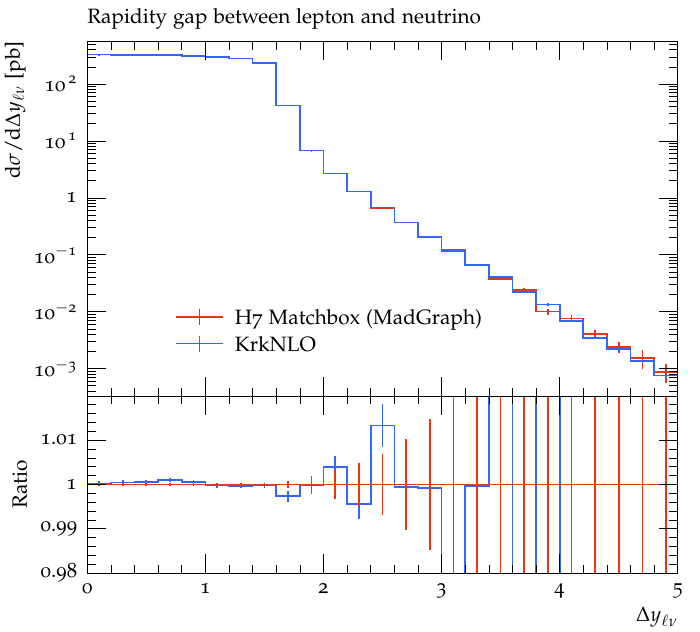}
    }
    \caption{$q \qbar \to \ell \nu$ \label{fig:validation_W_virtual_pp}}
    \end{subfigure}
	\begin{subfigure}[t]{\textwidth}
    \centering
    \makebox[\textwidth][c]{
	\includegraphics[width=.32\textwidth]{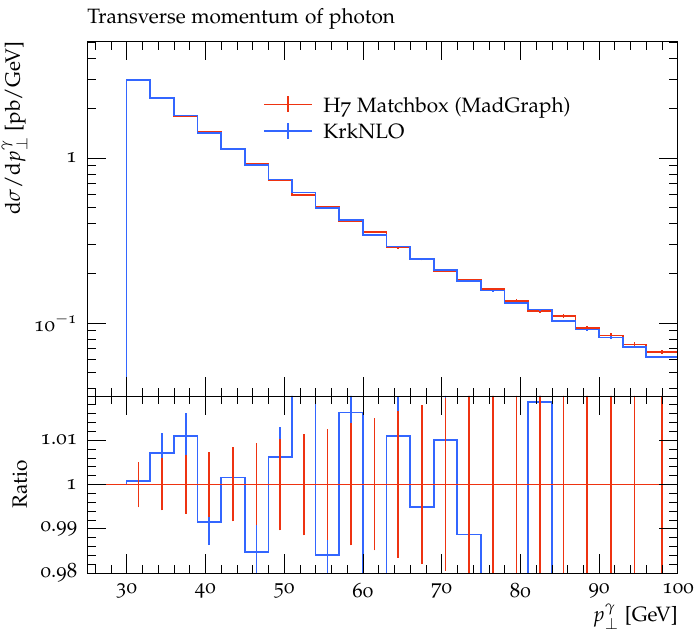}
	\includegraphics[width=.32\textwidth]{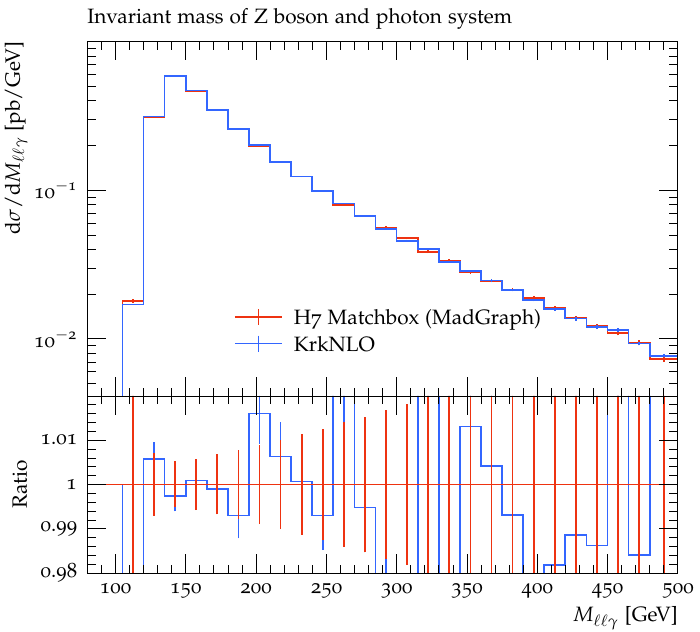}
	\includegraphics[width=.32\textwidth]{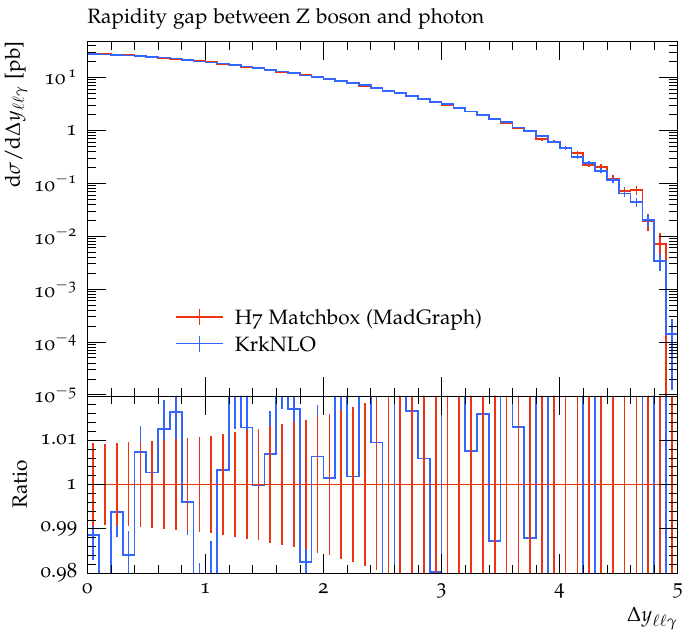}
    }
    \caption{$q \qbar \to \ell \ell \gamma $ \label{fig:validation_Zgam_virtual_pp}}
    \end{subfigure}
	\begin{subfigure}[t]{\textwidth}
    \centering
    \makebox[\textwidth][c]{
	\includegraphics[width=.32\textwidth]{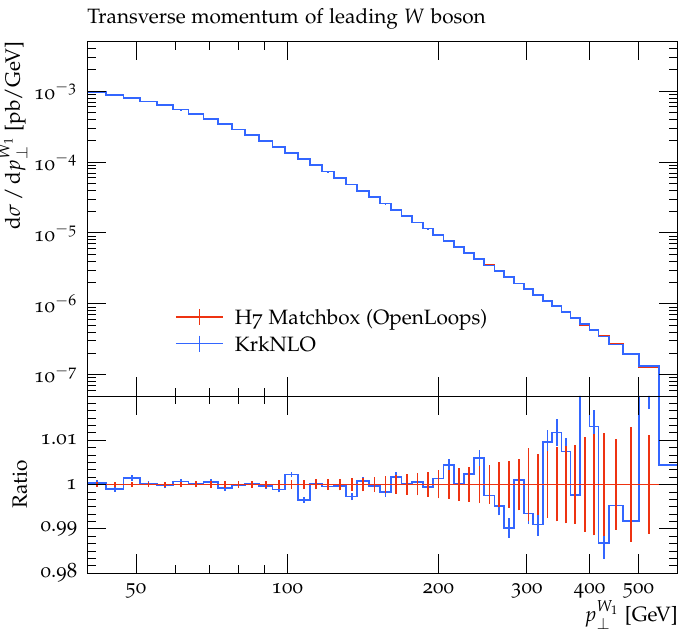}
	\includegraphics[width=.32\textwidth]{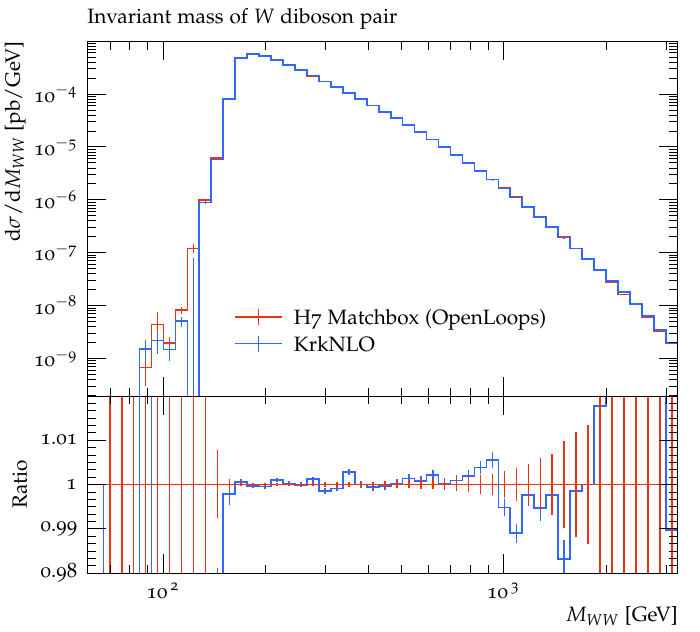}
	\includegraphics[width=.32\textwidth]{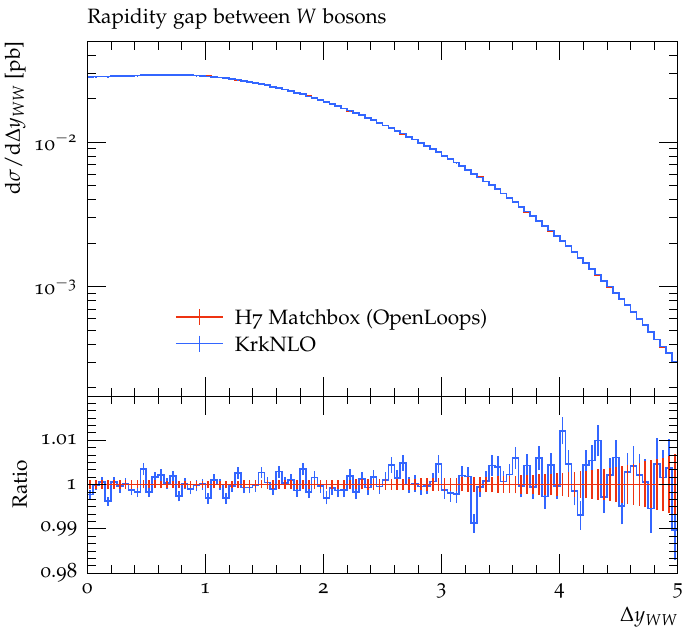}
    }
    \caption{$q \qbar \to e \mu \nu \nu $ \label{fig:validation_WW_virtual_pp}}
    \end{subfigure}
	\begin{subfigure}[t]{\textwidth}
    \centering
    \makebox[\textwidth][c]{
	\includegraphics[width=.32\textwidth]{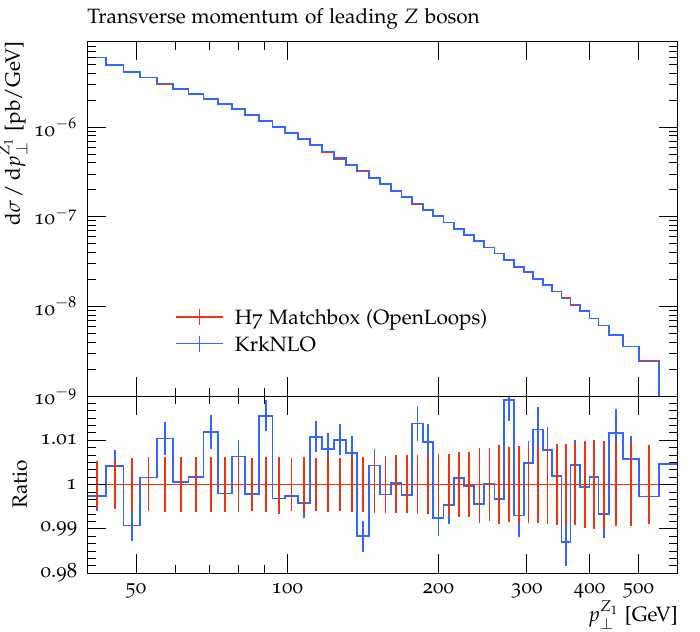}
	\includegraphics[width=.32\textwidth]{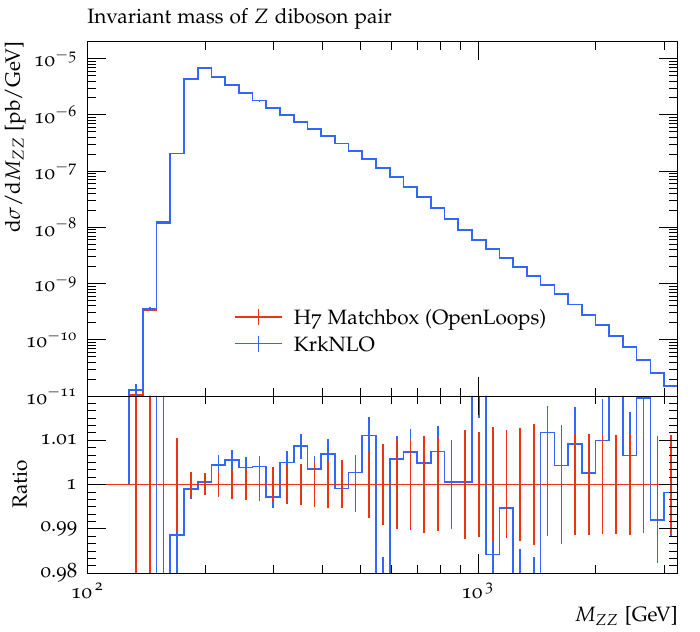}
	\includegraphics[width=.32\textwidth]{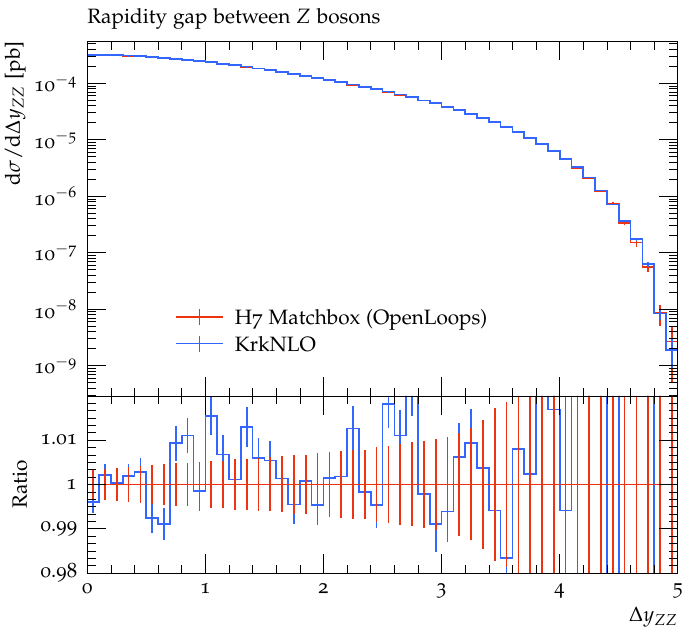}
    }
    \caption{$q \qbar \to e^+ e^- \mu^+ \mu^- $ \label{fig:validation_ZZ_virtual_pp}}
    \end{subfigure}
	\caption{Validation of the virtual weight $\rV + \rI$ for the $q\qbar$-channel
     		as described in \cite{Sarmah:2024hdk}.
		    By setting the shower IR cutoff to guarantee $t_0 > Q(\Phi_m)$,
		    thus disabling the shower, and manually
		    disabling the Born and $\Delta_0^{\krk}$ contributions within \krknlo,
		    the \krknlo implementation of the virtual terms can be isolated and compared with
		    those generated automatically by \matchbox within \herwigseven.
            Plots shown for the $W$, $Z\gamma$, $WW$ and $ZZ$ processes respectively.
			\label{fig:validation_virtual}}
\end{figure}

The resulting differential cross-sections are shown in 
\cref{fig:validation_virtual},
compared against those 
computed using the corresponding automated \matchbox implementation within \herwig,
calculated with one-loop matrix-elements provided by either 
\madgraph or \openloops as shown.
We see percent-level agreement in general,
and permille-level wherever the statistical convergence is sufficient to
resolve this level of precision.

\subsection{Non-diagonal CKM matrix}

\begin{figure}[t]
\centering
    \begin{subfigure}[t]{\textwidth}
		\centering
        \makebox[\textwidth][c]{
		\includegraphics[width=.32\textwidth]{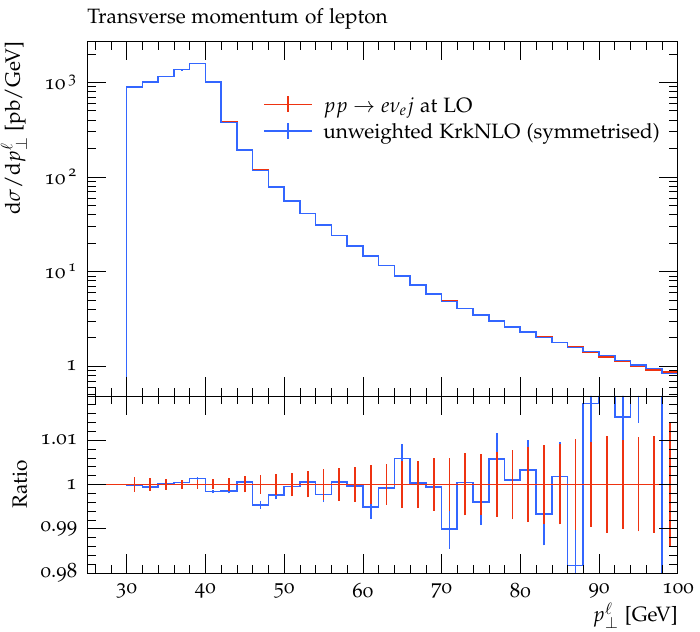}
		\includegraphics[width=.32\textwidth]{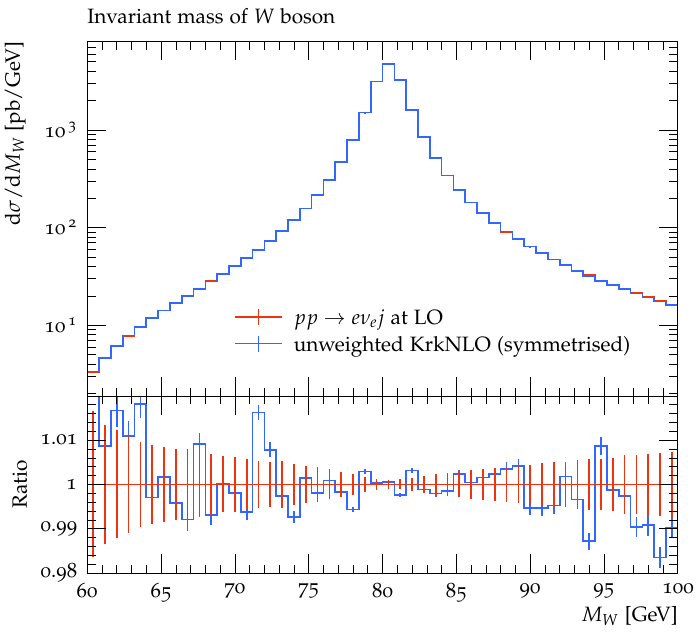}
        \includegraphics[width=.32\textwidth]{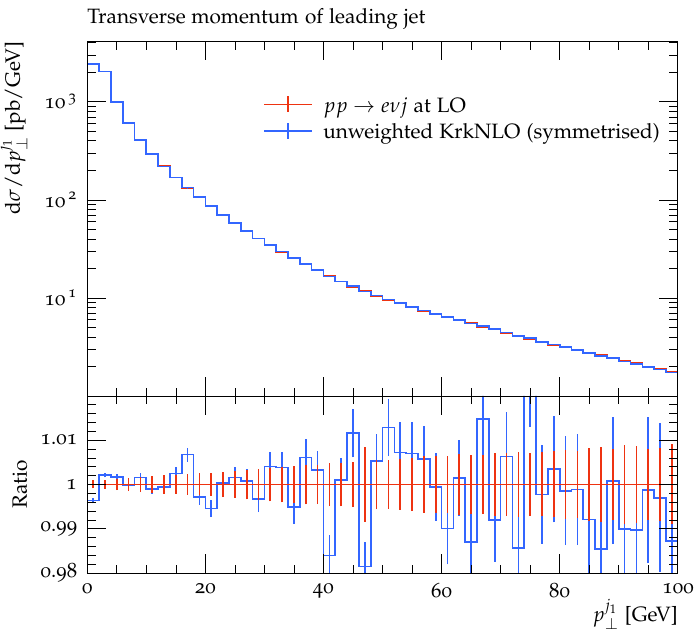}
		}
		\caption{real weight validation:
        $p p \to \ell \nu j$ \label{fig:validation_WtrueCKM_real_pp}}
    \end{subfigure}
	\begin{subfigure}[t]{\textwidth}
        \centering
        \makebox[\textwidth][c]{
    	\includegraphics[width=.32\textwidth]{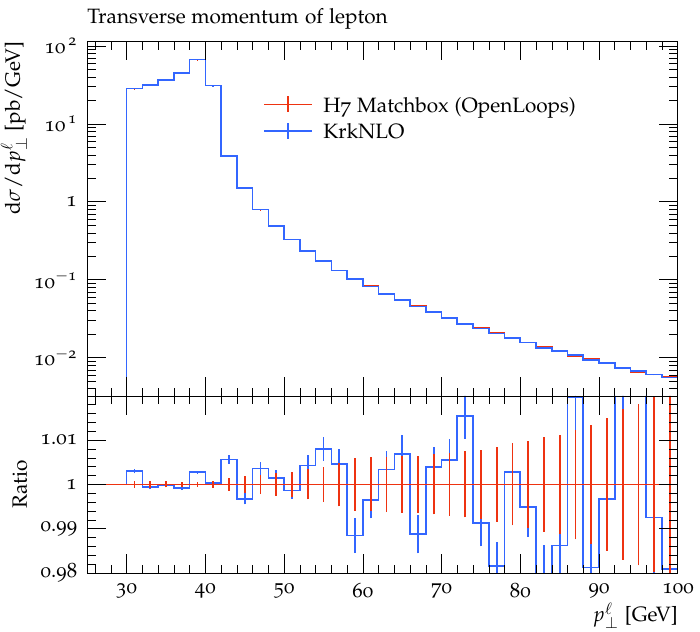}
    	\includegraphics[width=.32\textwidth]{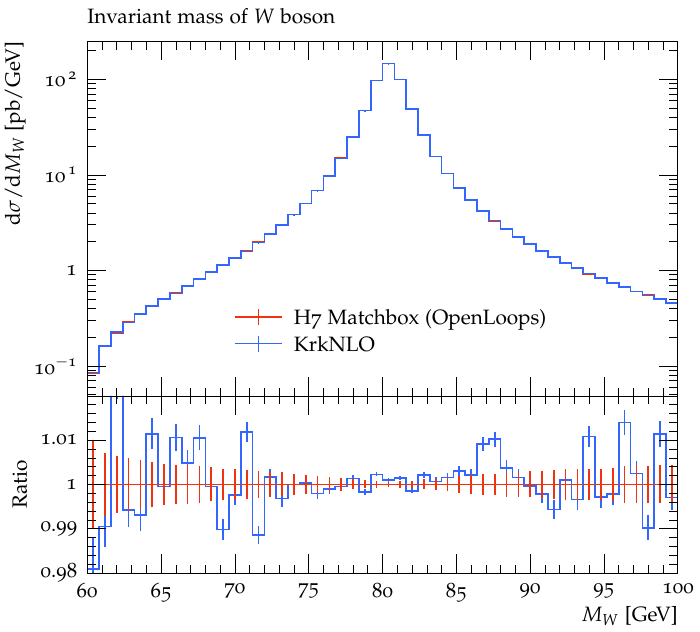}
    	\includegraphics[width=.32\textwidth]{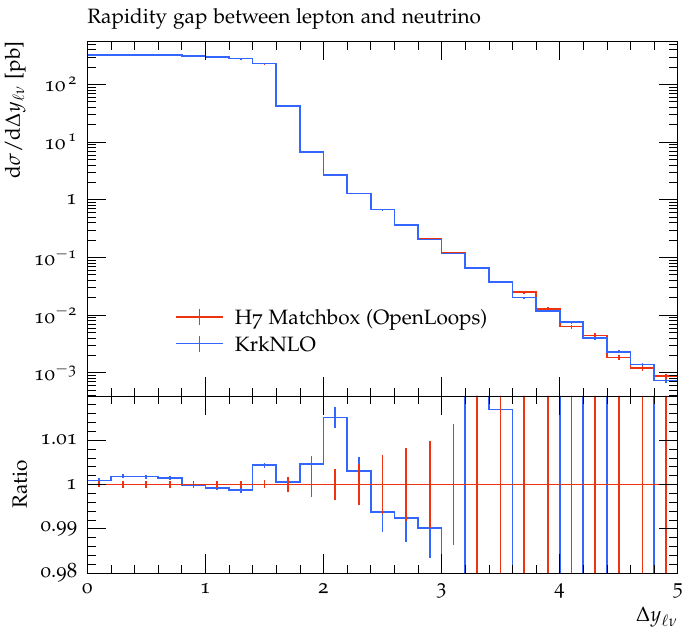}
        }
        \caption{virtual weight validation: $q \qbar \to e \nu_e$ \label{fig:validation_WtrueCKM_virtual_pp}}
    \end{subfigure}
	\caption{Validation of the \krknlo implementation using a non-diagonal
    CKM matrix, for the process $pp \to W \to \ell \nu$.
    This figure is a counterpart to \cref{fig:validation_W_real_pp,fig:validation_W_virtual_pp}, 
    which are identical save that the CKM matrix is there approximated by the identity matrix.
    \label{fig:validation_trueCKM}}
\end{figure}

The validation plots presented above were calculated using the flavour-diagonal
identity-matrix approximation for the Cabibbo--Kobayashi--Maskawa (CKM) mixing matrix,
which is used by default within \matchbox.
For processes involving $W$-bosons the effect of the true CKM is not guaranteed to be negligible.
In \cref{fig:validation_trueCKM} we therefore present the validation of the \krknlo code in the case of $W$-production for the full, non-diagonal, CKM
mixing matrix,
using the \herwig defaults (taken from the PDG \cite{ParticleDataGroup:2022pth})
for both the \krknlo calculation and the \matchbox reference.%
\footnote{Note that for the runs presented in \cref{sec:matching,sec:LHCpheno},
the default diagonal approximation has been used.}
Within the \krknlo implementation, the CKM matrix elements set within \herwig
are automatically passed to \openloops,
and may therefore be changed using the usual interface.

\section{Analysis of NLO matching uncertainty}
\label{sec:matching}

In general, we follow the same approach as for diphoton production in \cite{Sarmah:2024hdk},
to allow a systematic comparison of matching methods
and the associated matching uncertainties, across the processes.

We use \herwigseven for all predictions, and the \herwigseven implementation \cite{Platzer:2009jq}
of the
dipole shower \cite{Nagy:2005aa,Nagy:2006kb,Dinsdale:2007mf,Schumann:2007mg},
based on the Catani--Seymour subtraction formalism \cite{Catani:1996vz}.
Within the \krknlo method, the choice of factorisation scale is determined by the NLO matching condition.%
\footnote{\label{footnote:krknlomatching}See \cite{Sarmah:2024hdk} for details of the NLO matching condition and the \krknlo method.}
We therefore consistently adopt the renormalisation and factorisation scales
\begin{align}
	\mur(\Phi_m) = \sqrt{\hat{s}_{12}} & {} \equiv M = \muf(\Phi_m) \\
	\mur(\Phi_{m+1}) & {} = M = \muf(\Phi_{m+1}),
\end{align}
for \krknlo, \mcatnlo and the fixed-order NLO calculations,
where $M$ is the invariant mass of the $m$-particle colourless final state system which defines the process at leading-order,
and $\Phi_m$ is the $m$-particle Born phase-space (and $\Phi_{m+1}$ the real-emission phase-space).%
\footnote{To avoid cluttering the plots, we do not consider scale-variations, which like matching-uncertainty is formally-NNLO.
          Indicative estimates for the scale-uncertainty of the processes considered here are available in the
          literature cited for each process.}

Within the parton shower we use the scale $\mu^{(\alpha)} = \lVert \mathbf{k}_\rT^{(\alpha)} \rVert$,
the transverse-momentum of the generated splitting relative to the emitter-spectator dipole,
for both the ratios of PDFs and for the running of $\alphas$.

For \mcatnlo we consider several comparators with alternative choices of shower starting scale:
\begin{itemize}
	\item a `power-shower' with $Q(\Phi_m) = Q_{\max{}}(\Phi_m)$ and $Q(\Phi_{m+1}) = Q_{\max{}}(\Phi_{m+1})$;%
    \footnote{
Note that the `power-shower' choice for \mcatnlo is generally not recommended \cite{Bellm:2016rhh},
and is included to enable a direct comparison with the power-shower choice required by \krknlo.
For \mcatnlo, the power-shower provides a strict upper-bound for the shower starting-scale and so corresponds
to one edge of the uncertainty envelope.}
	\item a `default' shower with $Q(\Phi_m) = \sqrt{\hat{s}_{12}} \equiv M$ and $Q(\Phi_{m+1}) = \ptj{1}$;
	\item a `DGLAP-inspired' choice in which the shower starting-scale consistently matches the factorisation scale, here $Q(\Phi_m) = M$ and $Q(\Phi_{m+1}) = M$.
\end{itemize}
For the \krknlo method, the shower starting-scale is fixed to $Q_{\max{}} (\Phi_m)$, namely
the largest scale kinematically-accessible from the Born phase-space,
which is required to populate the full real-emission phase-space 
and thus to satisfy the NLO matching condition.\cref{footnote:krknlomatching}

Throughout we use \texttt{CT18NLO} PDFs \cite{Hou:2019efy}, either in the \msbar scheme
or transformed into the \krk scheme as described in \cite{Sarmah:2024hdk}.
Accordingly, we adopt $\alphas(M_Z) = 0.118$ as the input to the running of the strong coupling
throughout the hard process, shower, and the \krknlo code.
We use loose generator cuts, summarised in \cref{tab:generator_cuts}, to avoid excluding regions of phase-space
that would contribute to the fiducial region after showering.
We use the \herwigseven default dipole-shower cutoff scale $\ptcut = 1 \, \GeV$.
Within \herwig we disable both hadronisation and the \texttt{RemnantDecayer},
so the final-state of the hard-process is the only source of final-state QCD partons,
and the hard-process is the only input into the parton shower initial-conditions.

In contrast with the results for the diphoton process studied in \cite{Sarmah:2024hdk},
in this work we do not include the $gg$-channel loop-induced diagrams (formally NNLO).%
\footnote{These may be expected to make a difference to the overall
level of numerical agreement with experimental data,
but not to the differences between the alternative matching schemes.}
For each process we state the approximate magnitude of the missing NNLO corections
in \cref{sec:LHCpheno}.

\begin{table}[tp]
	\centering
	\begin{tabular}{ccc}
		\toprule
		Process & Centre-of-mass energy $\sqrt{s}$ & Generation cuts \\
		\midrule
		$p p \rightarrow W \rightarrow \ell \nu$ 
		& 7 \TeV 
		& \makecell[c]{
			$ p_{\rT}^{\ell, \nu} > 
			\begin{cases}
				15\,\GeV & \mcatnlo \\ 1\,\GeV & \krknlo
			\end{cases}$
		  \\ $M_{\ell \nu} > 45 \,\GeV$
		  }
		\\ \addlinespace[0.5cm]
		$p p \rightarrow Z \gamma \rightarrow \ell \ell \gamma$ 
		& 13 \TeV
		& \makecell[c]{
			$ p_{\rT}^{\ell} > 
				\begin{cases}
					5\,\GeV & \mcatnlo \\ 5\,\GeV & \krknlo
				\end{cases}$ \\
			$ p_{\rT}^{\gamma} > 
				\begin{cases}
					10\,\GeV & \mcatnlo \\ 5\,\GeV & \krknlo
				\end{cases}$ 
			\\ $ M_{\ell \ell} > 35 \,\GeV$ }
	 	\\ \addlinespace[0.5cm]
		$p p \rightarrow W W \rightarrow e \mu \nu_e \nu_\mu$
		& 13 \TeV
		& \makecell[c]{
			$\ptofp{\ell,\nu} > 1 \,\GeV,$ 
			\\ $\absyof{\ell,\nu} < 25$ }
		\\ \addlinespace[0.5cm]
		$p p \rightarrow Z Z \rightarrow \ell \ell \ell \ell$ 
		& 13 \TeV
		& \makecell[c]{
			$\ptofp{\ell,\nu} > 1 \,\GeV, $ 
			\\ $ \absyof{\ell,\nu} < 25 $ }
		\\ \bottomrule
	\end{tabular}
	\caption{Energy scale and generation cuts applied for each process.}
	\label{tab:generator_cuts}
\end{table}

In this section we perform a systematic
comparison of the matching schemes under consideration,
between
\krknlo and the three variants of \mcatnlo outlined in
\cref{sec:matching}
(`power-shower', `default', and `DGLAP').
We make two comparisons between matching schemes, which we label `first-emission' and `full-shower',
in \cref{sec:matching_1em} and \cref{sec:matching_full} respectively.

For `first-emission' comparisons, we consistently truncate the shower
in the $\Phi_{m+1}$ phase-space.
For \mcatnlo, this corresponds to `H'-events with no shower emissions,
and `S'-events with at most one shower emission.
At this point, the matching between the hard-process and the shower is complete,
and the subsequent evolution of each event is handled entirely by the parton shower
algorithm, which is the same in both cases.%
\footnote{This is subject to the caveat that the shower starting-scale for the
	first `post-matching' emission, $Q(\Phi_{m+1})$ within \mcatnlo,
	may not match the starting-scale for the second-emission continuation of the shower within \krknlo.}
In the case of the \krknlo method,
the soft--virtual reweight is not applied to the one-emission events.
    
For `full-shower' comparisons, we allow the shower to run
to its final cutoff scale, fully populating the multiple-emission phase-space.
This gives the full matched prediction,
which we further use to perform the comparisons with experimental data from the LHC
in \cref{sec:LHCpheno}.

We use generator cuts as given in
\cref{tab:generator_cuts}
and apply analysis cuts as summarised in
\cref{tab:analysis_cuts}.
These have been chosen to give acceptance regions with
similar kinematics across the different processes.
For the $Z\gamma$ process, in place of experimental photon isolation we use
smooth-cone (`Frixione') isolation \cite{Frixione:1998jh}
with the `tight' isolation parameters from the 2013 Les Houches Accords \cite{Andersen:2014efa},
which corresponds to
\begin{align}
	\chi(r; R) 
	= \left(  \frac{1 - \cos r}{1 - \cos R}  \right)
	\equiv \left( \frac{\sin \frac{1}{2} r}{\sin \frac{1}{2} R}  \right)^2,
\end{align}
where $\Etiso (r)$ is the cumulative transverse isolation energy within 
(rapidity-azimuth) radius $r$,
calculated
as the transverse magnitude of the total momentum of all
non-photon particles within a cone of radius $r$.

Where jet distributions are shown, jets are identified using the
anti-$\kt$ algorithm \cite{Cacciari:2008gp}
with a jet clustering radius of 0.4,
a $\pt$-cut of 1 GeV and a pseudo-rapidity cut $\lvert \eta \rvert < 4.5$.

\subsection{First-emission only}
\label{sec:matching_1em}

\begin{table}[tp]
	\centering
	\begin{tabular}{cc}
		\toprule
		Process & Analysis cuts \\
		\midrule
		$p p \rightarrow W \rightarrow \ell \nu$
		& \makecell[cc]{
			$\ptell{} > 30 \, \GeV $, $ \ptnu{} > 25 \, \GeV$ \\ 
			$\vert y^{\ell,\nu} \vert < 5$ \\ 
            $\Mlnu > 50 \;\GeV$ 
		}
		\\ \addlinespace[0.5cm]
		$p p \rightarrow Z \gamma \rightarrow \ell \ell \gamma$ 
		& \makecell[c]{
			 $\ptell{1} > 30 \,\GeV $, 
			 $\ptell{2} > 25 \,\GeV $,
			 $ \ptg{} > 30 \,\GeV$ \\
			 $ \vert y^{\mathrm{\ell,\gamma}} \vert < 2.5 $ \\
			 $\Mll > 40 \,\GeV$ \\
			 $\dRlg > 0.4 $ \\
			 $\Etiso (r) < 0.1 \, \ptg{} \, \chi(r; R) \;\text{within cone } r \leqslant R = 0.2$ 
	 	}
	 	\\ \addlinespace[0.5cm]
		$p p \rightarrow W W \rightarrow e \mu \nu_e \nu_\mu$ 
		& \makecell[c]{
			$\ptell{} > 30 \,\GeV, $ \\
			$\absyell < 2.5 $ \\
			$\pt^{\text{miss}} > 20 \,\GeV $ \\
			$ M_{\ell\nu} \in [60, 100] \,\GeV $
		}
	    \\ \addlinespace[0.5cm]
		$\qquad p p \rightarrow Z Z \rightarrow e^+ e^- \mu^+ \mu^-$%
		\tablefootnote{For the results presented in this section, we avoid issues related to the possible ambiguity
        of $ZZ$-reconstruction by restricting only to final-states containing different-flavour lepton-pairs.}
		$\qquad$
		&
		\makecell[c]{
			$\ptell{} > 30 \,\GeV $ \\
			$\absyell < 2.5 $ \\
			$ M_{\ell\ell} \in [65, 115] \,\GeV$
		}
		\\ \bottomrule
	\end{tabular}
	\caption{Summary of fiducial cuts applied for each process for the matching study in
    \cref{sec:matching}.
    The boson reconstruction from the leptonic final-state is performed at truth-level.}
	\label{tab:analysis_cuts}
\end{table}

Differential cross-sections for
the invariant mass of the colour-singlet system,
and the azimuthal angle and rapidity separation between the constituents of the colour-singlet system
(the two bosons for diboson processes; lepton and neutrino for $W$)
are shown in \cref{fig:observables_processes_oneemission}.
These observables have been chosen to characterise the imputed underlying two-particle final-state
of the Born-level $2\to 2$ hard-process, prior to subsequent boson decay or additional QCD radiation.

\begin{figure}[t]
	\centering
        \includegraphics[width=.32\textwidth]{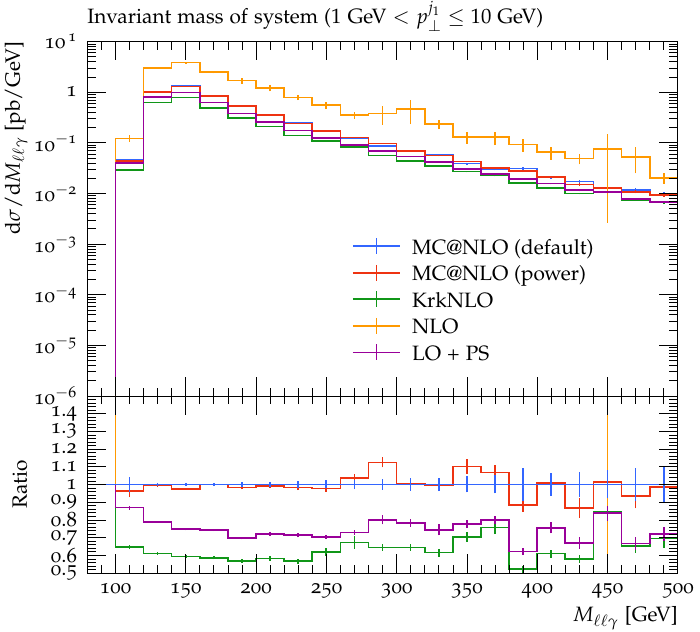}
        \includegraphics[width=.32\textwidth]{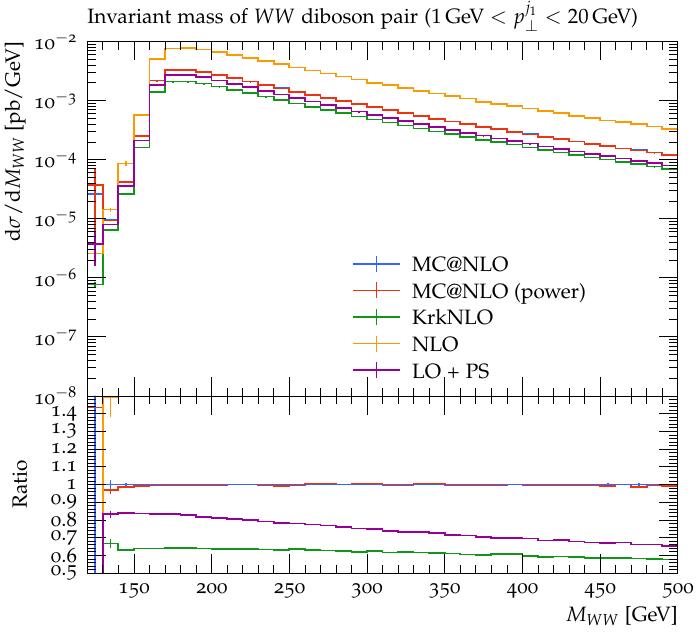}
        \includegraphics[width=.32\textwidth]{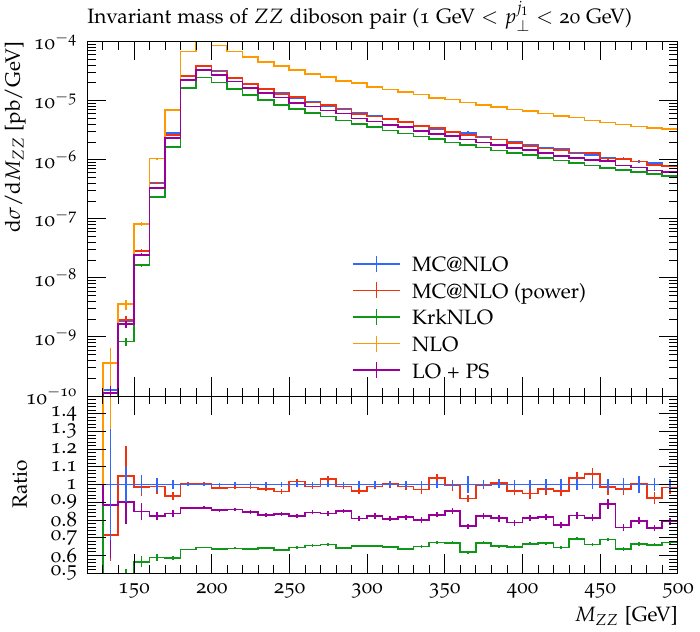}
        \begin{subfigure}[b]{.32\textwidth}
        \includegraphics[width=\textwidth]{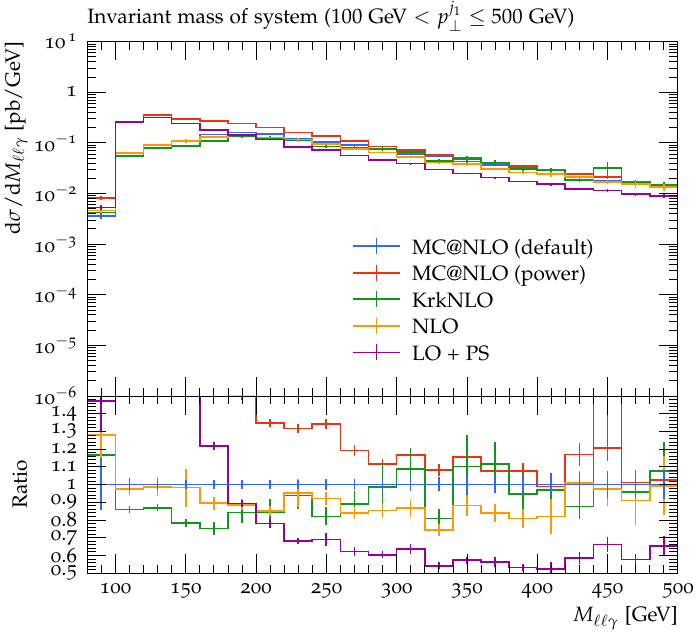}
        \caption{$Z \gamma$}
        \end{subfigure}
        \begin{subfigure}[b]{.32\textwidth}
        \includegraphics[width=\textwidth]{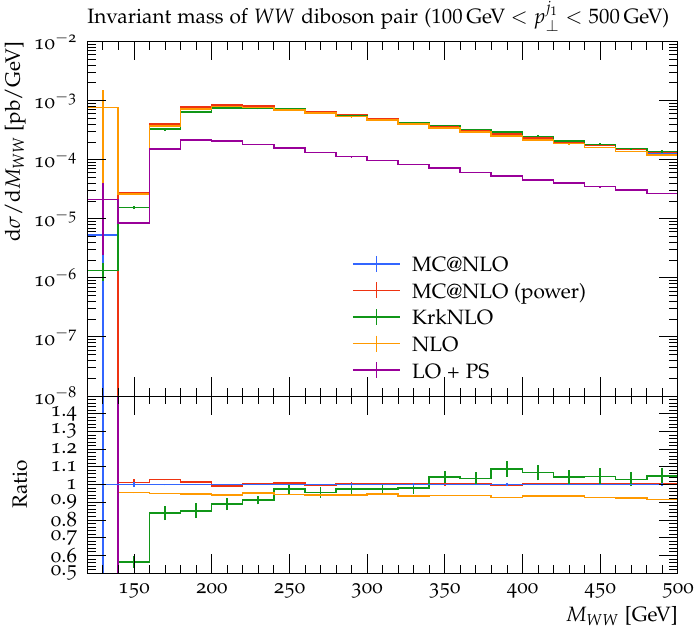}
        \caption{$WW$}
        \end{subfigure}
        \begin{subfigure}[b]{.32\textwidth}
        \includegraphics[width=\textwidth]{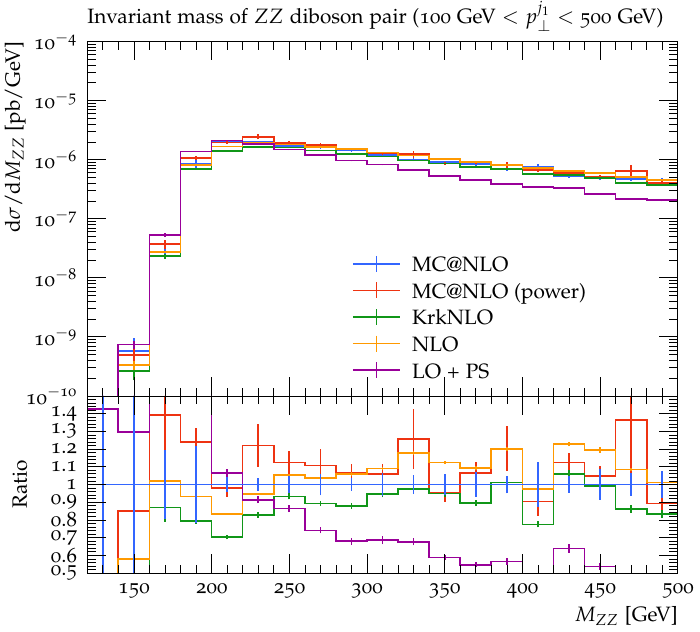}
        \caption{$ZZ$}
        \end{subfigure}
	\caption{`Parton-level' (first-emission) comparison of the invariant mass of the
		 colour-singlet system in `soft' and `hard' slices of the transverse momentum
		 of the leading jet.
     The full-shower counterpart is shown in \cref{fig:fullshower_dsigma_dM_dptj}.
     The complete array of double-differential slices is shown in \cref{fig:oneemission_dsigma_dM_dptj_full}.
    \label{fig:oneemission_dsigma_dM_dptj}}
\end{figure}

Differential cross-sections for the transverse momentum
and the rapidity of the leading jet are shown in \cref{fig:observables_processes_oneemission_j}.
To clarify the roles of the QCD radiation and corresponding Sudakov factors
across phase-space,
differential cross-sections for the invariant mass of the colour-singlet system
in slices of the transverse momentum of the leading jet
are shown in \cref{fig:oneemission_dsigma_dM_dptj}.%
\footnote{Throughout we will refer to such distributions as `double-differential';
note, however, that for ease of comparability between figures,
they have not been normalised to the bin-width of the second observable.
They are therefore single-differential distributions, `in slices of'
the second observable.}

\afterpage{
\begin{landscape}
\begin{figure}
    \vspace{-20mm}
	\centering
        \makebox[\linewidth][c]{%
        \includegraphics[width=.24\linewidth]{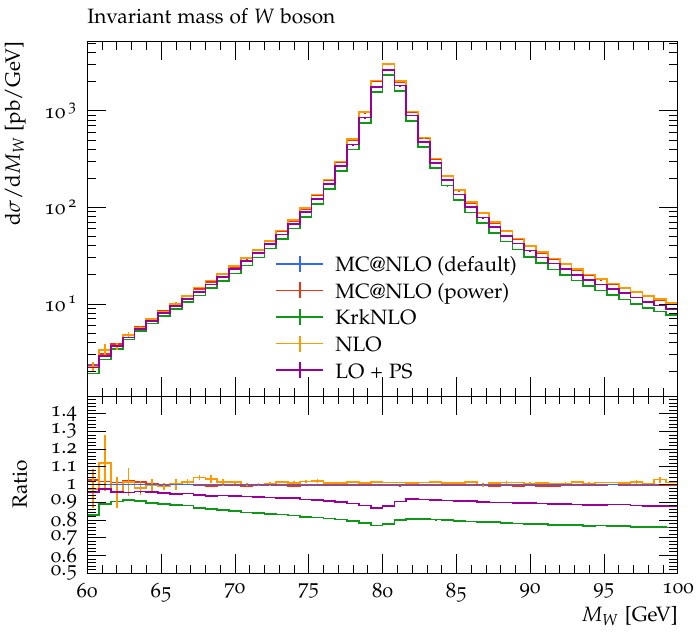}
        \includegraphics[width=.24\linewidth]{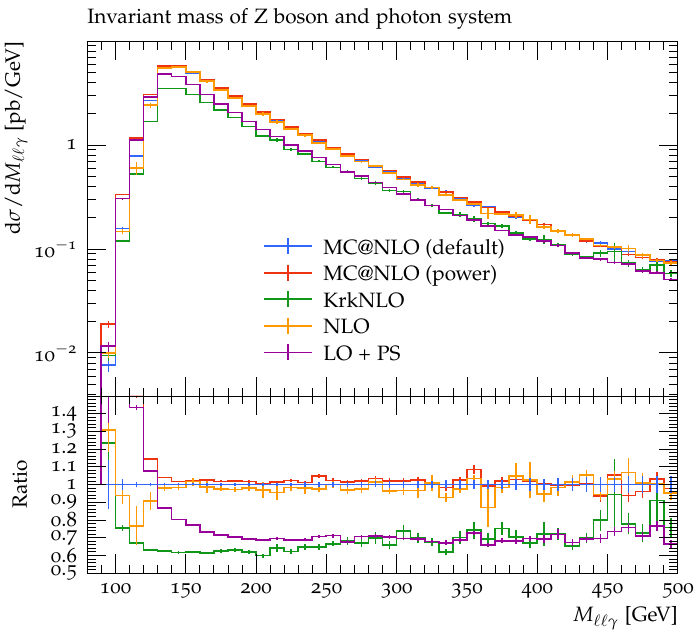}
        \includegraphics[width=.24\linewidth]{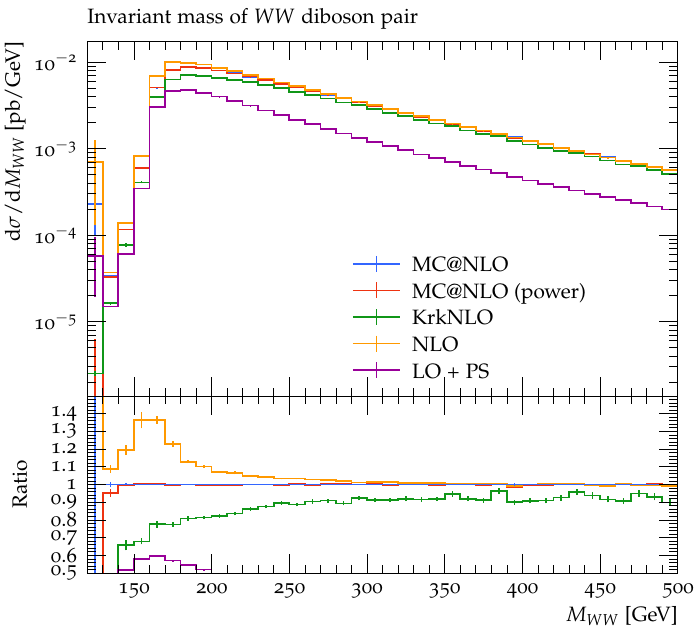}
        \includegraphics[width=.24\linewidth]{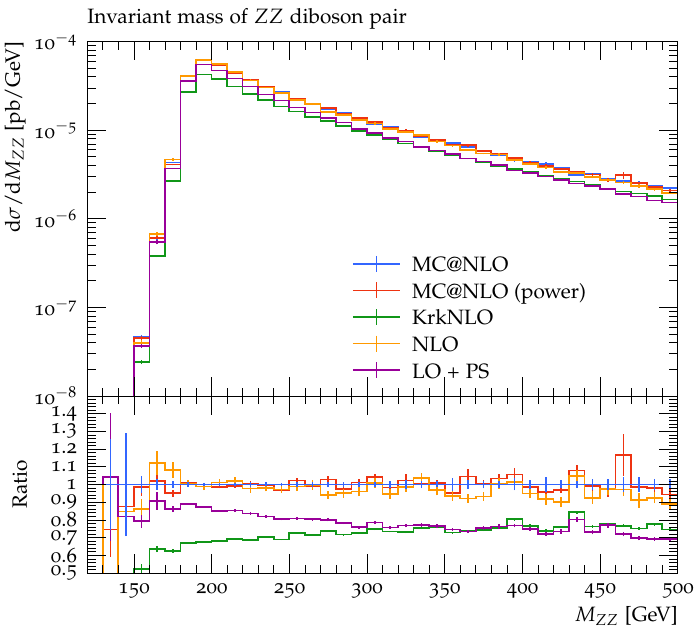}
        }
        \makebox[\linewidth][c]{%
        \includegraphics[width=.24\linewidth]{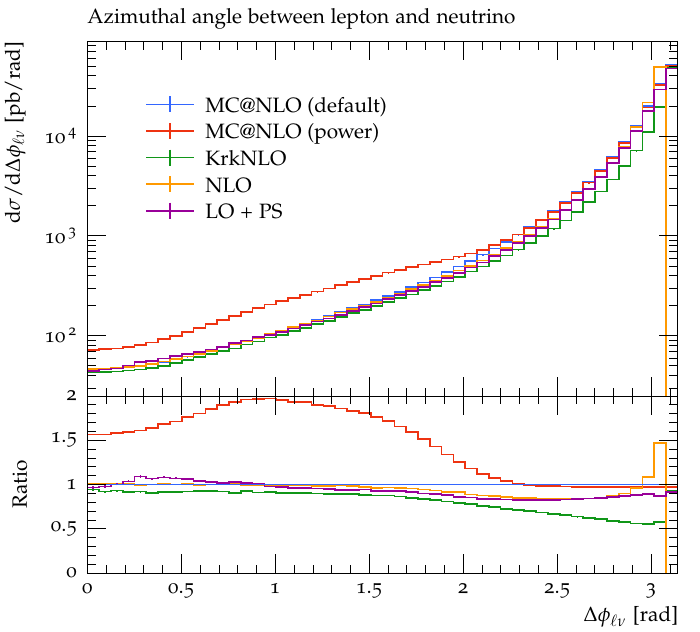}
        \includegraphics[width=.24\linewidth]{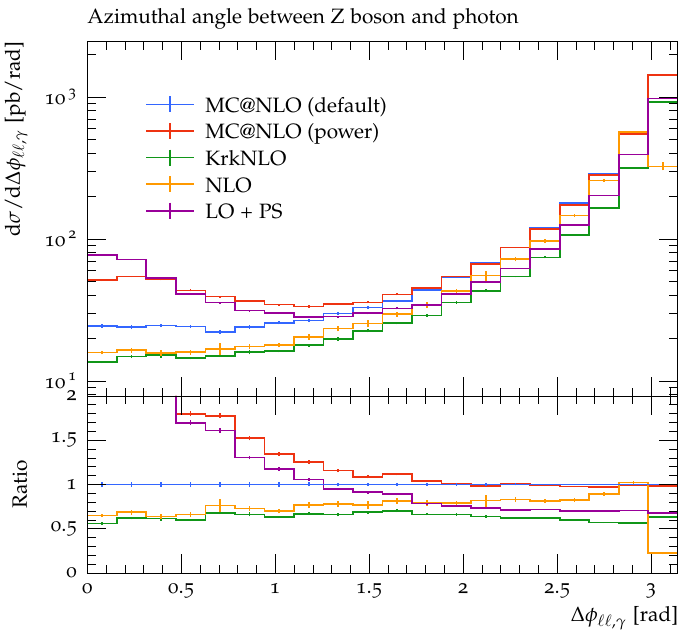}
        \includegraphics[width=.24\linewidth]{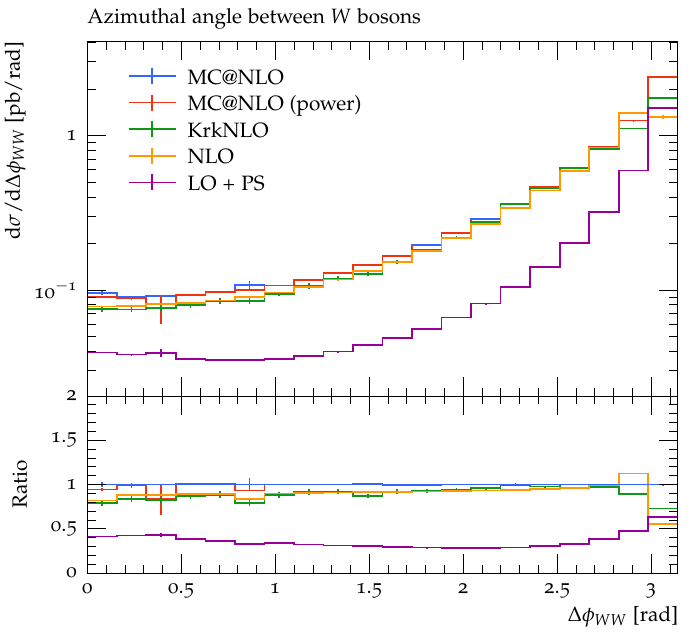}
        \includegraphics[width=.24\linewidth]{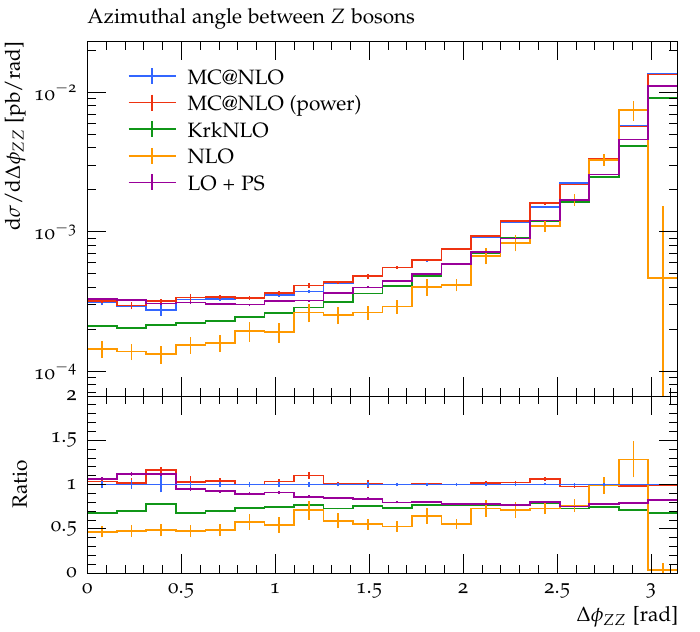}
        }
        \makebox[\linewidth][c]{%
        \begin{subfigure}[b]{.24\linewidth}
        \includegraphics[width=\linewidth]{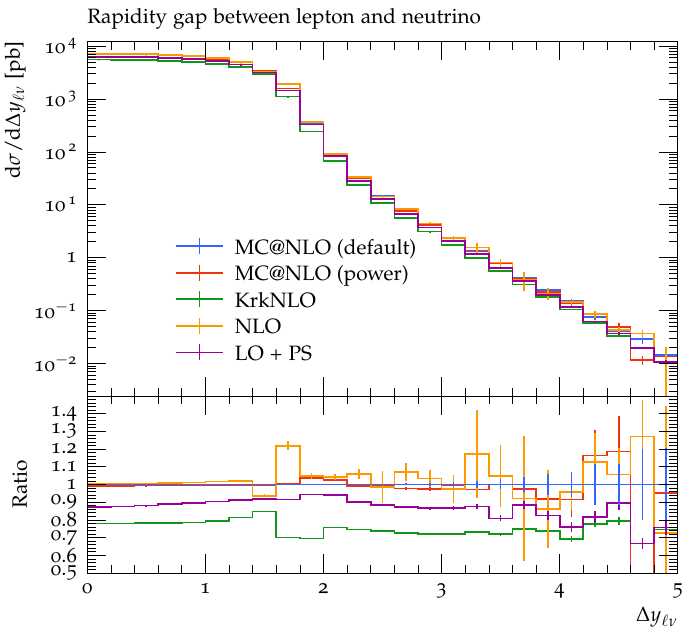}
        \caption{$W$}
        \end{subfigure}
        \begin{subfigure}[b]{.24\linewidth}
        \includegraphics[width=\linewidth]{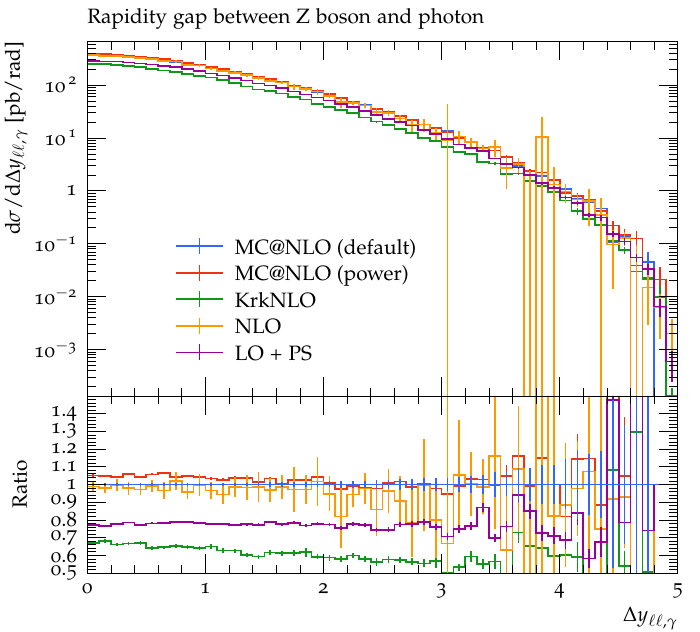}
        \caption{$Z\gamma$}
        \end{subfigure}
        \begin{subfigure}[b]{.24\linewidth}
        \includegraphics[width=\linewidth]{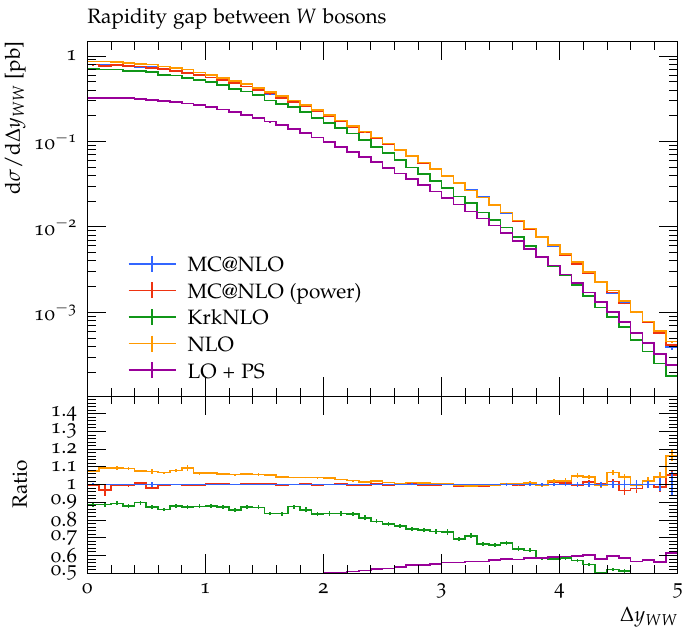}
        \caption{$WW$}
        \end{subfigure}
        \begin{subfigure}[b]{.24\linewidth}
        \includegraphics[width=\linewidth]{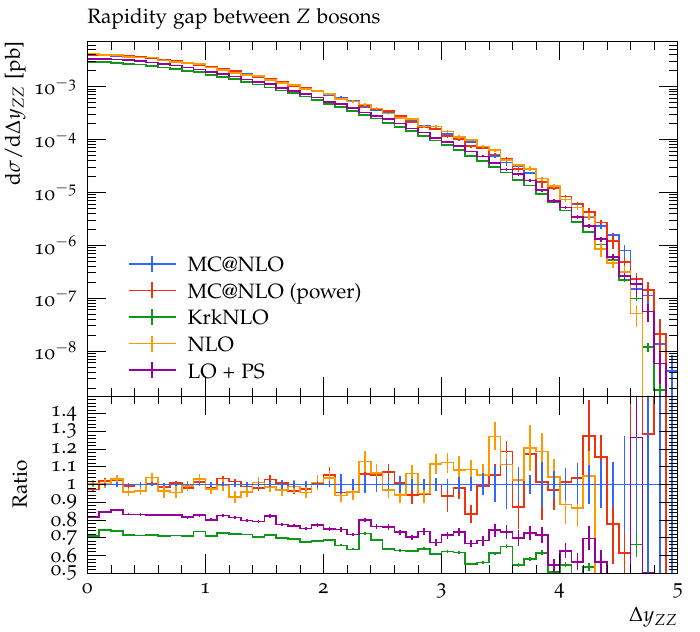}
        \caption{$ZZ$}
        \end{subfigure}
        }
	\caption{`Parton level' (first-emission) comparison of KrkNLO with MC@NLO, NLO fixed-order,
            and the corresponding first-emission distributions generated by the parton shower from a leading-order calculation.}
    \label{fig:observables_processes_oneemission}
    \vspace{-10mm}
\end{figure}
\end{landscape}
}

\afterpage{
\begin{landscape}
\begin{figure}
    \vspace{-10mm}
	\centering
        \makebox[\linewidth][c]{%
        \includegraphics[width=.24\linewidth]{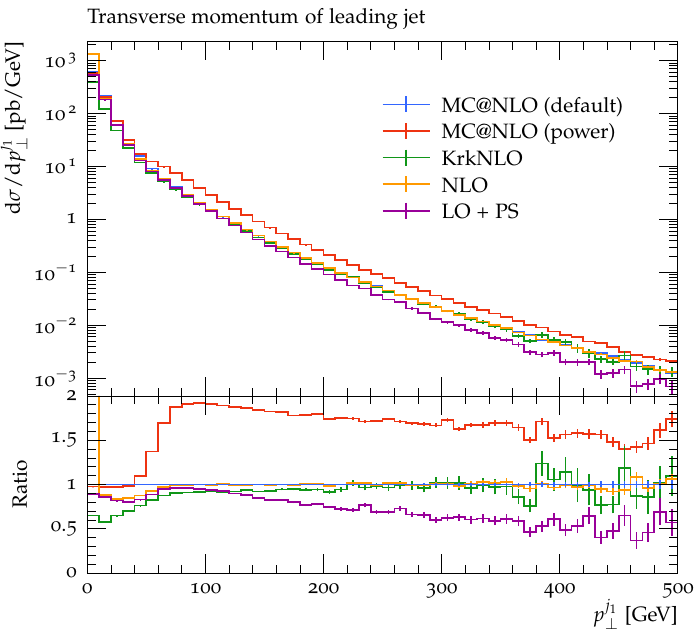}
        \includegraphics[width=.24\linewidth]{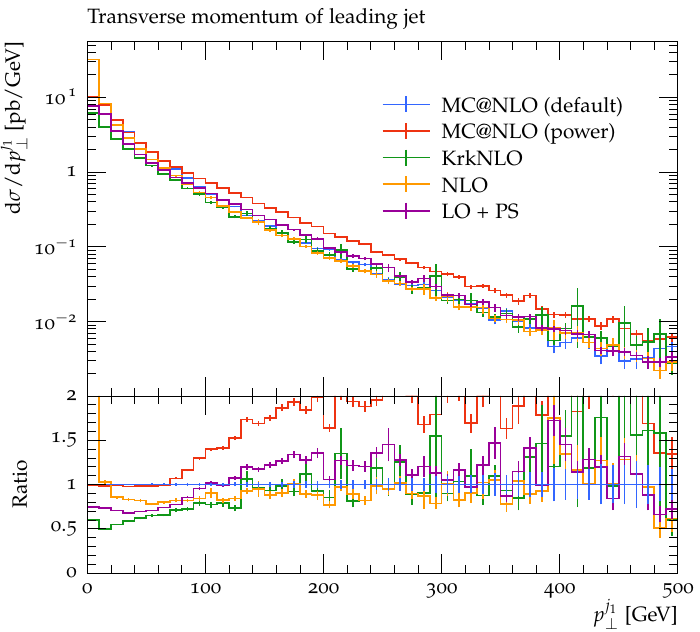}
        \includegraphics[width=.24\linewidth]{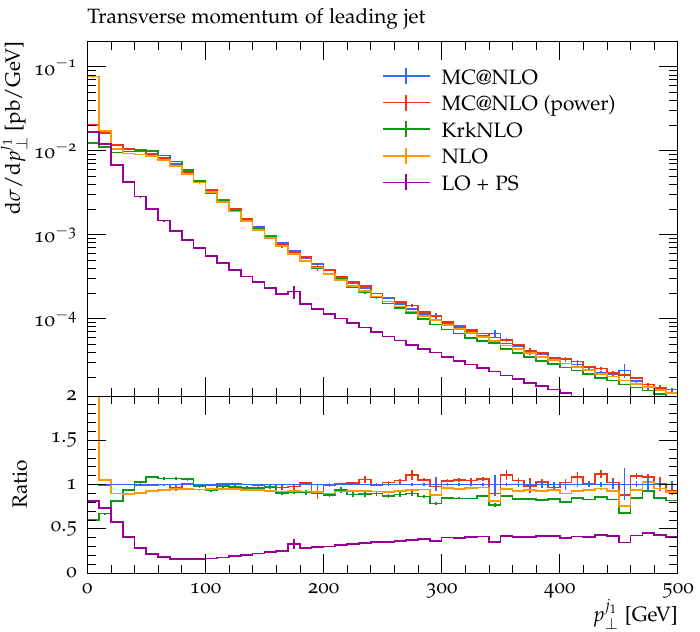}
        \includegraphics[width=.24\linewidth]{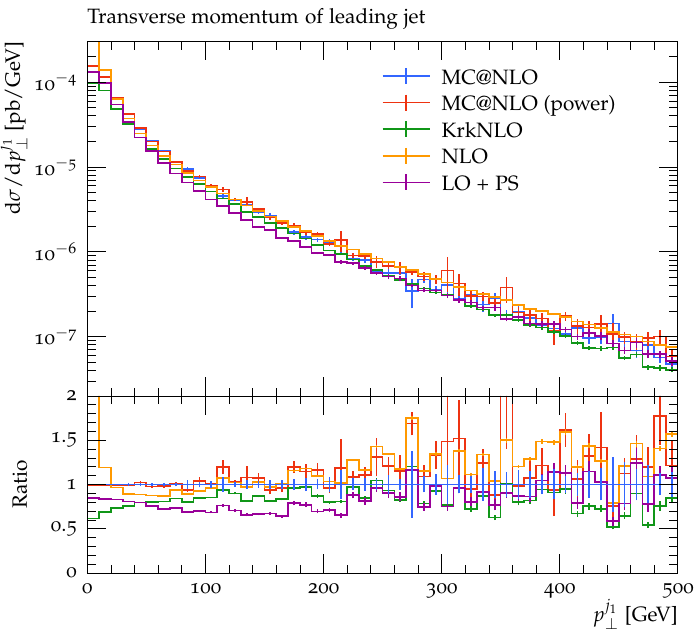}
        }
        \makebox[\linewidth][c]{%
        \begin{subfigure}[b]{.24\linewidth}
        \includegraphics[width=\linewidth]{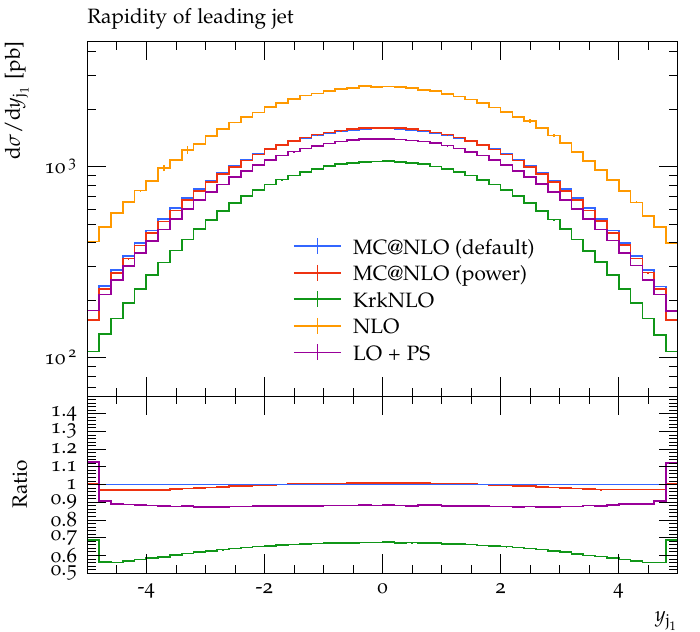}
        \caption{$W$}
        \end{subfigure}
        \begin{subfigure}[b]{.24\linewidth}
        \includegraphics[width=\linewidth]{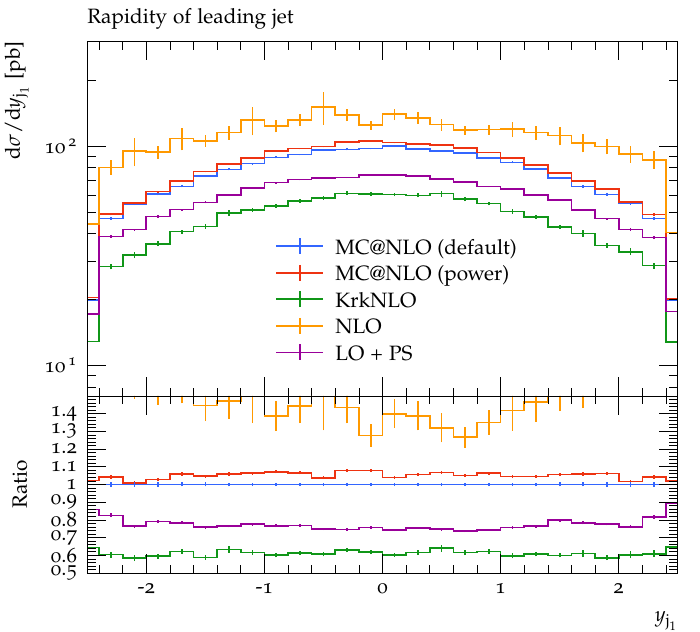}
        \caption{$Z\gamma$}
        \end{subfigure}
        \begin{subfigure}[b]{.24\linewidth}
        \includegraphics[width=\linewidth]{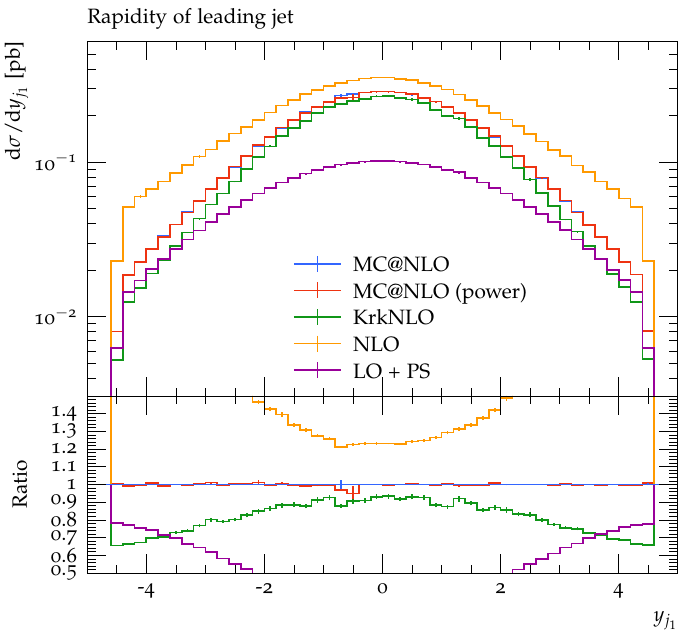}
        \caption{$WW$}
        \end{subfigure}
        \begin{subfigure}[b]{.24\linewidth}
        \includegraphics[width=\linewidth]{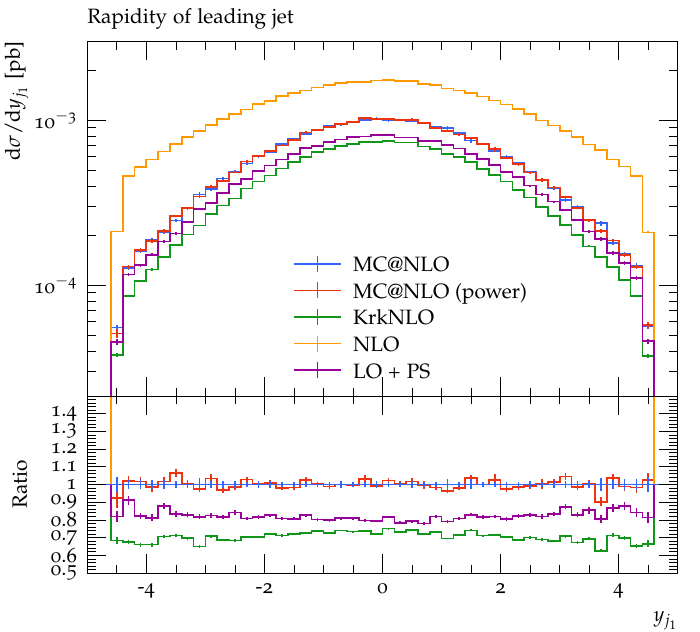}
        \caption{$ZZ$}
        \end{subfigure}
        }
	\caption{`Parton level' (first-emission) comparison of 
            the QCD radiation arising from
            KrkNLO with MC@NLO, NLO fixed-order,
            and the first-emission generated by the parton shower from a leading-order calculation.
            Note that here, due to the multiplicity of the final-state, there is at most one jet, and it comprises a single parton.
    \label{fig:observables_processes_oneemission_j}}
\end{figure}
\end{landscape}
}

The transverse momentum of the leading jet, shown in \cref{fig:observables_processes_oneemission_j},
is consistently-described at high-$\pt$ between fixed-order NLO, the \krknlo method and \mcatnlo (default).
This is largely as expected, due to the role of the real-emission
matrix-element in providing (leading-order) perturbative accuracy in this region.
For processes with a massless particle in the $2 \to 2$ final-state, the matching
uncertainty within the \mcatnlo method due to the choice of shower starting-scale
is significant.%
\footnote{This difference is attributable to `S'-events in which the transverse
momentum of the first-emission exceeds the invariant mass of the colour-singlet system,
as can be seen from \cref{fig:oneemission_dsigma_dM_dptj}.}
From the double-differential plots of \cref{fig:oneemission_dsigma_dM_dptj},
we see that the agreement in this high-$\pt$ region is
also reached double-differentially in invariant mass, save for
$Z\gamma$ which deviates only for the \mcatnlo power-shower prediction 
(which agrees with the showered leading-order prediction) at low-$M$.

At low-$\ptj{}$, the Sudakov factor $\Delta \bigr\vert_{\ptj{1}}^{Q(\Phi_m)} (\Phi_m)$
can be seen in \cref{fig:observables_processes_oneemission_j,fig:oneemission_dsigma_dM_dptj}
to tame the characteristic divergence of the real-emission matrix element in all cases.
This is the limit in which perturbation theory breaks down and the
all-order resummation provided by the shower (here, truncated to a single emission) becomes relevant.
The variants of $\mcatnlo$ converge here, due to the low sensitivity of
emissions at this low scale to the upper-limit of shower-radiation,
and as a result the \krknlo prediction lies outwith the implied \mcatnlo matching-uncertainty envelope.
This suppression dominates the overall differences seen in the inclusive distributions,
as shown in further detail in \cref{fig:fullshower_dsigma_dM_dptj} and discussed below.
Note that the magnitude of the suppression induced by the Sudakov factor is a function
of the choice of shower cutoff-scale $\ptcut$, here set to 1 GeV.
The difference between the \krknlo and leading-order predictions in this region is attributable
to the use of the \krk-scheme PDFs for the former.

Due to the low jet-identification $\pt$-cut used, and the falling $\dd \sigma / \dd \ptj{}$ distribution,
the $\dd \sigma / \dd y_{j}$ distribution is dominated by the normalisation differences implied by the
discussion of the low-$\ptj{}$ region above.
The \krknlo distribution again lies outside the \mcatnlo matching uncertainty envelope, for the same reason,
due to the relative suppression in the low-$\ptj{1}$ which dominates the distribution.
The shapes of the rapidity distributions are broadly comparable, with the exception of the $WW$ process.

Turning to the inclusive distributions shown in \cref{fig:observables_processes_oneemission},
the differential cross-section with respect to the invariant mass of the colour-singlet system
is dominated by the low-$\pt$ region (as shown in \cref{fig:oneemission_dsigma_dM_dptj})
in which the Sudakov factor and \krk PDFs suppress the \krknlo prediction relative to the others
(see also the discussion of $\dd \sigma / \dd \ptj{1}$ above).%
\footnote{See also the corresponding full-shower plots of \cref{fig:fullshower_dsigma_dM_dptj}
and the accompanying discussion for the effect of the subsequent radiation.}
At intermediate-$\pt$, the lower peak of the invariant-mass distribution for the $Z\gamma$
process allows the difference in shower-scale within the \mcatnlo alternatives
to generate greater radiation below the $\dd \sigma / \dd M_{\ell\ell\gamma}$-peak.
At high-$\pt$ the invariant-mass distributions generally converge between the NLO-accurate
predictions.

Overall, due to the shape of the $\dd \sigma / \dd M$ distributions,
especially for the $ZZ$ and $WW$ processes, there is relatively
limited opportunity for the starting-scale 
difference between the invariant mass of the colour-singlet
system (in the `DGLAP'/`default' case) and the kinematic upper-limit (in the `power' case)
to contribute to matching uncertainty at this level.
The uncertainty envelope obtained by shower-scale variation within the \mcatnlo method is therefore negligible,
for the three choices considered here,
and so underestimates the true matching-uncertainty at one-emission level.

Considering the azimuthal angle between particles within the colour-singlet system,
shown in \cref{fig:observables_processes_oneemission},
the \krknlo method best reproduces the NLO fixed-order distributions away from the
back-to-back limit in which perturbation theory breaks down ($\Delta \phi \to \pi$),
while the effect of the low-$\pt$ suppression of the \krknlo method is again visible
close to the back-to-back region.
The high-$\pt$ effect of the power-shower discussed above,
due to recoil against a hard emission, can again here be seen to affect
the $\Delta \phi$ distribution, generating a large matching-uncertainty
within the \mcatnlo method at low-$\Delta \phi$.

The rapidity-separation, by contrast,
shows good agreement within the \mcatnlo predictions and between the \mcatnlo
predictions and fixed-order NLO.
The low-$\pt$ Sudakov suppression discussed above
again leads to a normalisation difference across the distribution
for the \krknlo method, which increases slightly at large rapidities.

\subsection{Full shower}
\label{sec:matching_full}

\afterpage{
\begin{landscape}
\begin{figure}
	\centering
    \vspace{-15mm}
        \makebox[\linewidth][c]{%
        \includegraphics[width=.24\linewidth]{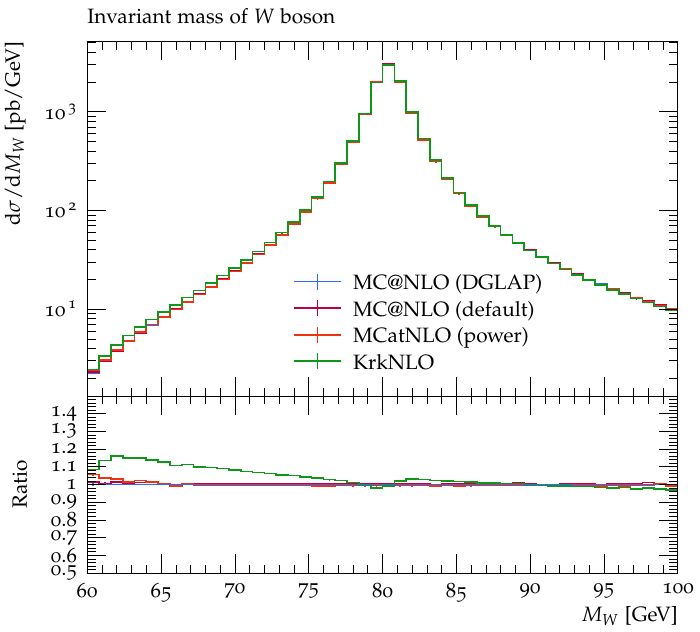}
        \includegraphics[width=.24\linewidth]{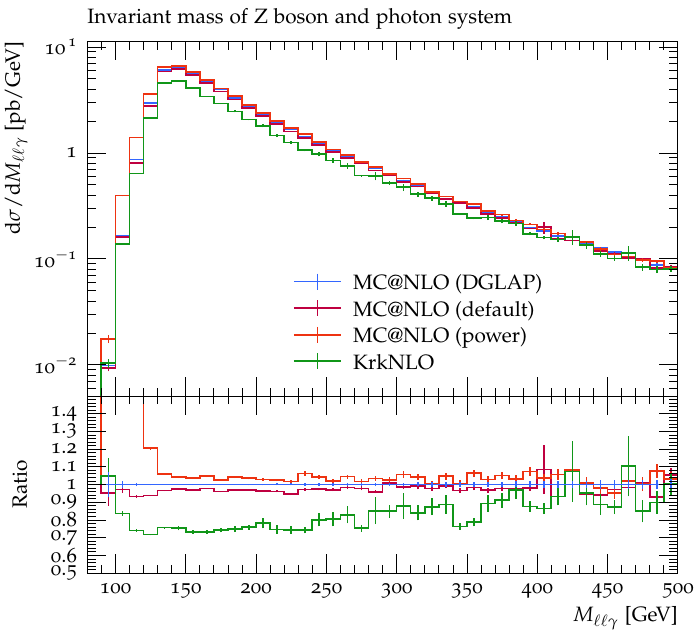}
        \includegraphics[width=.24\linewidth]{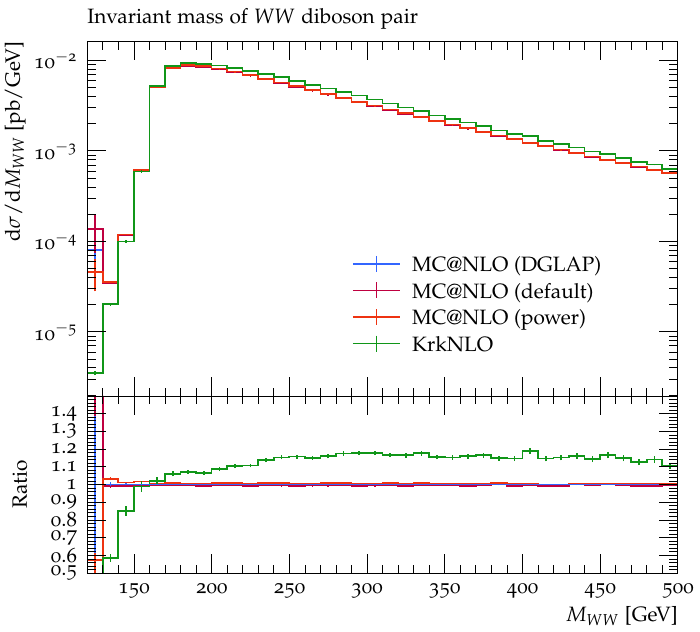}
        \includegraphics[width=.24\linewidth]{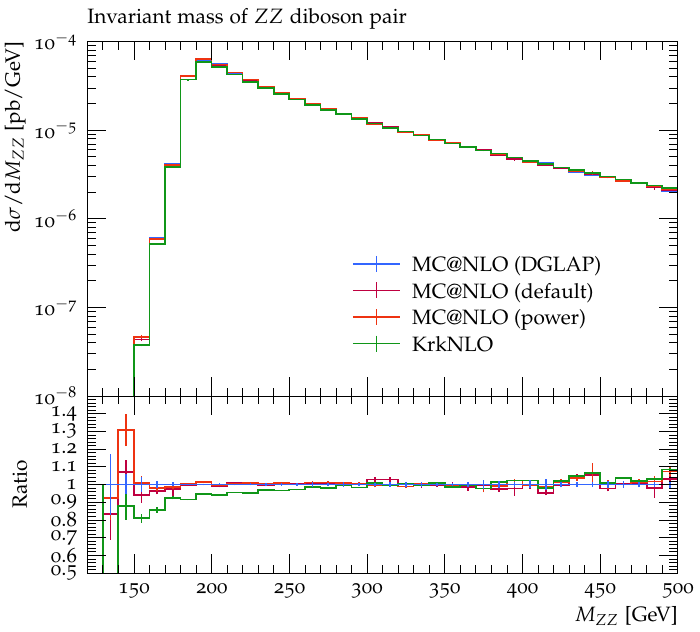}
        }
        \makebox[\linewidth][c]{%
        \includegraphics[width=.24\linewidth]{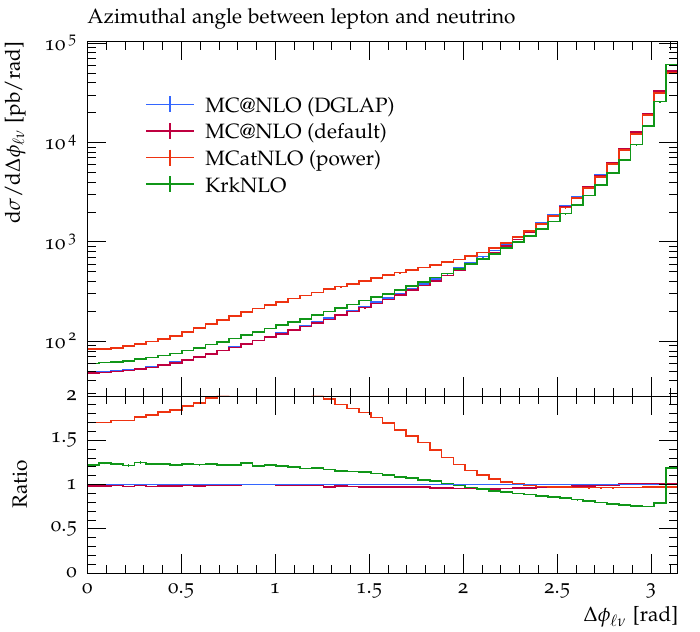}
        \includegraphics[width=.24\linewidth]{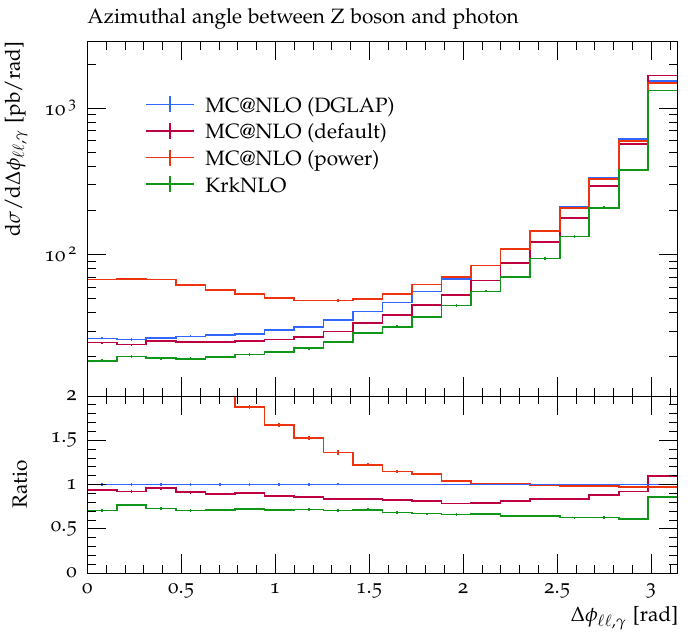}
        \includegraphics[width=.24\linewidth]{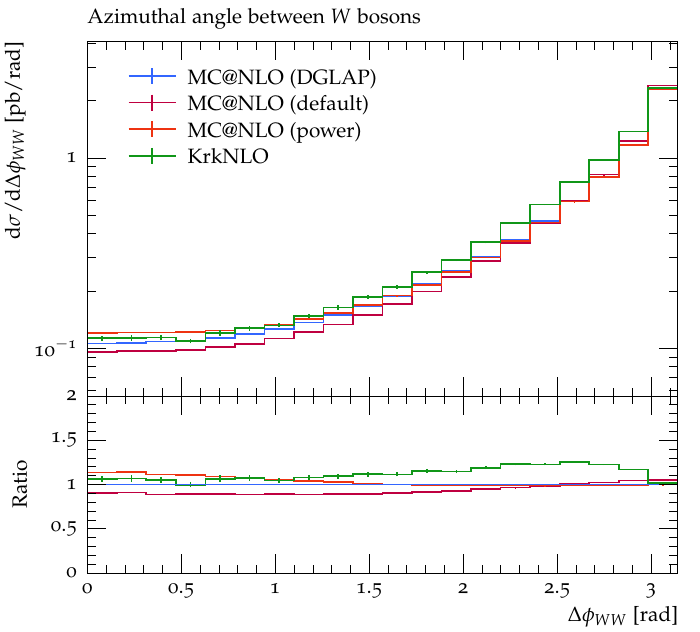}
        \includegraphics[width=.24\linewidth]{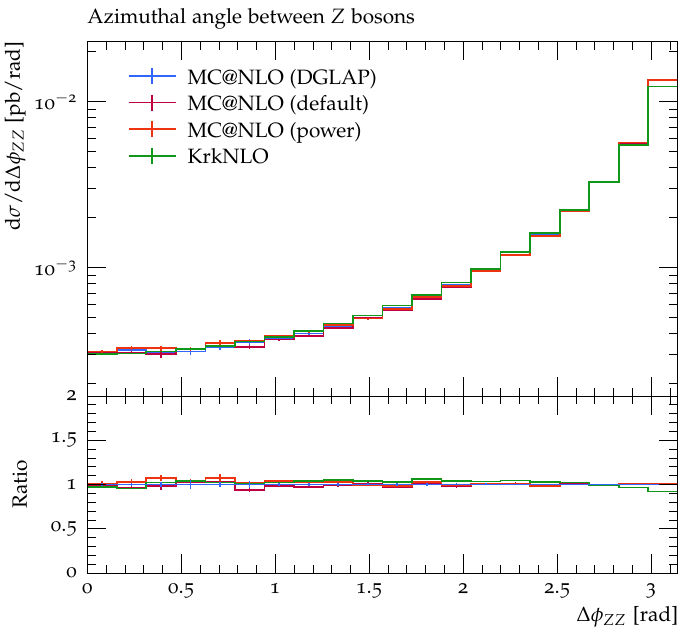}
        }
        \makebox[\linewidth][c]{%
        \begin{subfigure}[b]{.24\linewidth}
        \includegraphics[width=\linewidth]{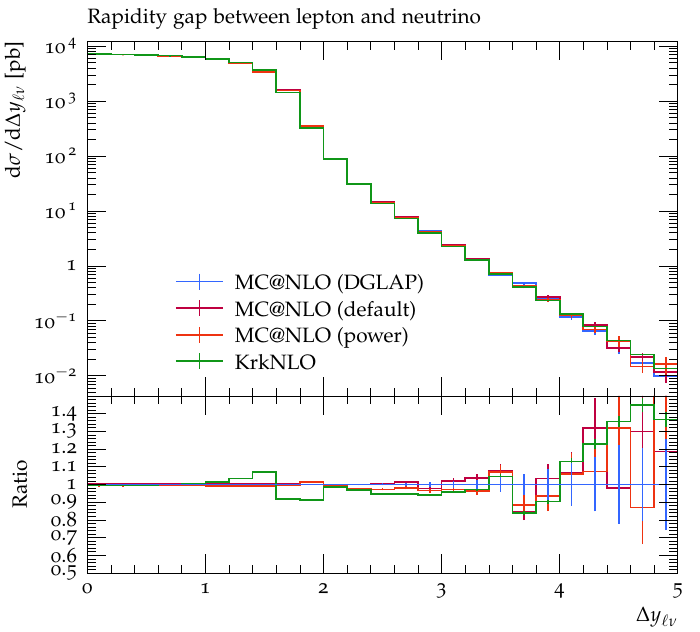}
        \caption{$W$}
        \end{subfigure}
        \begin{subfigure}[b]{.24\linewidth}
        \includegraphics[width=\linewidth]{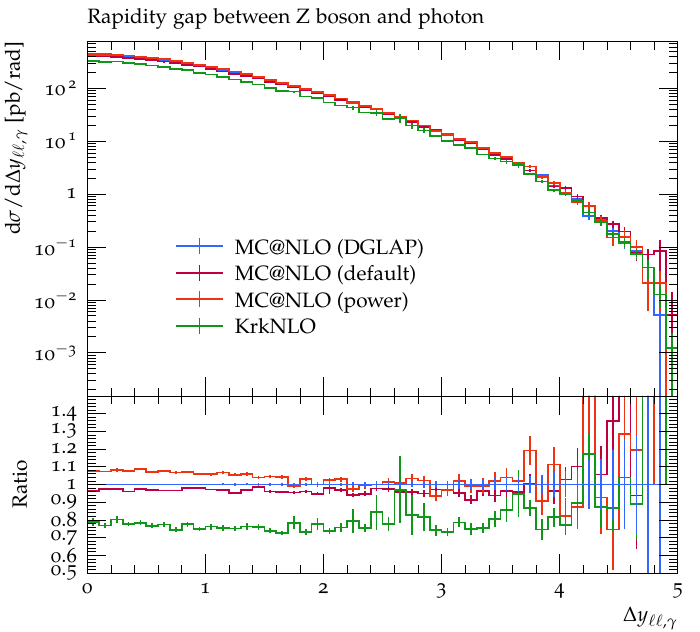}
        \caption{$Z \gamma$}
        \end{subfigure}
        \begin{subfigure}[b]{.24\linewidth}
        \includegraphics[width=\linewidth]{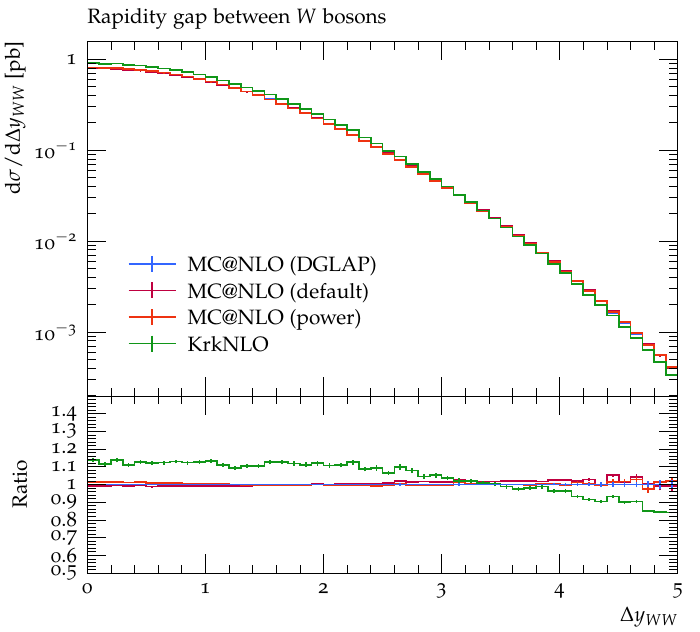}
        \caption{$WW$}
        \end{subfigure}
        \begin{subfigure}[b]{.24\linewidth}
        \includegraphics[width=\linewidth]{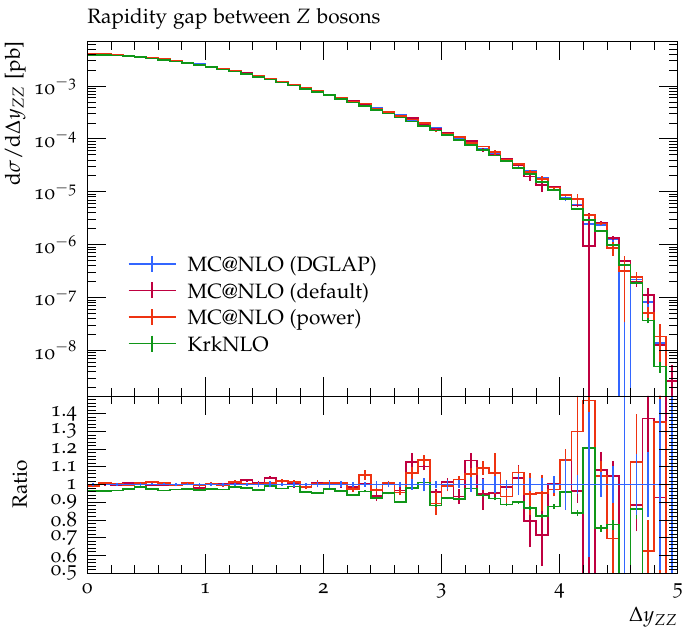}       
        \caption{$ZZ$}
        \end{subfigure}
        }
	\caption{Full-shower comparison of the inclusive distributions.
            The one-emission counterpart is shown in \cref{fig:observables_processes_oneemission}.}
    \label{fig:observables_processes_fullshower}
    \vspace{-10mm}
\end{figure}
\end{landscape}
}

\afterpage{
\begin{landscape}
\begin{figure}
	\centering   
        \vspace{-10mm}
        \makebox[\linewidth][c]{%
        \includegraphics[width=.24\linewidth]{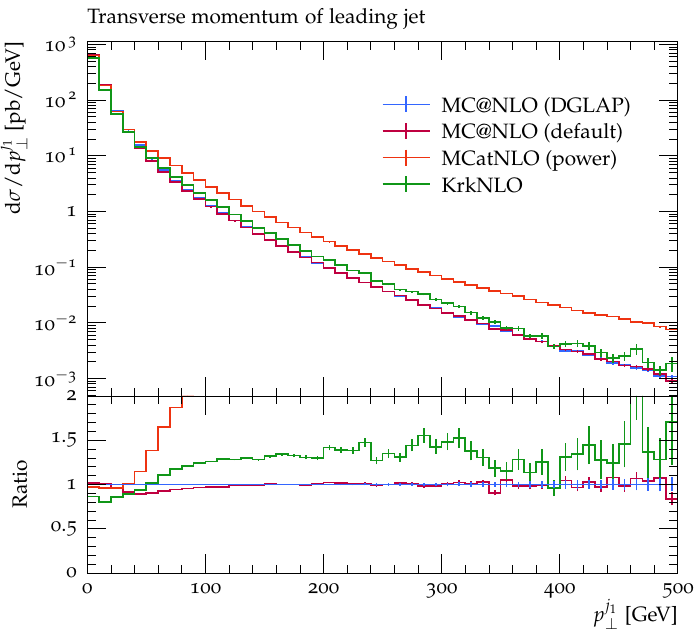}
        \includegraphics[width=.24\linewidth]{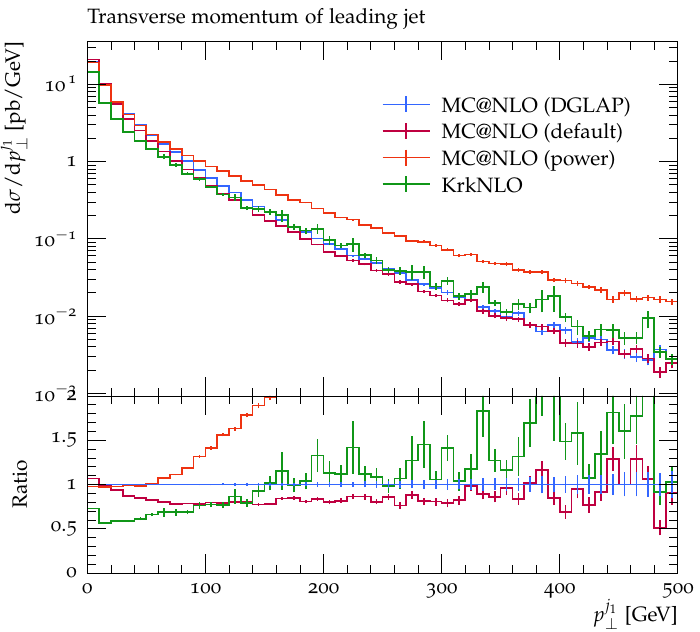}
        \includegraphics[width=.24\linewidth]{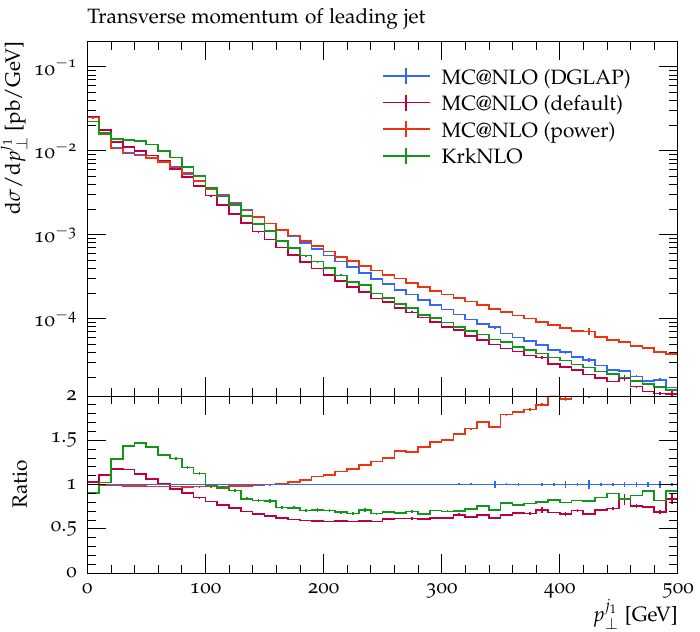}
        \includegraphics[width=.24\linewidth]{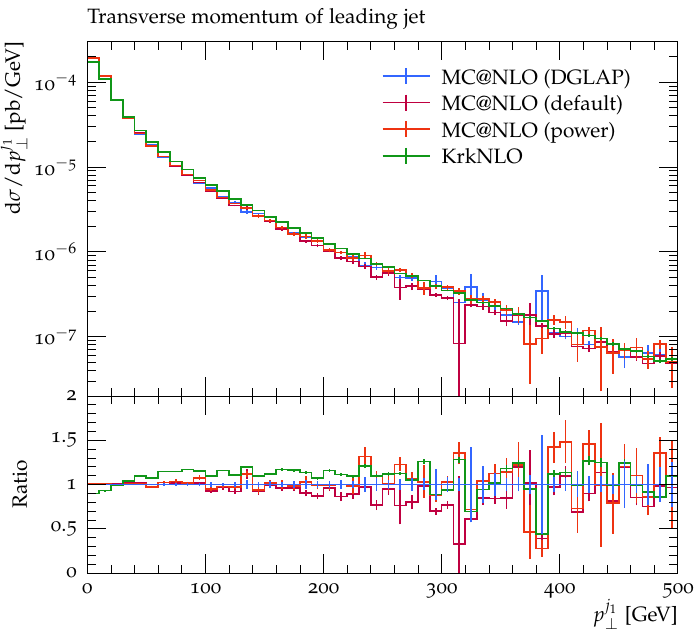}
        }
        \makebox[\linewidth][c]{%
        \begin{subfigure}[b]{.24\linewidth}
        \includegraphics[width=\linewidth]{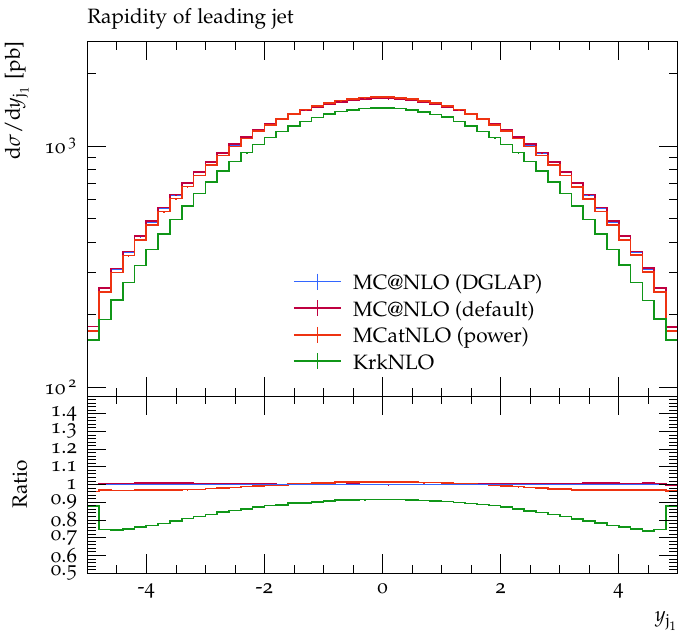}
        \caption{$W$}
        \end{subfigure}
        \begin{subfigure}[b]{.24\linewidth}
        \includegraphics[width=\linewidth]{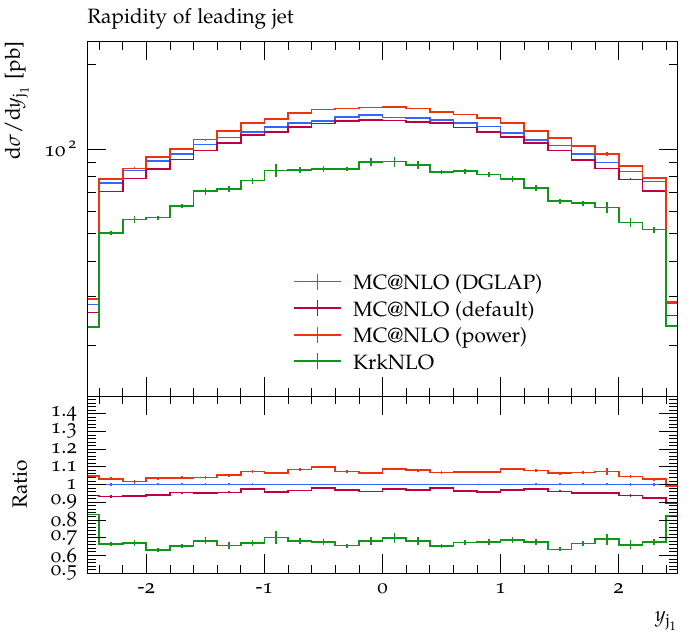}
        \caption{$Z \gamma$}
        \end{subfigure}
        \begin{subfigure}[b]{.24\linewidth}
        \includegraphics[width=\linewidth]{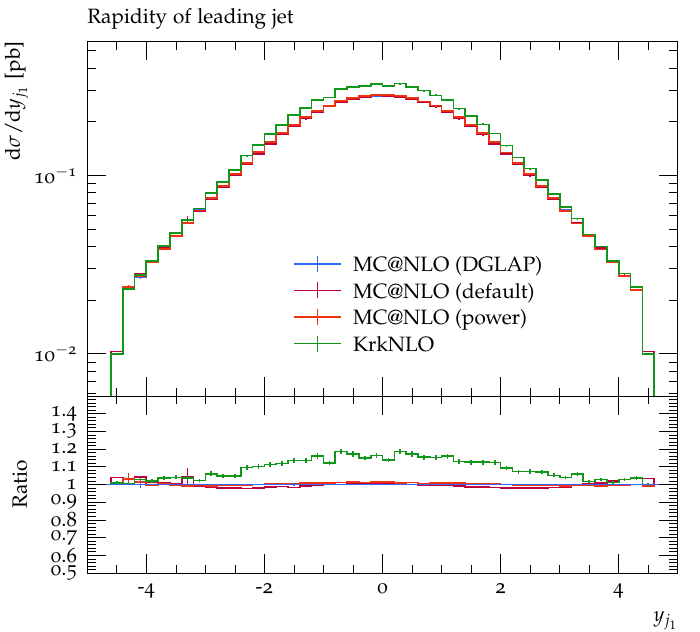}
        \caption{$WW$}
        \end{subfigure}
        \begin{subfigure}[b]{.24\linewidth}
        \includegraphics[width=\linewidth]{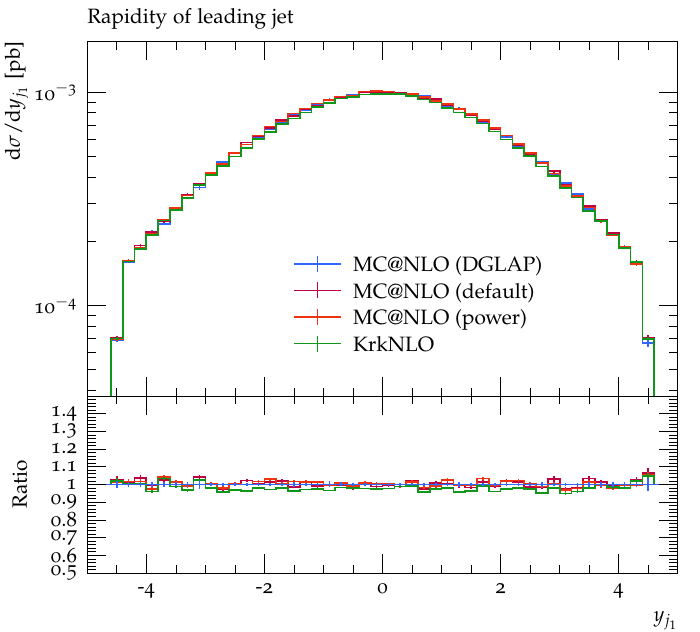}
        \caption{$ZZ$}
        \end{subfigure}
        }
	\caption{Full-shower comparison of the leading-jet distributions.
            The one-emission counterpart is shown in \cref{fig:observables_processes_oneemission_j}.}
    \label{fig:observables_processes_fullshower_j}
\end{figure}
\end{landscape}
}

\begin{figure}[tp]
	\centering
        \includegraphics[width=.32\textwidth]{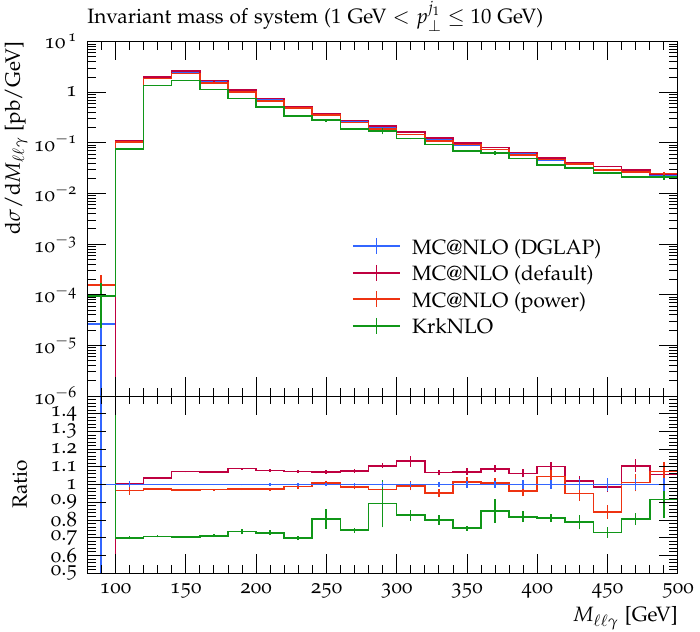}
        \includegraphics[width=.32\textwidth]{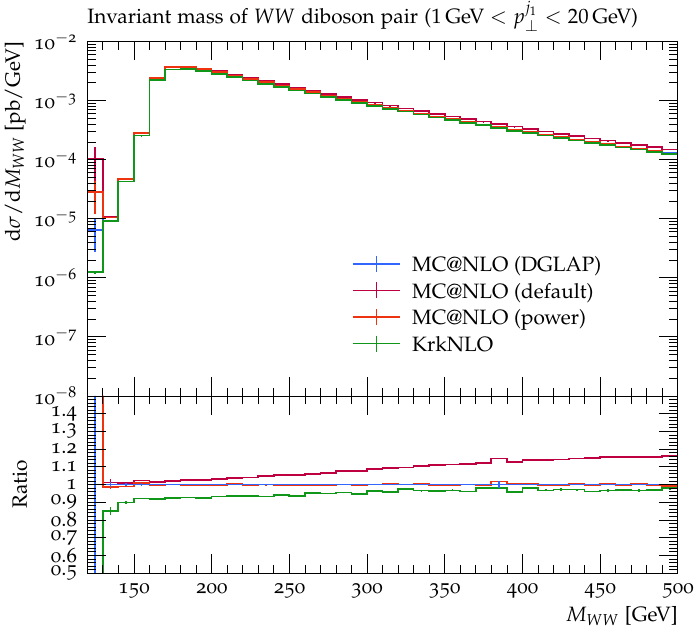}
        \includegraphics[width=.32\textwidth]{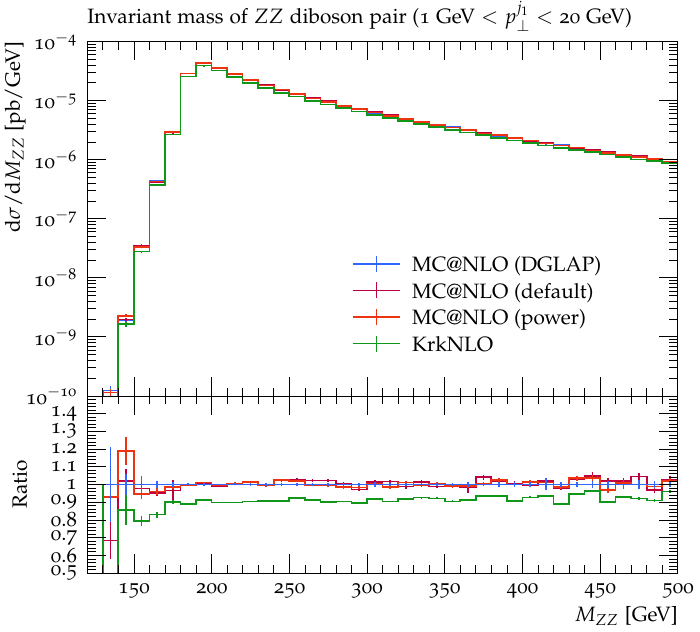}
        \begin{subfigure}[b]{.32\textwidth}
        \includegraphics[width=\textwidth]{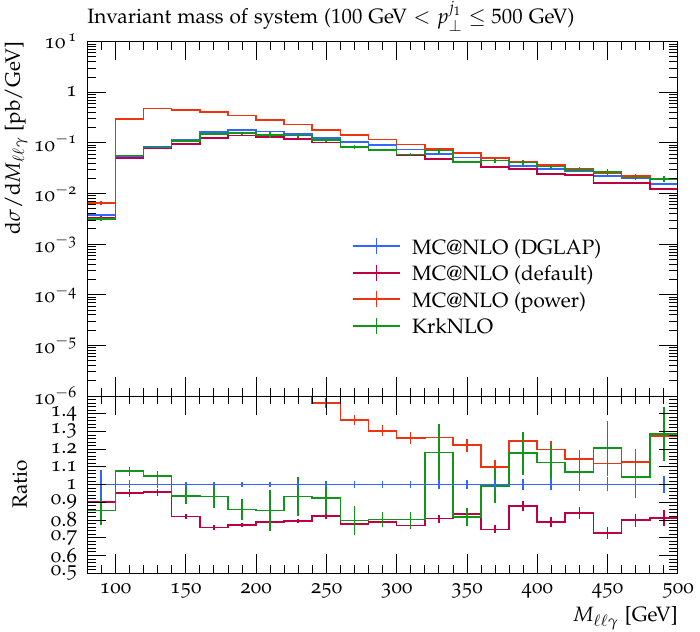}
        \caption{$Z \gamma$}
	    \end{subfigure}
        \begin{subfigure}[b]{.32\textwidth}
        \includegraphics[width=\textwidth]{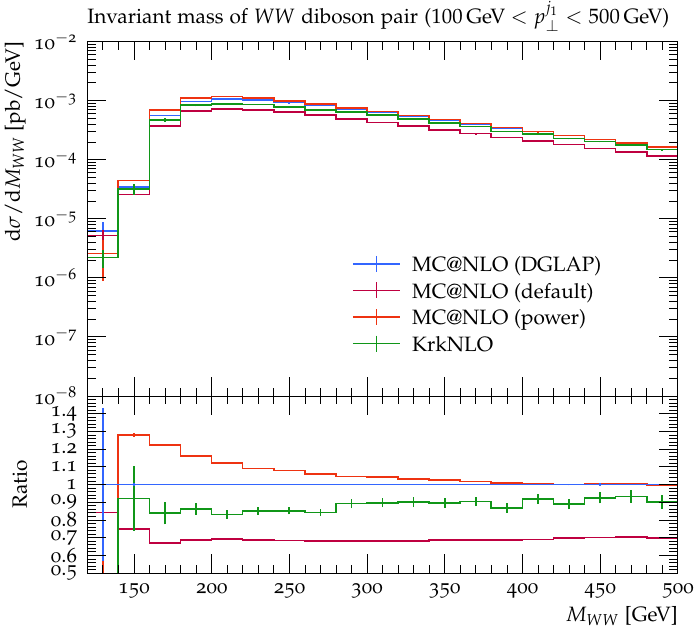}
        \caption{$WW$}
	    \end{subfigure}
        \begin{subfigure}[b]{.32\textwidth}
        \includegraphics[width=\textwidth]{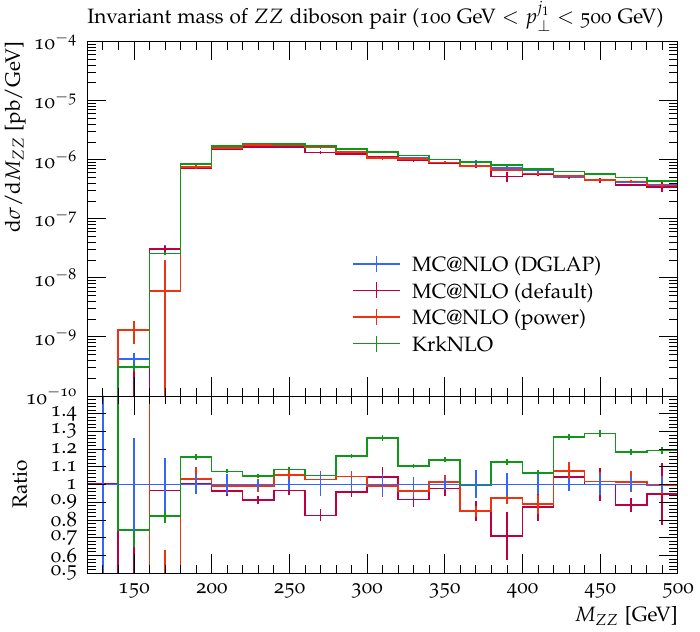}
        \caption{$ZZ$}
        \end{subfigure}
	\caption{Full-shower comparison of the invariant mass of the
		 colour-singlet system for `soft' and `hard' slices of the transverse momentum
		 of the leading jet.
      The one-emission counterpart is shown in \cref{fig:oneemission_dsigma_dM_dptj}.
      The complete array of double-differential slices is shown in \cref{fig:fullshower_dsigma_dM_dptj_full}.
    \label{fig:fullshower_dsigma_dM_dptj}}
\end{figure}

\begin{figure}[tp]
	\centering
        \makebox[\textwidth][c]{%
        \includegraphics[width=0.2\paperwidth]{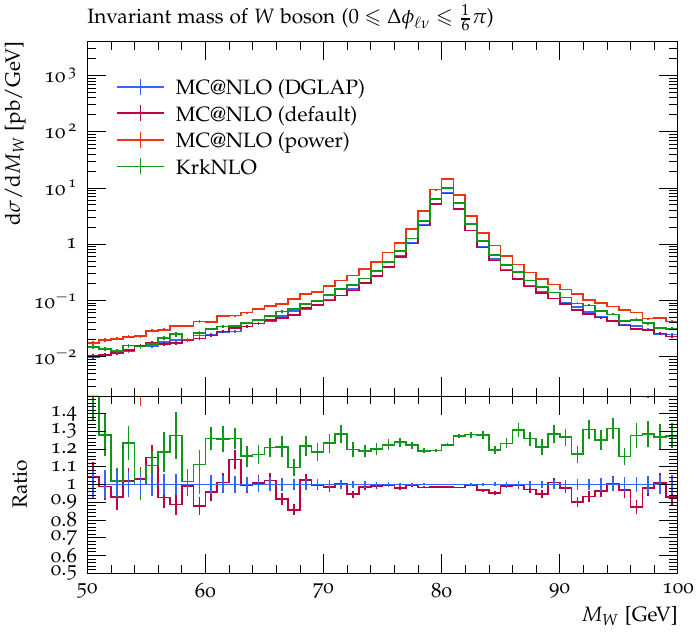}
        \includegraphics[width=0.2\paperwidth]{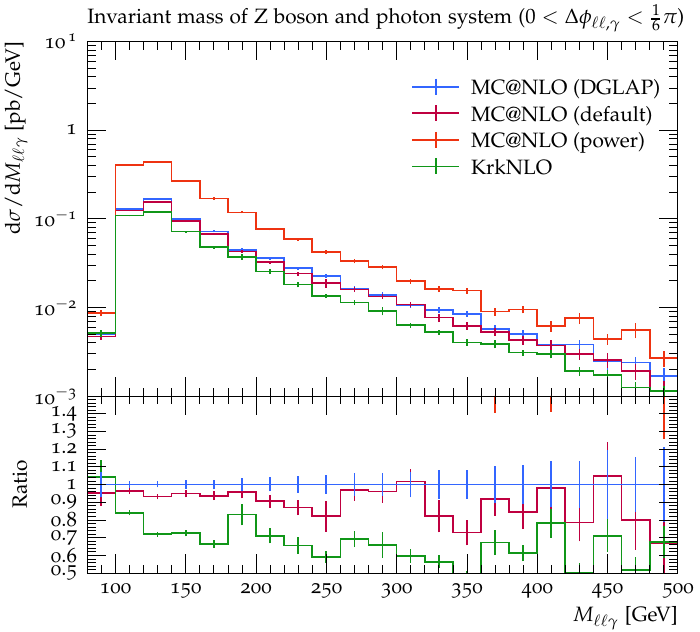}
        \includegraphics[width=0.2\paperwidth]{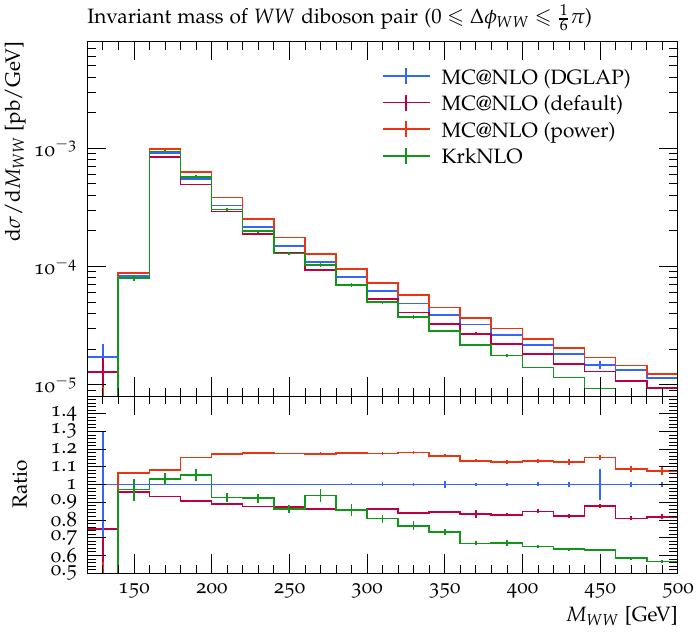}
        \includegraphics[width=0.2\paperwidth]{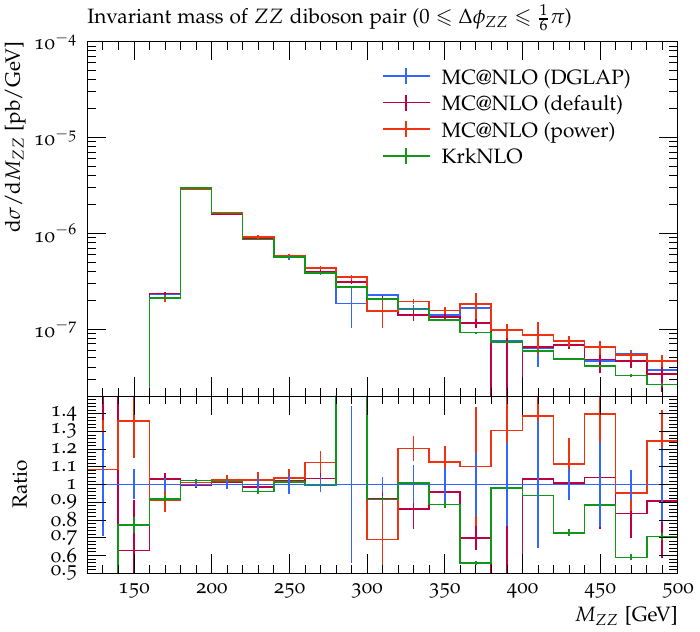}
        } \makebox[\textwidth][c]{%
        \begin{subfigure}[b]{0.2\paperwidth}
        \includegraphics[width=\textwidth]{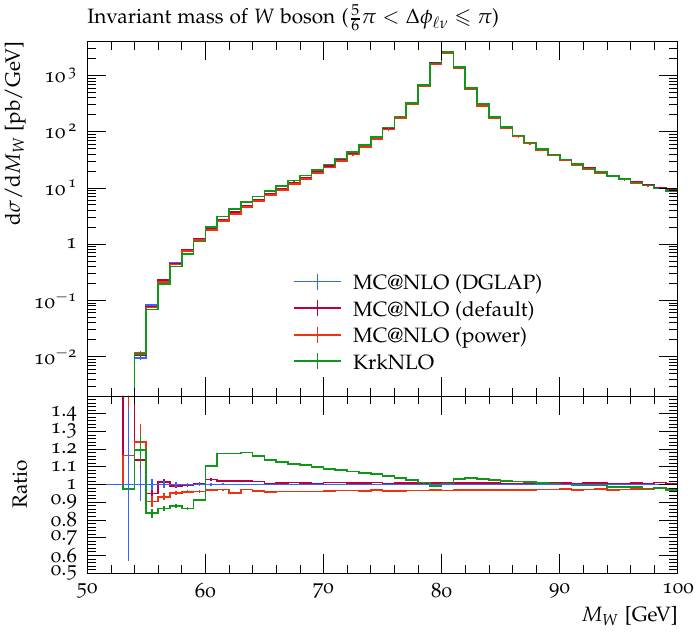}
        \caption{$W$}
        \end{subfigure}
        \begin{subfigure}[b]{0.2\paperwidth}
        \includegraphics[width=\textwidth]{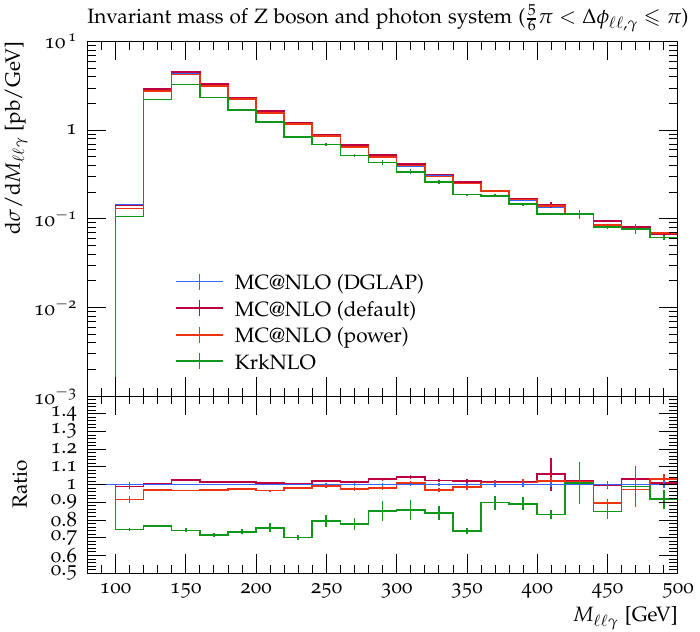}
        \caption{$Z\gamma$}
        \end{subfigure}
        \begin{subfigure}[b]{0.2\paperwidth}
        \includegraphics[width=\textwidth]{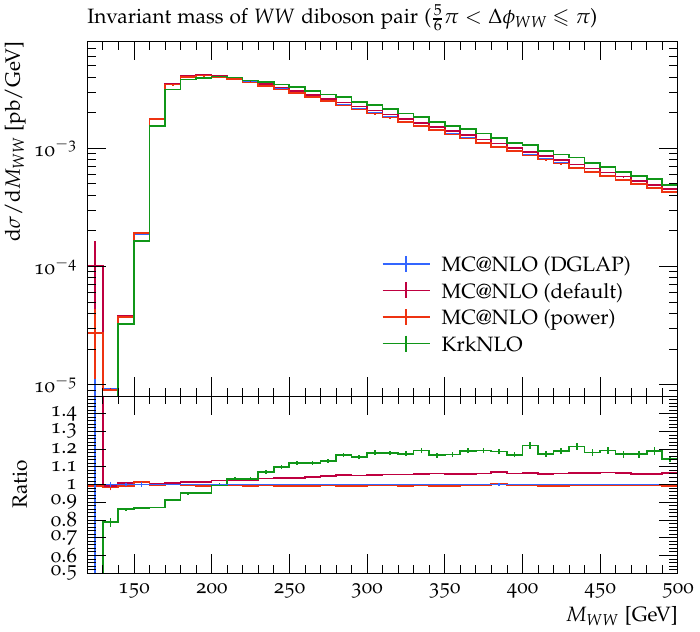}
        \caption{$WW$}
        \end{subfigure}
        \begin{subfigure}[b]{0.2\paperwidth}
        \includegraphics[width=\textwidth]{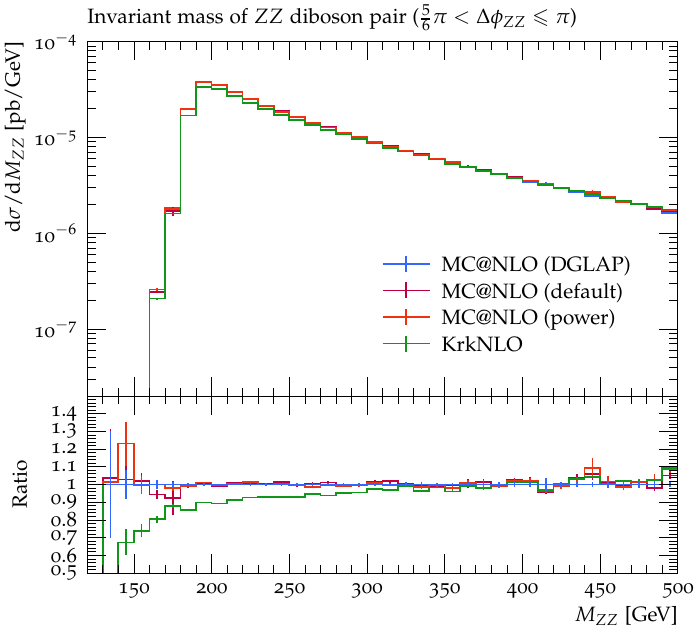}
        \caption{$ZZ$}
        \end{subfigure}
        }
	\caption{Full-shower comparison of the invariant mass of the
		 colour-singlet system in slices of the azimuthal angle between the
         (reconstructed)
         two-particle constituents of the colour-singlet system,
         for `hard' and `soft' (respectively) radiation.
         The complete array of double-differential slices is shown in \cref{fig:fullshower_dsigma_dM_dphi_full}.
    \label{fig:fullshower_dsigma_dM_dphi}}
\end{figure}

\begin{figure}[tp]
	\centering
        \makebox[\textwidth][c]{%
        \includegraphics[width=0.2\paperwidth]{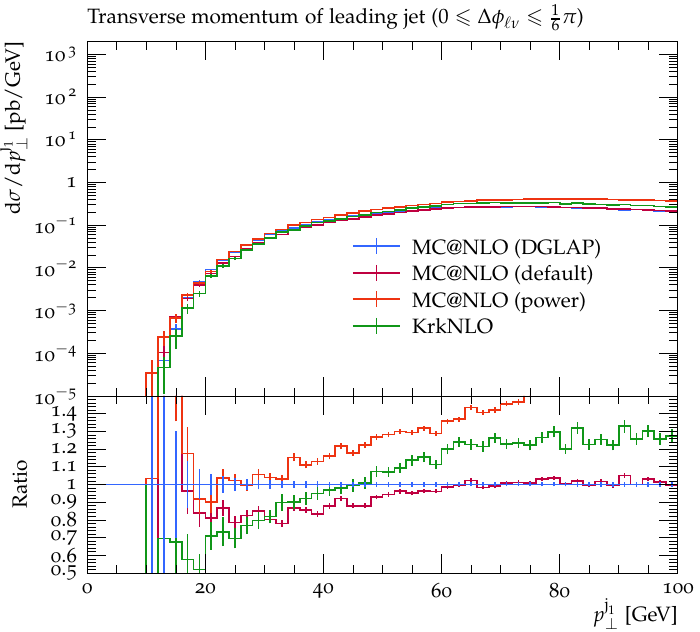}
        \includegraphics[width=0.2\paperwidth]{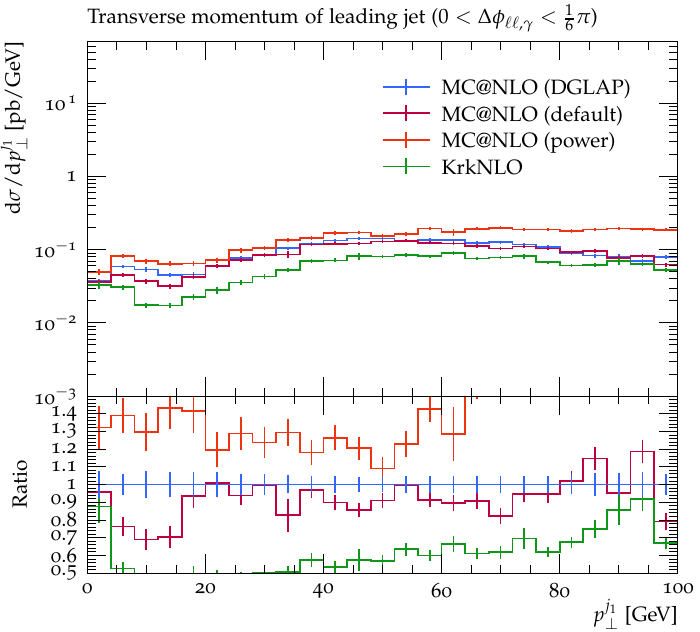}
        \includegraphics[width=0.2\paperwidth]{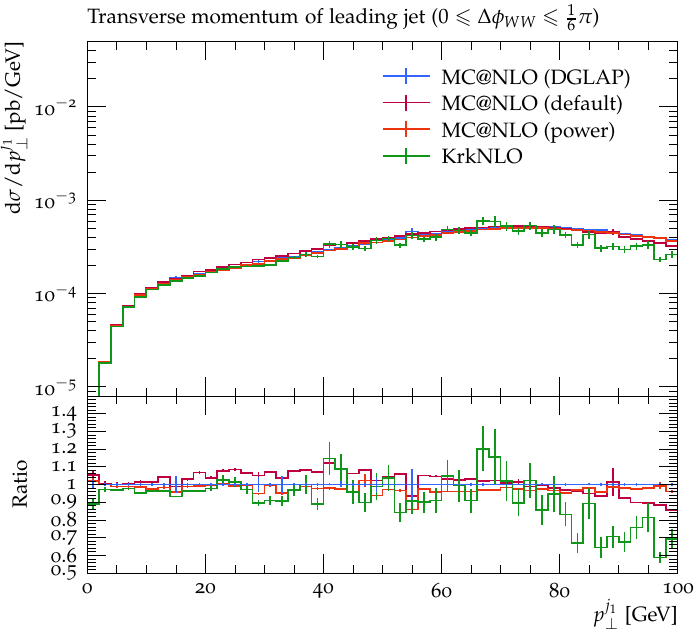}
        \includegraphics[width=0.2\paperwidth]{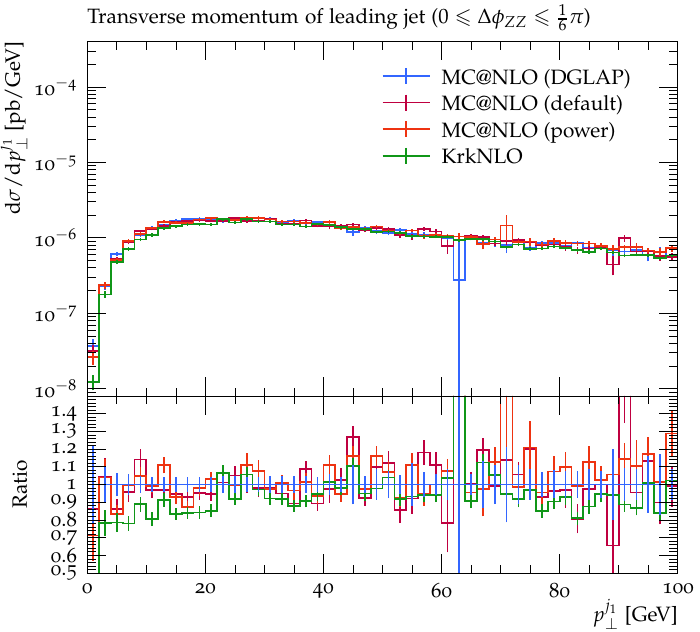}
        } \makebox[\textwidth][c]{
        \begin{subfigure}[b]{0.2\paperwidth}
        \includegraphics[width=\textwidth]{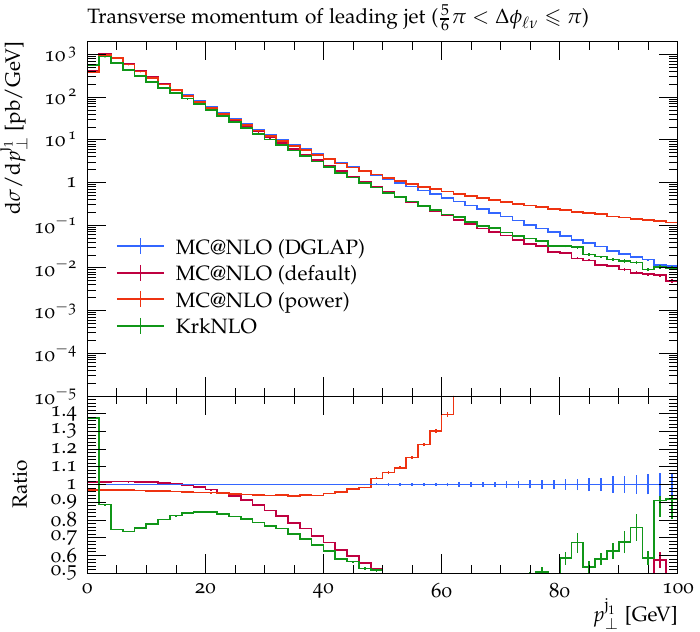}
        \caption{$W$}
        \end{subfigure}
        \begin{subfigure}[b]{0.2\paperwidth}
        \includegraphics[width=\textwidth]{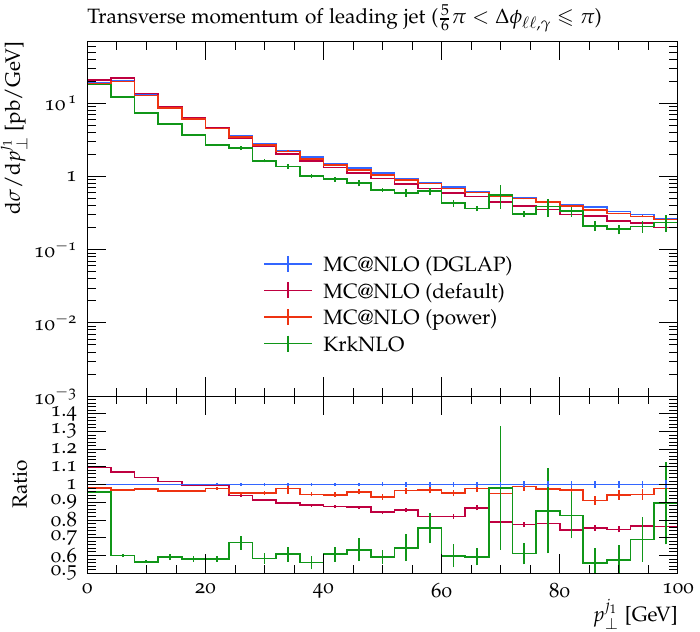}
        \caption{$Z \gamma$}
        \end{subfigure}
        \begin{subfigure}[b]{0.2\paperwidth}
        \includegraphics[width=\textwidth]{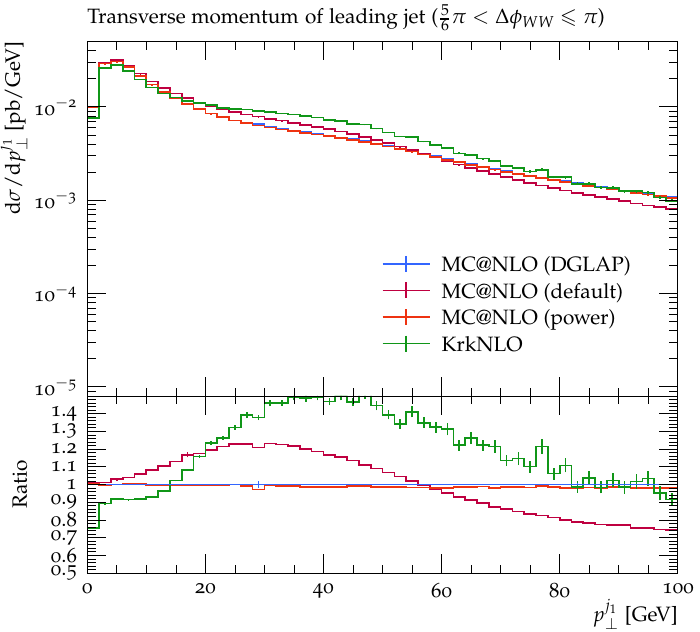}
        \caption{$WW$}
        \end{subfigure}
        \begin{subfigure}[b]{0.2\paperwidth}
        \includegraphics[width=\textwidth]{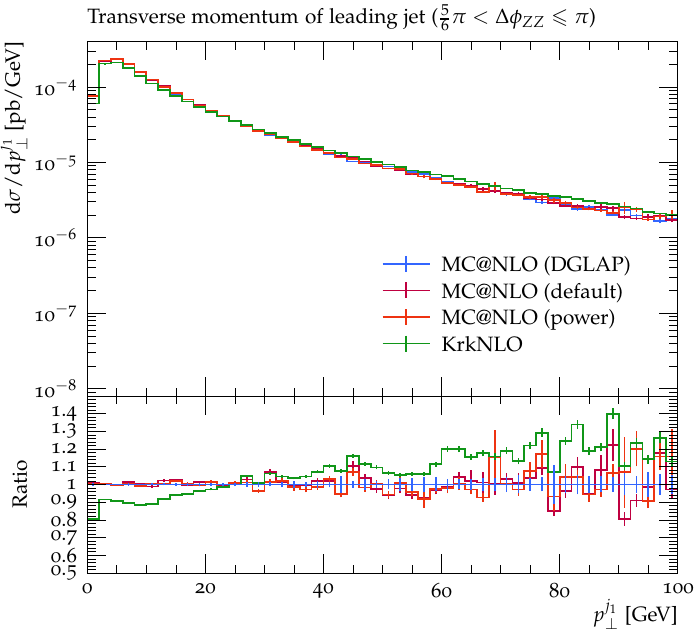}
        \caption{$ZZ$}
        \end{subfigure}
        }
	\caption{Full-shower comparison of the transverse momentum of the leading jet
    in slices of the azimuthal angle between the (reconstructed) two-particle constituents of the colour-singlet system,
    for `hard' and `soft' (respectively) radiation.
    The complete array of double-differential slices is shown in \cref{fig:fullshower_dsigma_dptj_dphi_full}.
    \label{fig:fullshower_dsigma_dptj_dphi}}
\end{figure}

Differential cross-sections for the invariant mass of the colour-singlet system, and the azimuthal angle
and rapidity separations between the two particles in the imputed $2\to 2$ final-state, are shown in 
\cref{fig:observables_processes_fullshower}; differential cross-sections for the transverse momentum
and rapidity of the leading jet are shown in \cref{fig:observables_processes_fullshower_j}.
As in the one-emission case, to elucidate the role of QCD radiation across the inclusive phase-space,
differential cross-sections $\dd \sigma / \dd M$ in slices of $\ptj{1}$ are presented in 
\cref{fig:fullshower_dsigma_dM_dptj}.
Additionally, following the approach of \cite{Sarmah:2024hdk},
$\dd \sigma / \dd M$ and $\dd \sigma / \dd \ptj{1}$ further divided into
six equal slices of the azimuthal separation between the particles within the imputed two-body
colour-singlet final-state
are shown in \cref{fig:fullshower_dsigma_dM_dphi} and \cref{fig:fullshower_dsigma_dptj_dphi}
respectively.

The Born-like kinematics corresponds to $\ptj{1} \to 0$ and $\Delta \phi \to \pi$,
and so these limits parametrise the break-down of perturbation theory
for fixed-order calculations.
In this limit, the parton shower provides all-orders resummation.
Far from this limit, the colour-singlet system recoils against hard QCD radiation.
Close to it, the kinematic configuration is effectively Born-like, with only
relatively soft QCD radiation.

Overall, the distributions in \cref{fig:observables_processes_fullshower}
show good agreement between the methods,
with the deviation between \krknlo and the conventional
\mcatnlo variants (`default' and `DGLAP')
typically smaller than 20\% and often significantly smaller.
This is generally smaller than the uncertainty within the 
\mcatnlo method due to the choice of shower starting-scale,
as illustrated by the large deviations for the power-shower
variant of \mcatnlo.
The deviations appear to shrink with effective mass-scale, with
the $ZZ$ process showing very little matching uncertainty
in any distribution.

\Cref{fig:fullshower_dsigma_dM_dptj,fig:fullshower_dsigma_dM_dphi}
show that in the hard-emission region,
the \krknlo method generally gives results lying within the \mcatnlo uncertainty envelope,
and generally lying close to the `default' and DGLAP \mcatnlo variants.
A similar picture can be seen from the small-azimuthal-separation regime
in \cref{fig:fullshower_dsigma_dptj_dphi}.

In the resummation region, the differences are much smaller than at one-emission,
with relatively close agreement between the predictions shown in
\cref{fig:fullshower_dsigma_dM_dphi,fig:fullshower_dsigma_dM_dptj}.
However, larger differences are visible in the 
$\dd^2 \sigma / \dd \phi \, \dd \ptj{1}$ distributions, which isolate the resummation
region simultaneously in angle and in transverse momentum.
Here the \krknlo method consistently lies outwith the envelope of the other predictions,
but with distinctive shape similarities between the `default' \mcatnlo variant
and \krknlo for both $W$ and $WW$,
also observed in the diphoton case in \cite{Sarmah:2024hdk}.
The low-$\ptj{1}$, large-$\Delta\phi$ region dominates each process;
for many experimental analyses with typical jet cuts, this region would 
correspond to (and dominate) events with no identified jets.
The scale below which the difference in Sudakov suppression in the low-$\ptj{1}$ region
emerges is approximately
$\ptj{1} < 20 \; \GeV$ for $WW$ and $ZZ$,
$\ptj{1} < 50 \; \GeV$ for $W$,
and $\ptj{1} < 80 \; \GeV$ for $Z\gamma$.
That the matching uncertainty is concentrated in the low-$\pt$,
back-to-back region is perhaps unsurprising,
since this is the region in which the perturbative constraints on higher-order
terms are confronted with large logarithms arising from the parton shower.
However, this is also the region of phase-space in which
the uncertainty envelope arising from shower-scale variation
within the \mcatnlo method collapses, due to the low emission-scale
being permitted irrespective of the functional-form chosen for the starting-scale.

Since the overall matching uncertainty is dominated
by the soft-emission phase-space region in which shower-scale variation is inconsequential,
variation within the \mcatnlo method is insufficient and the comparison of alternative methods is necessary.

\section{LHC phenomenology}
\label{sec:LHCpheno}

In this section we use the same runs as in \cref{sec:matching},
and apply analyses previously used for LHC data.
These are therefore histograms of the same events from the same 
simulations, but with different cuts and observable binnings.
In particular, rather than cuts and observables chosen for
consistent inter-process comparison,
they reflect choices made for experimental purposes for each process independently.

For each process, we summarise the current theoretical status and the
expected size of the missing, but known, higher-order-corrections.
We then contextualise 
The differences between the matching methods, and resulting matching uncertainty,
can therefore be contextualised with respect to the missing higher-order corrections,
the observed level of agreement with data, and the experimental uncertainty.

\subsection{Charged-current Drell--Yan}
\label{sec:LHCpheno_W}

The charged-current Drell--Yan process has been calculated to
\nnnlo accuracy in QCD \cite{Duhr:2020sdp,Chen:2022cgv,Chen:2022lwc,Campbell:2023lcy},
and at NNLO+PS in the 
UN\textsuperscript{2}LOPS \cite{Hoche:2014uhw},
NNLOPS post-hoc reweighting \cite{Karlberg:2014qua},
\textsc{Geneva} \cite{Alioli:2015toa}
and
MiNNLO\textsubscript{PS} \cite{Monni:2019whf}
formalisms.
The magnitude of QCD corrections beyond NLO is relatively small.
The $W + j$ process has been calculated to NNLO accuracy in QCD
\cite{Boughezal:2015dva,Boughezal:2016dtm,Gehrmann-DeRidder:2017mvr,NNLOJET:2025rno};
the NLO local $K$-factor for $\dd \sigma / \dd \ptj{1}$ is approximately 50\%,
while the NNLO corrections are much smaller in magnitude.

\begin{figure}[tp]
	\centering
    \vspace{-10mm}
    \begin{subfigure}[t]{\textwidth}
        \includegraphics[width=.49\textwidth]{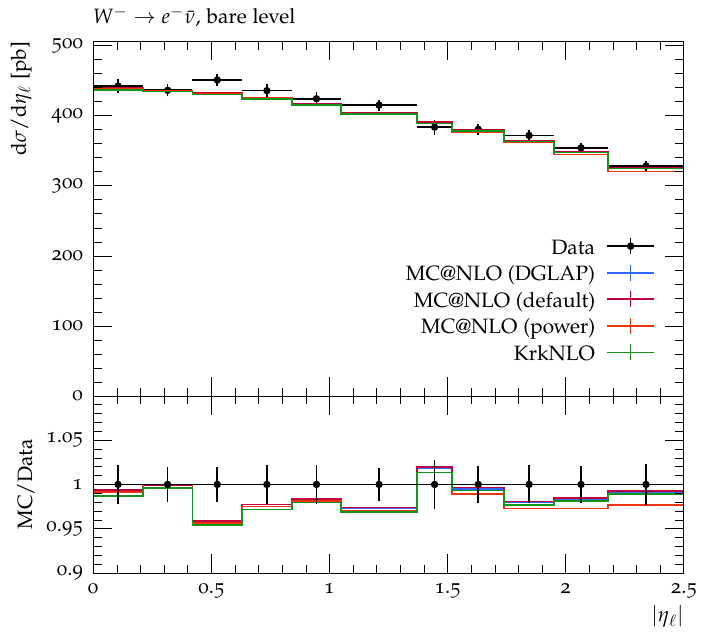}
        \hfill
        \includegraphics[width=.49\textwidth]{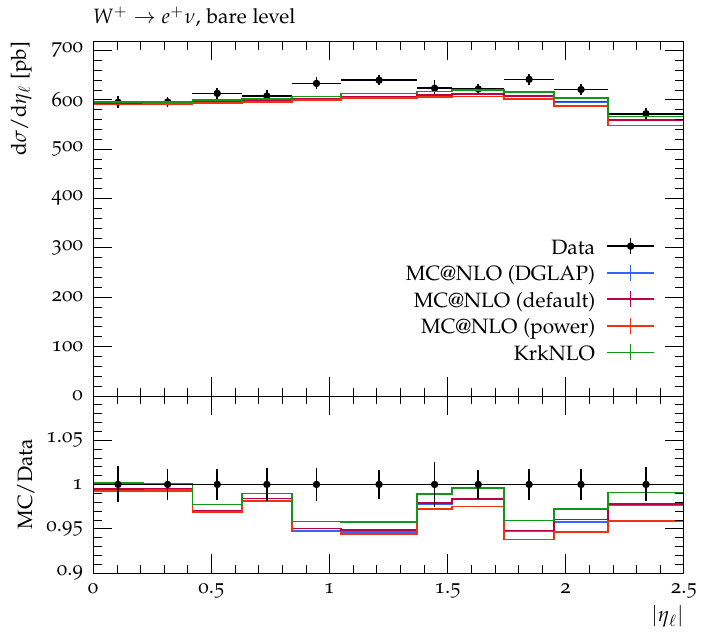}
        \hfill
        \\ \centering
        \includegraphics[width=.49\textwidth]{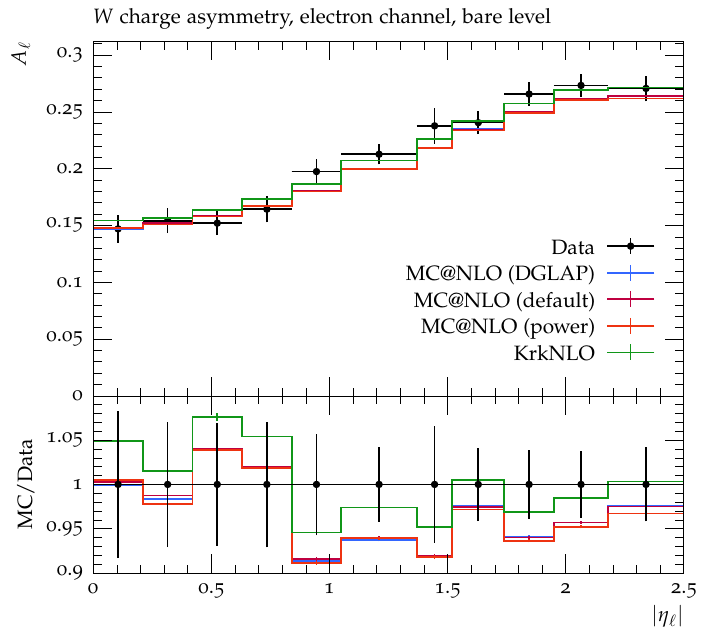}
	\caption{{\texttt{ATLAS\_2011\_I928289}} \cite{ATLAS:2011qdp} \label{fig:W_pheno_data_fullshower_inc}}
    \end{subfigure}
    \begin{subfigure}[t]{\textwidth}
        \includegraphics[width=.49\textwidth]{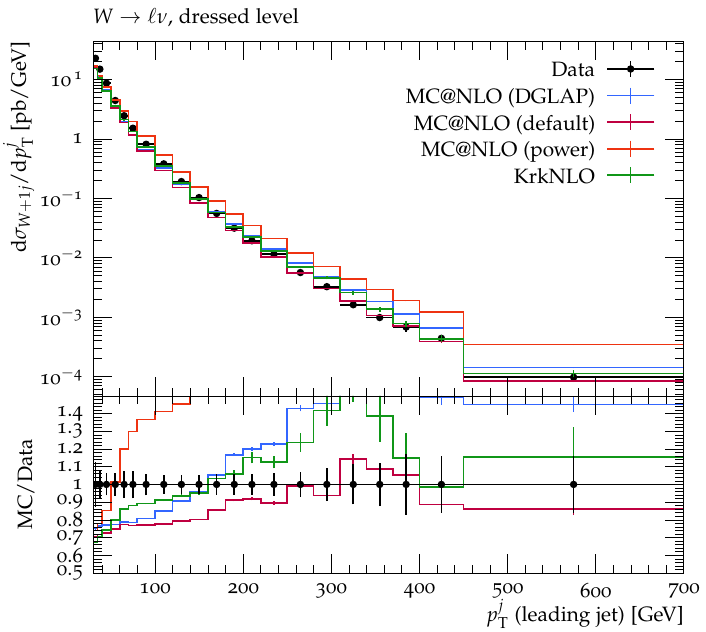}
        \hfill
        \includegraphics[width=.49\textwidth]{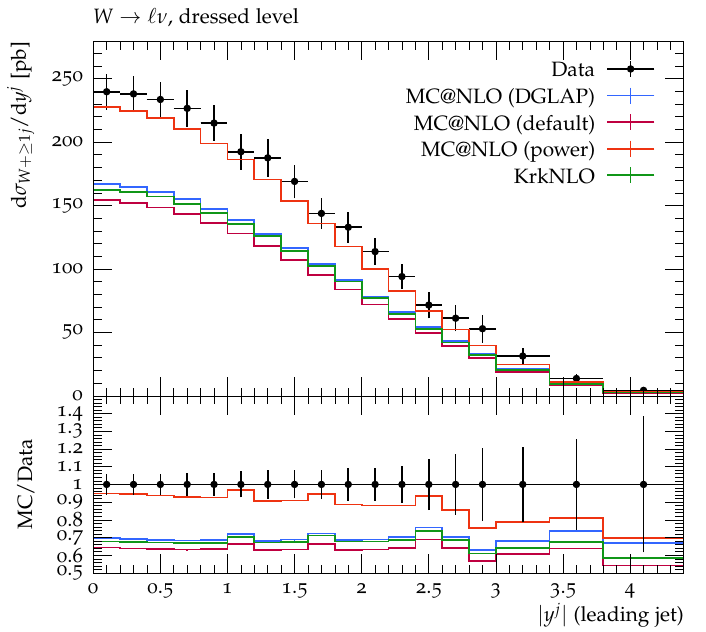}
	\caption{{\texttt{ATLAS\_2014\_I1319490}} \cite{ATLAS:2014fjg} \label{fig:W_pheno_data_fullshower_jet}}
    \end{subfigure}
	\caption{$W$:
    Comparison of matched NLO-plus-parton-shower differential cross-sections
    as generated by \krknlo and \mcatnlo,
    to 7 TeV \atlas data from \cite{ATLAS:2011qdp,ATLAS:2014fjg}.
    Note that the experimental results of \cref{fig:W_pheno_data_fullshower_inc} are averaged
    over lepton-flavour, and only the unfolding is flavour-specific.
    The jet distributions in \cref{fig:W_pheno_data_fullshower_jet} are formally of LO accuracy.
    \label{fig:W_pheno_data_fullshower}}
\end{figure}

We use the generation parameters of \cref{tab:generator_cuts} and the 
experimental cuts of \cite{ATLAS:2011qdp}
\begin{subequations}
	\label{eqn:ATLAScuts_W_1}
	\begin{align}
		\ptell{} &> 20 \;\GeV , 
		& \vert \eta^{\ell} \vert &< 2.5, \\
		\Etmiss &> 25 \; \GeV,
		& \mt^W &> 40 \;\GeV,
	\end{align}
\end{subequations}
and \cite{ATLAS:2014fjg}
\begin{subequations}
	\label{eqn:ATLAScuts_W_2}
	\begin{align}
		\ptell{} &> 25 \;\GeV \,, 
		& \vert \eta^{\ell} \vert &< 2.5, \\
		\Etmiss &> 25 \; \GeV,
		& \mt^W &> 40 \;\GeV,  \\
		\ptj{} &> 30\;\GeV, 
		& \vert y^{\mathrm{jet}} \vert &< 4.4,
		& \dRlj &> 0.5,
	\end{align}
\end{subequations}
as implemented in the \texttt{ATLAS\_2011\_I928289\_W} and \texttt{ATLAS\_2014\_I1319490}
analyses within \rivet \cite{Bierlich:2019rhm} respectively.
The resulting plots are shown in \cref{fig:W_pheno_data_fullshower}.

We observe very little matching uncertainty in the inclusive distributions
of \cref{fig:W_pheno_data_fullshower_inc},
all of which describe the data
to a comparable level (approximately 5\%).
The \krknlo method consistently describes the charge-asymmetry
within experimental uncertainties across the pseudorapidity
distribution; all methods agree within 10\%.
The jet distributions of \cref{fig:W_pheno_data_fullshower_jet} show greater
matching uncertainty, as expected from their formal LO-accuracy and 
the discussion in \cref{sec:matching}.

The \krknlo predictions can again be seen to lie within the matching uncertainty envelope of the
\mcatnlo method,
and generally between the `default' and `DGLAP' shower-scale variants.
Although the `power'-shower variant appears to achieve the correct normalisation in the
jet-rapidity distribution,
this is an artefact of the tail of the $\ptj{1}$-distribution,
as in \cref{sec:matching}.

\subsection{\texorpdfstring{${Z\gamma}$ production}{Zγ production}}
\label{sec:LHCpheno_Zgam}

The $Z\gamma$ process has been calculated to
NNLO accuracy in QCD \cite{Grazzini:2013bna,Grazzini:2015nwa,Campbell:2017aul},
and at NNLO+PS in the 
MiNNLO\textsubscript{PS} \cite{Lombardi:2020wju}
formalism.
The NNLO corrections are of the order of 20\%.
The NLO EW corrections \cite{Denner:2015fca} 
become significant at large values of the invariant mass $M_{\ell\ell\gamma}$.

\begin{figure}[htp]
	\centering
        \includegraphics[width=.49\textwidth]{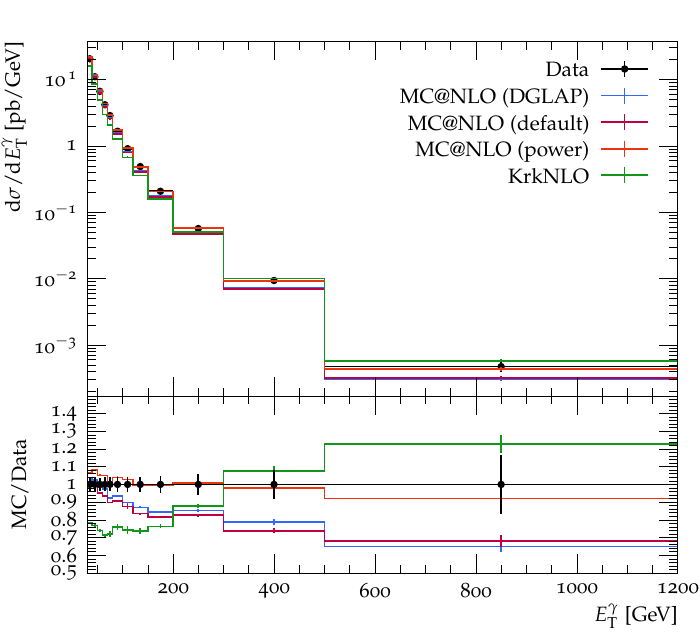}
        \hfill
        \includegraphics[width=.49\textwidth]{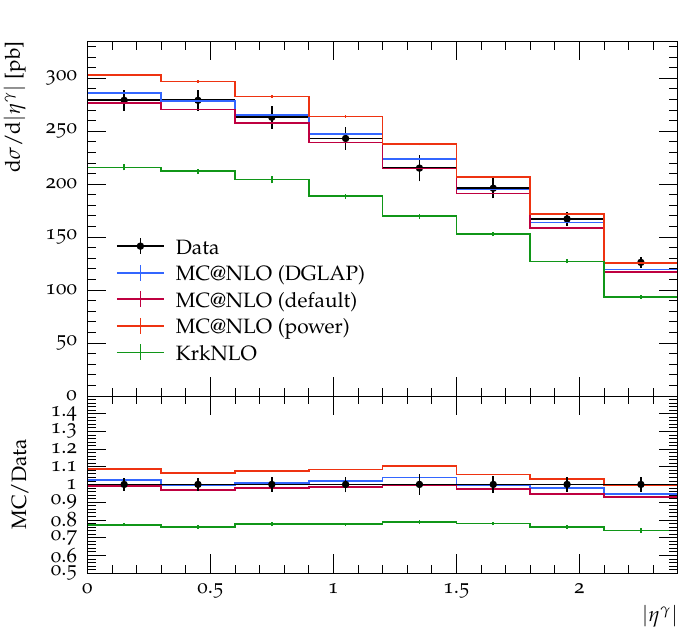} \\
        \includegraphics[width=.49\textwidth]{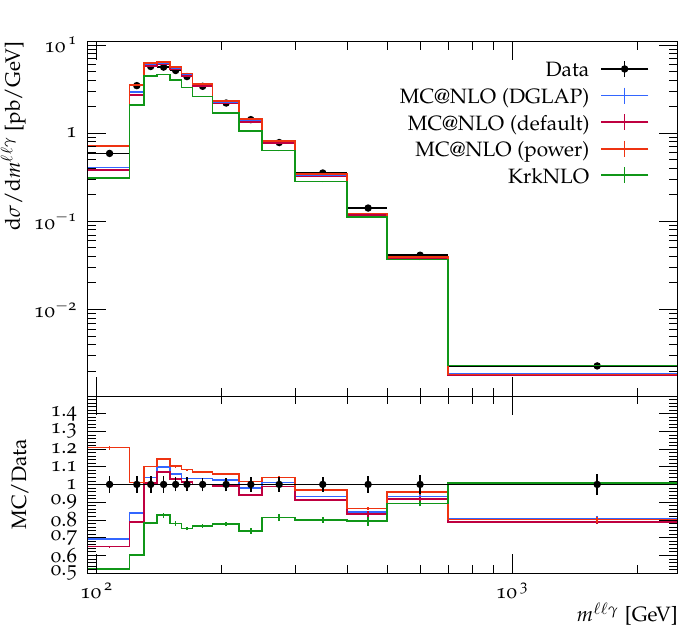}
        \hfill
        \includegraphics[width=.49\textwidth]{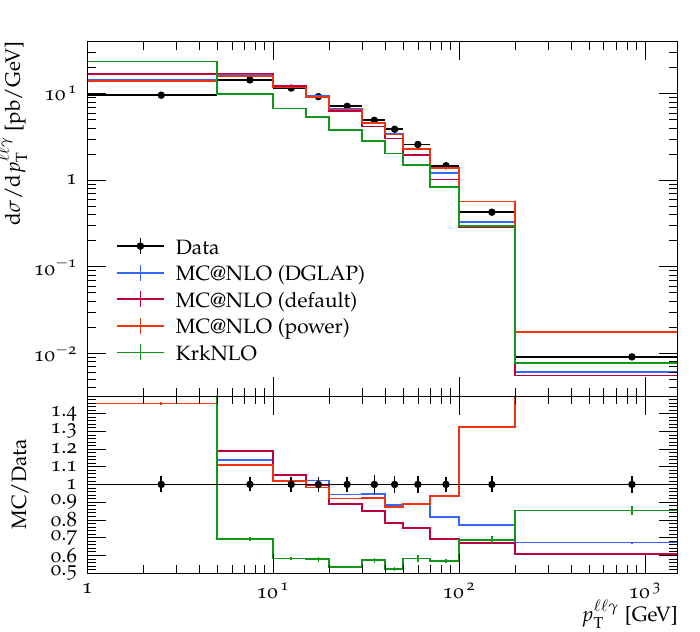} \\
        \includegraphics[width=.49\textwidth]{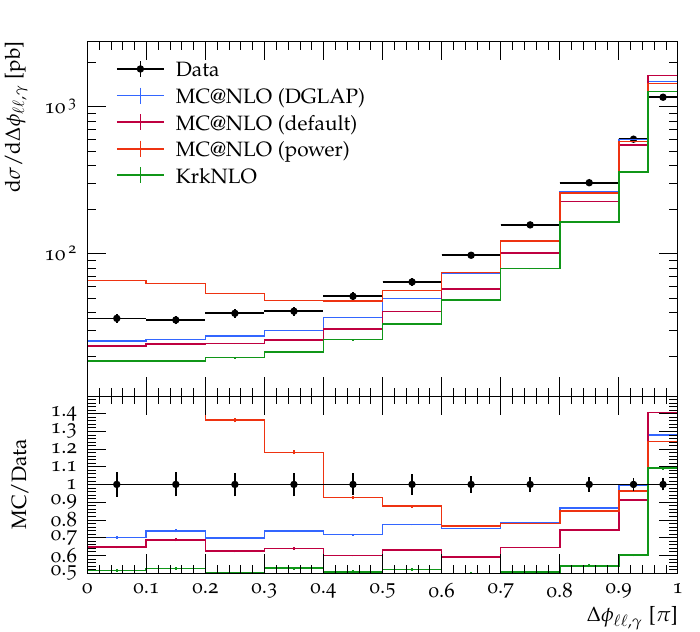}
        \hfill
        \includegraphics[width=.49\textwidth]{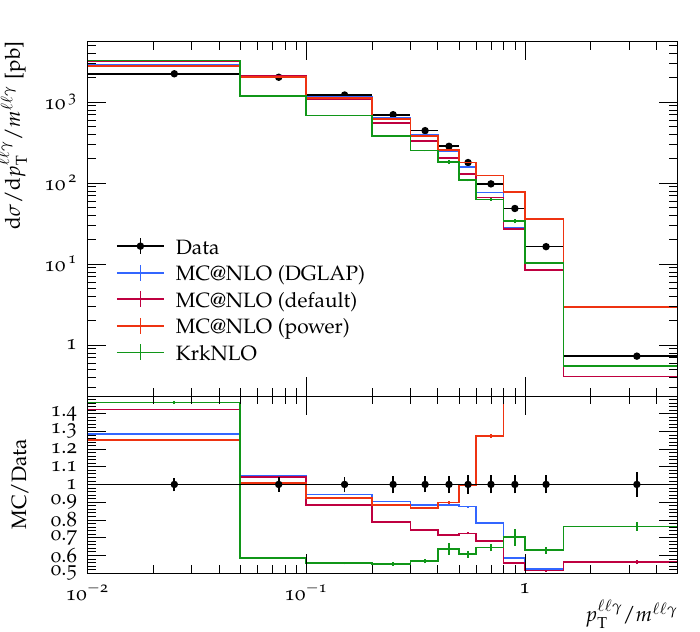}
	\caption{$Z\gamma$: Comparison of matched NLO-plus-parton-shower differential cross-sections
    as generated by \krknlo and \mcatnlo,
    to 13 TeV \atlas data from \cite{ATLAS:2019gey}.
    Note that the first three distributions are formally calculated to NLO accuracy, while the last three are LO.
    \label{fig:Zgam_pheno_data_fullshower}}
\end{figure}

We again use the generation parameters of \cref{tab:generator_cuts} and the experimental
cuts of \cite{ATLAS:2019gey}
\begin{subequations}
	\label{eqn:ATLAScuts_Zgamma}
	\begin{align}
		\ptell{1} &> 30 \;\GeV ,
		& \ptell{2} &> 25 \;\GeV ,
		& \vert \eta^{\ell} \vert &< 2.47, \\
		\pt^{\gamma} &> 30 \;\GeV ,
        &&& \vert \eta^{\gamma} \vert &< 2.37,  \\
		\Mll &> 40 \;\GeV ,
		& \dRlg &> 0.4 \, ,
        & \Mll + \Mllg &> 182 \;\GeV , \\
		\Etiso (r) &< 0.07 \, \ptg{} \;
		& \text{within cone } r &\leqslant R = 0.2,
	\end{align}
\end{subequations}
as implemented in the \rivet analysis \texttt{ATLAS\_2019\_I1764342}.
The resulting plots are shown in \cref{fig:Zgam_pheno_data_fullshower}.

As expected from the general observations of \cref{sec:matching},
in the $Z\gamma$ case we see substantial matching uncertainty.
Between the methods, neither \mcatnlo nor \krknlo is consistent with the
data across phase-space.
The data lies within the overall matching uncertainty envelope save for the
intermediate-$\ptof{\ell\ell\gamma}$ region where additional real-radiation is expected to be significant.
As in \cref{sec:matching}, but unlike the other processes studied here,
the \krknlo prediction generally lies outwith the \mcatnlo uncertainty envelope,
likely again due to the Sudakov factor accompanying the
(abundant) soft emissions.
In regions of phase-space characterised by hard recoil,
this is absent.

Again, although the power-shower variant of \mcatnlo 
agrees with data in certain distributions,
the shape difference observed in the
$\dd \sigma / \dd \Delta \phi_{\ell\ell, \gamma}$ distribution in \cref{sec:matching}
leads to a substantial overestimate relative to the data at low-$\Delta \phi_{\ell\ell,\gamma}$.
The apparent agreement in the normalisation of the \mcatnlo predictions
is in tension with the known magnitude of the missing NNLO corrections.

\subsection{\texorpdfstring{$WW$ production}{WW production}}
\label{sec:LHCpheno_WW}

The $WW$ process has been calculated to
NNLO accuracy in QCD \cite{Gehrmann:2014fva,Grazzini:2016ctr}
including with
polarised $W$-bosons \cite{Poncelet:2021jmj},
and at NNLO+PS in the 
NNLOPS post-hoc reweighting \cite{Re:2018vac},
MiNNLO\textsubscript{PS} \cite{Lombardi:2021rvg},
and
\textsc{Geneva} \cite{Gavardi:2023aco}
formalisms.
The \nnlo correction changes the total cross-section by approximately 15\% relative to \nlo \cite{Grazzini:2016ctr},
around half of which is attributable to the $gg$-channel.
The NLO EW corrections \cite{Grazzini:2019jkl} 
become significant at large values of the invariant mass $M_{WW}$.

\begin{figure}[tp]
	\centering
        \includegraphics[width=.49\textwidth]{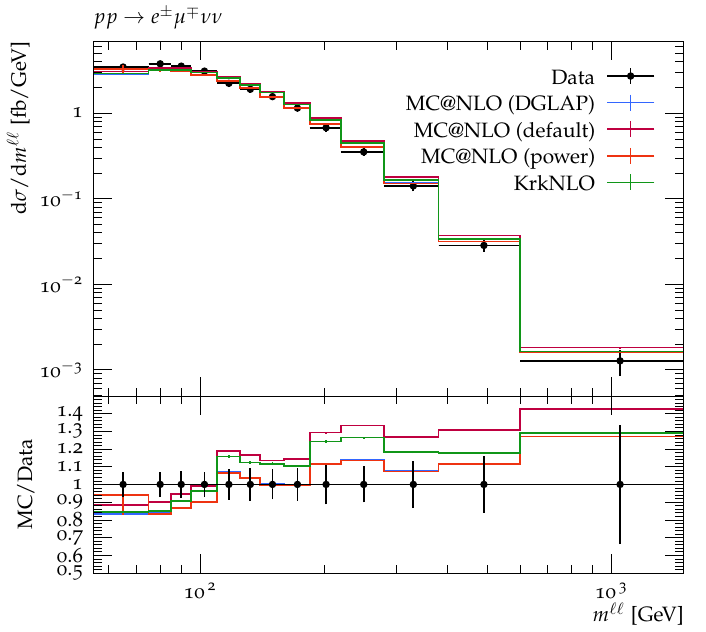}
        \hfill
        \includegraphics[width=.49\textwidth]{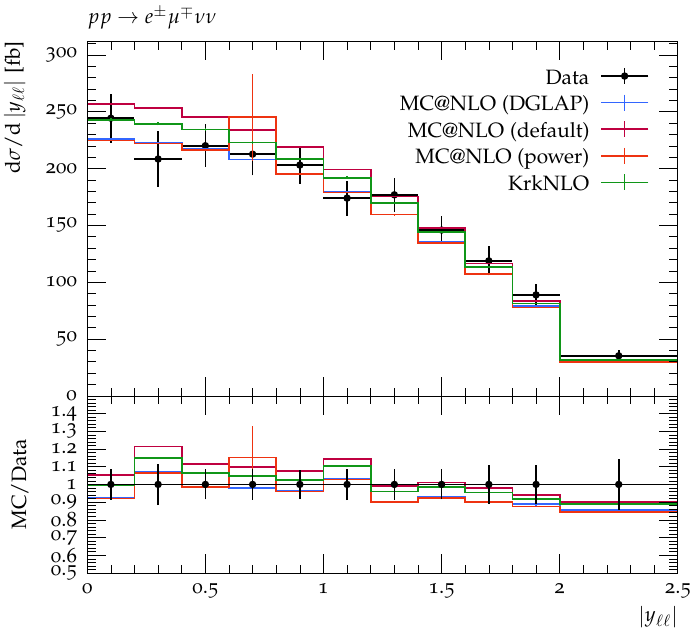} \\
        \includegraphics[width=.49\textwidth]{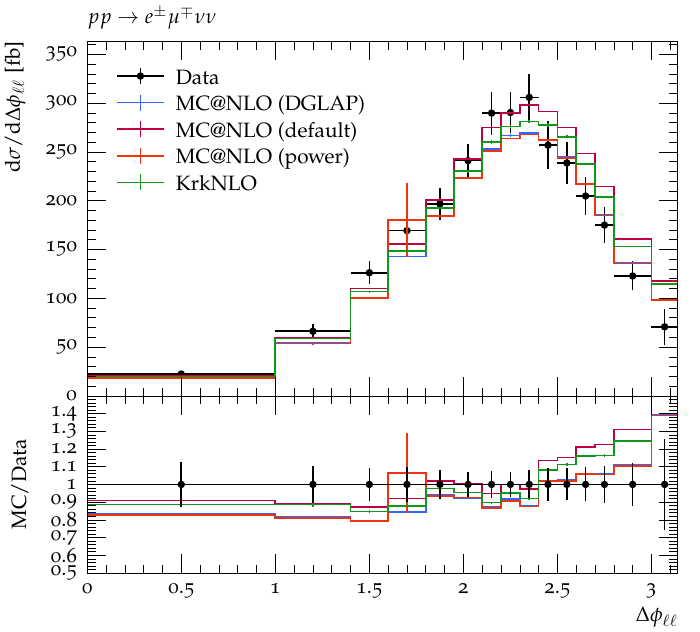}
        \hfill
        \includegraphics[width=.49\textwidth]{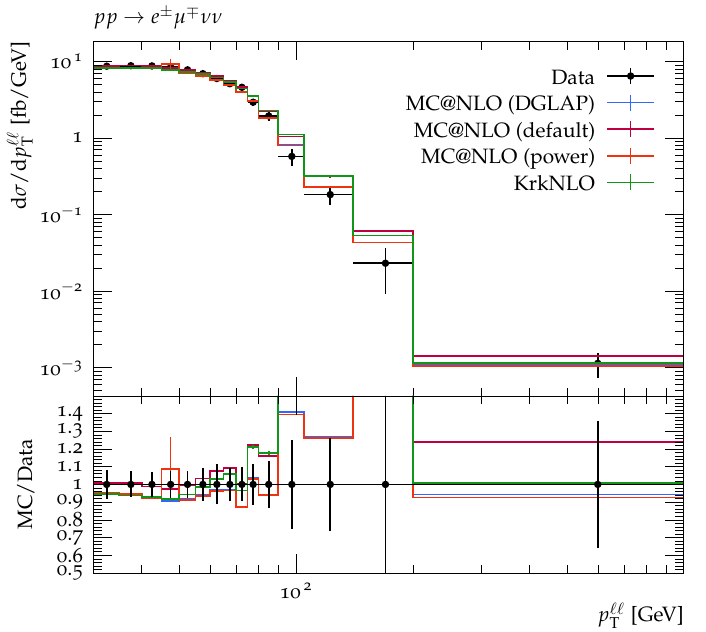} \\
        \includegraphics[width=.49\textwidth]{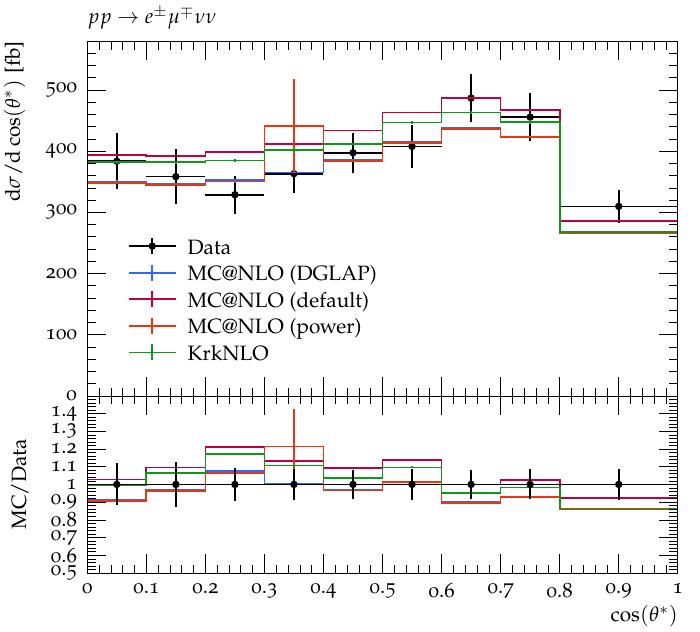}
        \hfill
        \includegraphics[width=.49\textwidth]{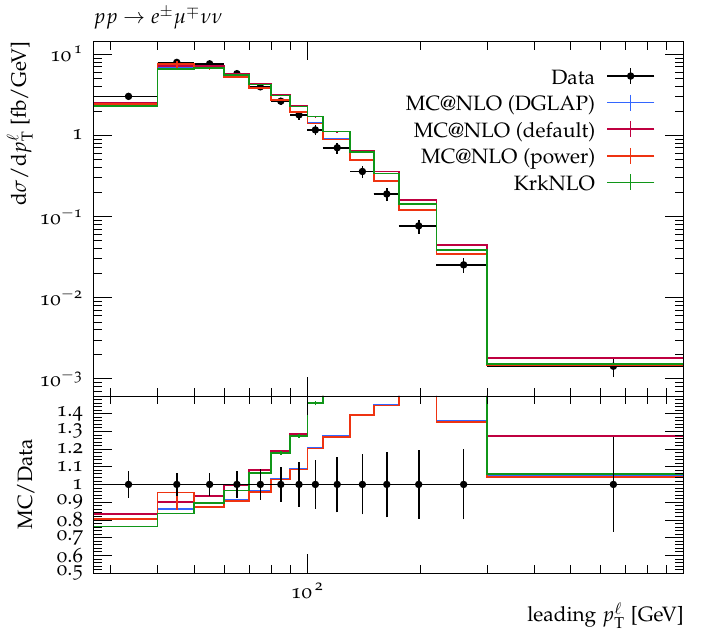}
	\caption{$WW$: Comparison of matched NLO-plus-parton-shower differential cross-sections
    as generated by \krknlo and \mcatnlo,
    to 13 TeV \atlas data from \cite{ATLAS:2019rob}.
    \label{fig:WW_pheno_data_fullshower}}
\end{figure}

We use the generation parameters of \cref{tab:generator_cuts} and the experimental
cuts of \cite{ATLAS:2019rob}, which select for two opposite-charge, different-flavour
leptons within fiducial cuts
\begin{subequations}
	\label{eqn:WWatlascuts}
	\begin{align}
		\ptell{} &> 27 \;\GeV , 	&	\absetaell &< 2.5, \\
		\pt^{\text{miss}} &> 20 \;\GeV, & \\
		M_{\ell\ell'} & > 55 \;\GeV, & \pt^{\ell\ell'} &> 30 \;\GeV,
	\end{align}
\end{subequations}
as implemented in the \rivet analysis \texttt{ATLAS\_2019\_I1734263}.
The resulting plots are shown in \cref{fig:WW_pheno_data_fullshower}.

For the $WW$ process, we observe a reduced matching uncertainty, as in \cref{sec:matching}.
The \krknlo distributions once again lie within the \mcatnlo envelope,
and generally closest to the `default'-shower \mcatnlo variant.
The data is consistent with the matched NLO predictions within the matching-uncertainty envelope,
save for the $\dd \sigma / \dd \ptof{\ell_1}$ distribution at intermediate-$\ptof{\ell_1}$.

\subsection{\texorpdfstring{$ZZ$ production}{ZZ production}}
\label{sec:LHCpheno_ZZ}

The $ZZ$ process has been calculated to
NNLO accuracy in QCD \cite{Cascioli:2014yka,Grazzini:2015hta,Kallweit:2018nyv,Heinrich:2017bvg}
including recently with
polarised $Z$-bosons \cite{Carrivale:2025mjy},
and at NNLO+PS in the 
\textsc{Geneva} \cite{Alioli:2021egp}
and
MiNNLO\textsubscript{PS} \cite{Buonocore:2021fnj} formalisms.
The \nnlo correction changes the total cross-section by approximately 20\% relative to \nlo \cite{Heinrich:2017bvg},
around half of which is attributable to the $gg$-channel which opens up at NNLO.
In turn, the gluon-initiated channel itself receives significant NLO (respectively, \nnnlo) corrections
\cite{Agarwal:2024pod}.
The NLO EW corrections \cite{Grazzini:2019jkl} 
become significant at large values of the invariant mass $M_{ZZ}$.

\begin{figure}[tp]
	\centering
        \includegraphics[width=.49\textwidth]{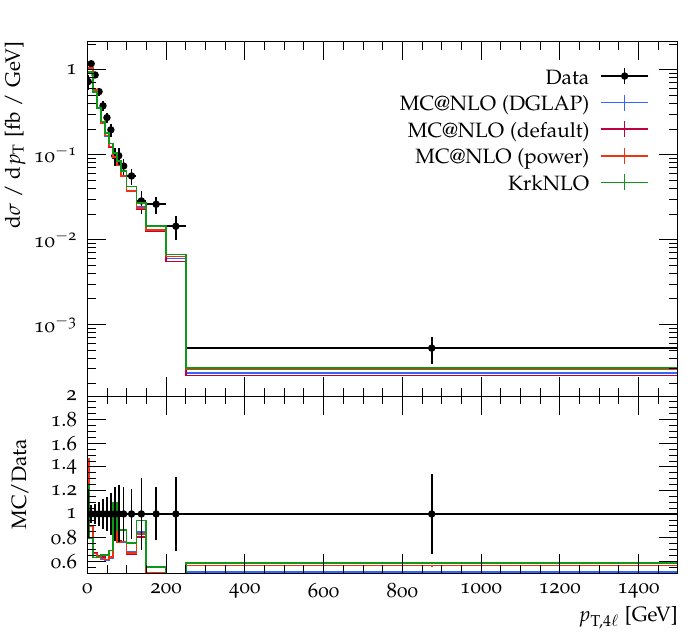}
        \hfill
        \includegraphics[width=.49\textwidth]{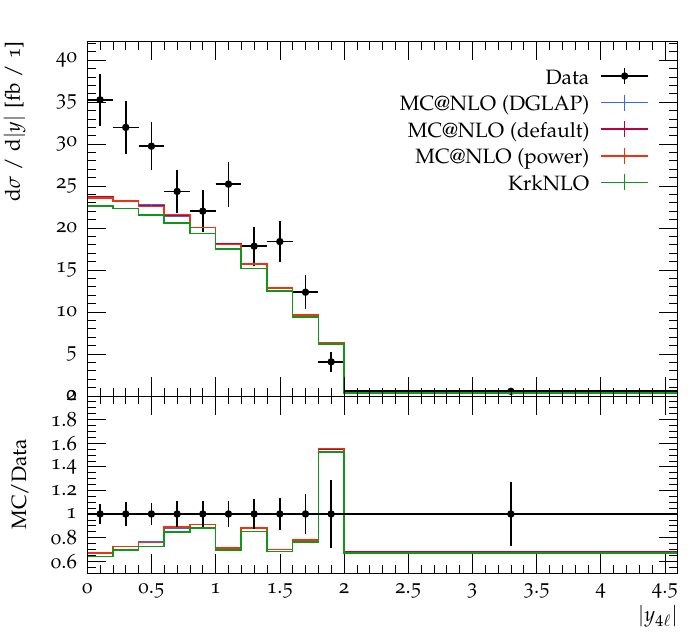} \\
        \includegraphics[width=.49\textwidth]{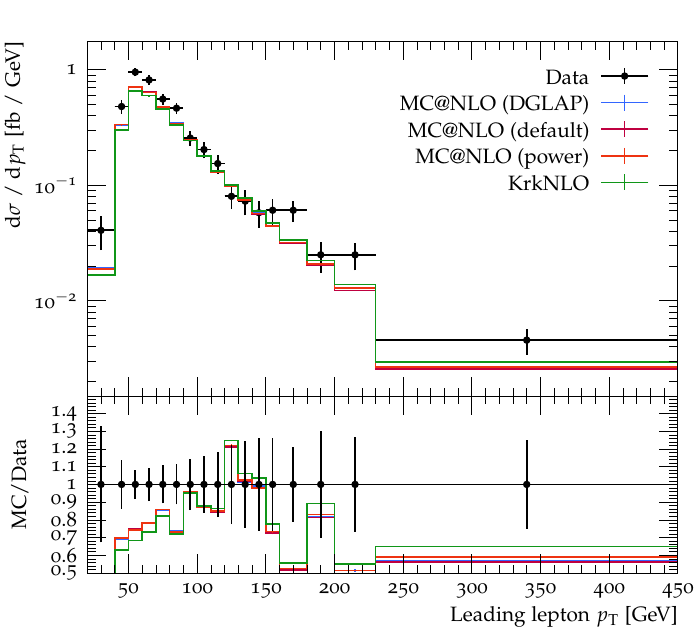}
        \hfill
        \includegraphics[width=.49\textwidth]{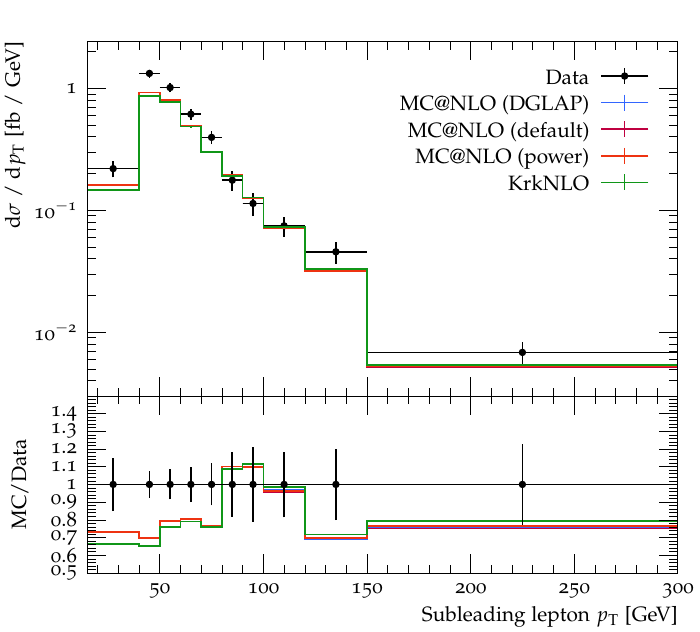} \\
        \includegraphics[width=.49\textwidth]{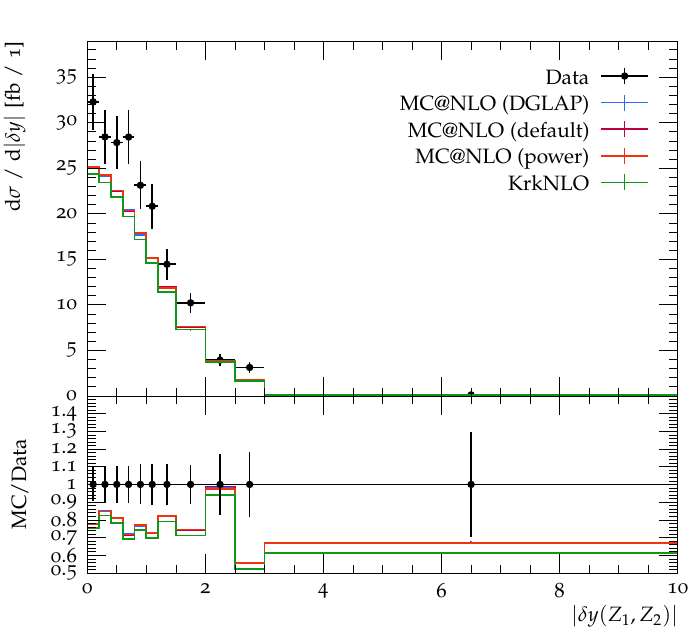}
        \hfill
        \includegraphics[width=.49\textwidth]{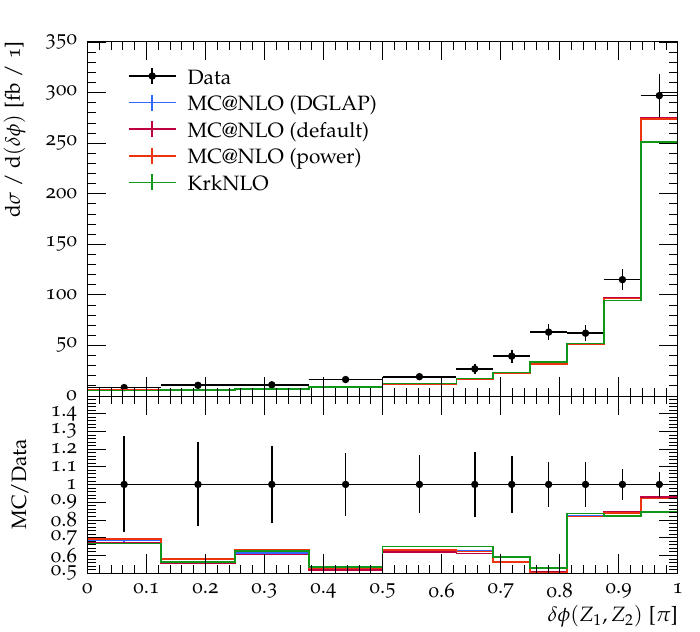}
	\caption{$ZZ$: Comparison of matched NLO-plus-parton-shower differential cross-sections
    as generated by \krknlo and \mcatnlo,
    to 13 TeV \atlas data from \cite{ATLAS:2017bcd}.
    \label{fig:ZZ_pheno_data_fullshower}}
\end{figure}

We use the generation parameters of \cref{tab:generator_cuts} and the experimental cuts of \cite{ATLAS:2017bcd},
which selects for two pairs of same-flavour, opposite-charge leptons within fiducial cuts%
\footnote{Full details of the fiducial phase-space and boson reconstruction
	procedure from the four-lepton final-state
	are given in \cite{ATLAS:2017bcd}.}
\begin{subequations}
	\label{eqn:ZZatlascuts}
	\begin{align}
		(\pt^{\ell_1}, \pt^{\ell_2}, \pt^{\ell_3}, \pt^{\ell_4}) &> (20, 15, 10, 5) \;\GeV , 	&	\absetaell &< 2.7, \\
		M_{Z_1Z_2} & \in (66, 116) \;\GeV, & M_{\ell^+\ell^-} &> 5 \;\GeV,
	\end{align}
\end{subequations}
implemented in the \rivet analysis \texttt{ATLAS\_2017\_I1625109}.
The resulting plots are shown in \cref{fig:ZZ_pheno_data_fullshower}.

Again, as discussed in \cref{sec:matching} we observe very limited matching-uncertainty between our matched-NLO predictions,
none of which describes the data,
as might be expected from the NNLO $K$-factor of
\cite{Cascioli:2014yka,Grazzini:2015hta,Kallweit:2018nyv,Heinrich:2017bvg}.
The results of \cite{Grazzini:2015hta} imply that the shape of distributions changes relatively
little between NLO and NNLO, only the normalisation.

\section{Conclusions}
\label{sec:conclusion}

In this paper we have presented the extension of the \krknlo implementation within \herwig to support
the full class of processes calculable with the method.
We applied this to four characteristic LHC processes, and investigated the matching uncertainty
of the resulting predictions, as compared with three variants of the standard \mcatnlo method.
By performing double-differential comparisons consistently across processes, we can assess the numerical effect of the formal differences between the alternative
matching algorithms.

We find generally consistent predictions between the methods, with the \krknlo predictions
mostly lying within the \mcatnlo uncertainty envelope, and closest to the `default' shower-scale choice
within \herwig.
For processes and observables very sensitive to additional radiation, the `power'-shower variant of the \mcatnlo
method inflates the \mcatnlo matching uncertainty,
leading to a wide matching-uncertainty band for inclusive observables.
In other regions of phase-space the \krknlo method provides a complementary prediction, indicating that
shower-scale variation within the \mcatnlo method alone underestimates the overall matching uncertainty.
The generally good level of agreement between the `default' \mcatnlo scale choice
and the predictions of the \krknlo method
provides empirical justification for its use as the default.

For many of the processes considered, \nlo predictions are insufficient 
to describe the full range of LHC data, due to missing contributions
including the neglected $gg$-channel.
Additionally, within the \krknlo method,
there is
an additional suppression arising from the parton-shower Sudakov factor.
As we pursue the further extension of the \krknlo method to higher levels of formal accuracy,
the numerical significance of this Sudakov suppression should be reduced.
We expect that the insights gleaned from this detailed process-by-process comparison will prove useful
and inform this work.

\acknowledgments

\enlargethispage{\baselineskip}

The authors wish to thank 
Wiesław Płaczek
and 
the late Stanisław Jadach
for their work on, and for many fruitful discussions about, the KrkNLO method.
We are grateful to Wiesław Płaczek for comments on the manuscript.

This work was supported by grant 2019/34/E/ST2/00457 of the National Science Centre, Poland.
AS is also supported by the Priority Research Area Digiworld under the program 
`Excellence Initiative -- Research University'
at the Jagiellonian University in Krakow.
AS thanks the CERN Theoretical Physics department for hospitality while part of this research
was being carried out.
We gratefully acknowledge Polish high-performance computing infrastructure PLGrid (HPC Centre: ACK Cyfronet AGH) for providing computer facilities and support within computational grants PLG/2024/017934 and PLG/2025/018065.

\appendix

\section{Positivity of KrkNLO event-weights}
\label{sec:app_pos}

The \krknlo method is formulated to generate only positive-weights by construction,
subject to the \krk-scheme PDFs being positive,%
\footnote{See \cite{Delorme:2025teo} for a study of the positivity
properties of alternative factorisation schemes, including the \krk scheme.}
and the positivity of the virtual reweight.
Because it does not use subtraction, there is no risk of negativity caused by over-subtraction,
which is a common source of negative event-weights in \mcatnlo.
As outlined for example in \cite{vanBeekveld:2025lpz}, in regions of phase-space in which perturbation
theory breaks down, large and negative virtual matrix-elements can, despite their formal suppression
within the $\alphas$-expansion, render the $\Phi_m$-contribution locally-negative.

Within the \krknlo method, as summarised in \cref{sec:nlomatching_krknlo},
the virtual reweight includes an additional virtual-like contribution $\Delta_0^\rFS$
compensating the $\delta(1-x)$ terms introduced within the PDF transformation from the \msbar to the \krk-scheme,%
\begin{align}
    1 + \frac{\alphas(\mur)}{2\pi}\left(\frac{\rV(\Phi_m;\, \mur)}{\rB(\Phi_m)} + \frac{\rI(\Phi_m;\, {\mu}_\rR)}{\rB (\Phi_m)} + \Delta_0^\rFS \right).
\end{align}
For the \krk-scheme, these terms are fixed by the choice to impose the \msbar momentum-conservation sum-rule \cite{Collins:1981uw}
\begin{align}
	\label{eq:momsumrule_PDFs}
	\sum_a \int_0^1 \xi \, f_{a}^{\rFS} (\xi, \mu) \; \dd \xi = 1 
\end{align}
on the \krk-scheme PDFs,%
\footnote{The significance of this choice, for a number of alternative
factorisation schemes, is explored in \cite{Delorme:2025teo}.}
which leads to \cite{Sarmah:2024hdk}
\begin{align}
	\Delta_0^{\krk}
	=
	\cf \left( 2 \pi^2 - \frac{3}{2} \right)
    \approx
    24.3189.
\end{align}
Since this is large and positive, it has the effect of further mitigating negativity even in regions of phase-space
where a negative virtual matrix-element alone would `overpower' the positivity of the Born matrix-element in isolation.
As an indicative value, $\Delta_0^\krk \, \alphas(M_Z) / 2\pi \approx 0.46$.

\begin{figure}[p]
	\centering
    \begin{subfigure}[t]{0.41\textwidth}
        \includegraphics[width=\textwidth]{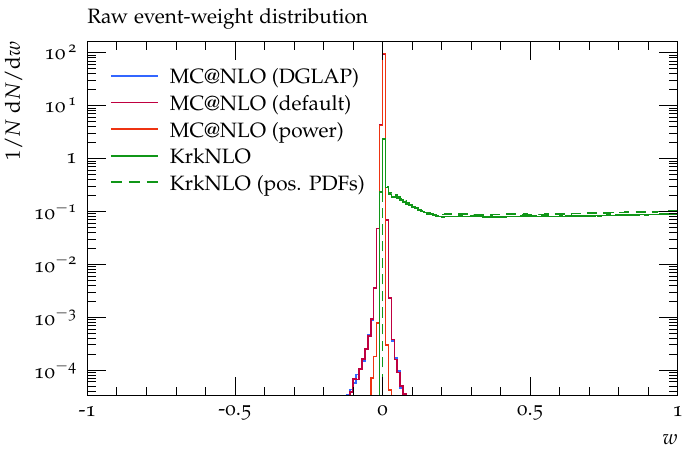}
        \caption{$W$ \label{fig:weight_distributions_W}}
    \end{subfigure}
    \begin{subfigure}[t]{0.41\textwidth}
        \includegraphics[width=\textwidth]{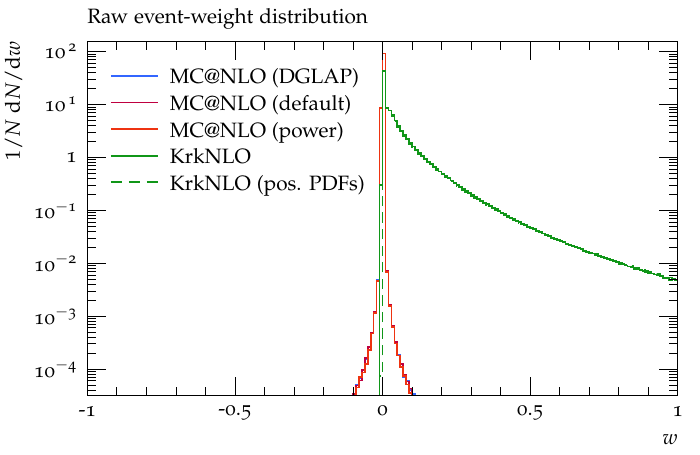}
        \caption{$Z\gamma$ \label{fig:weight_distributions_Zgam}}
    \end{subfigure}
    \begin{subfigure}[t]{0.41\textwidth}
        \includegraphics[width=\textwidth]{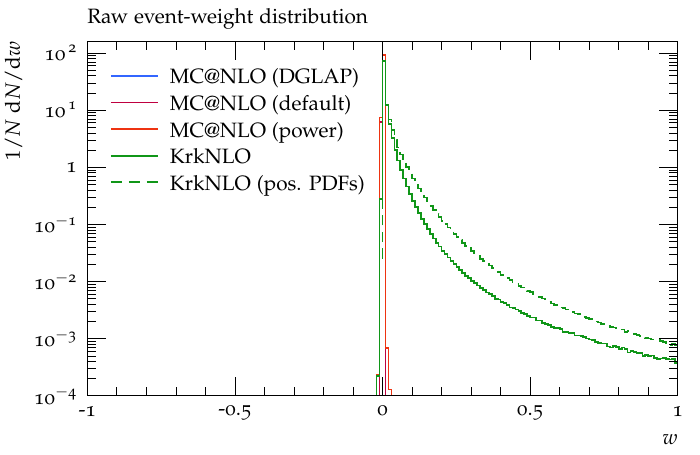}
        \caption{$WW$ \label{fig:weight_distributions_WW}}
    \end{subfigure}
    \begin{subfigure}[t]{0.41\textwidth}
        \includegraphics[width=\textwidth]{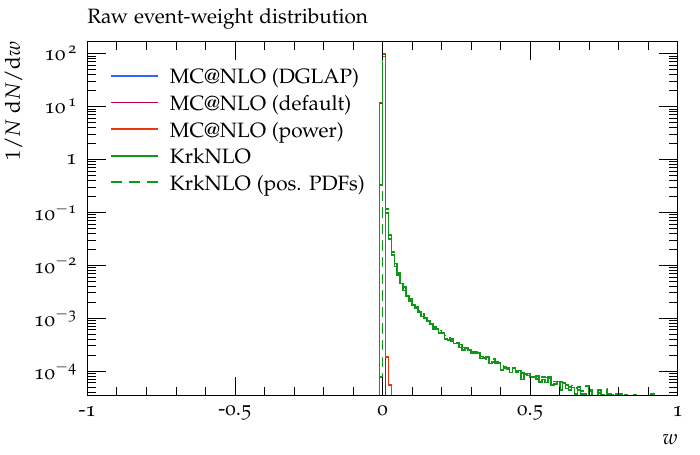}
        \caption{$ZZ$ \label{fig:weight_distributions_ZZ}}
    \end{subfigure}        
	\caption{Event-weight distributions of the \krknlo and \mcatnlo methods,
    for the generator set-up used for the predictions in \cref{sec:matching_full,sec:LHCpheno}.
    Note that only the generation cuts of \cref{tab:generator_cuts} have been applied, and no analysis cuts.
    \label{fig:weight_distributions}}
\end{figure}

\begin{figure}[p]
	\centering
	\begin{subfigure}[t]{0.41\textwidth}
		\includegraphics[width=\textwidth]{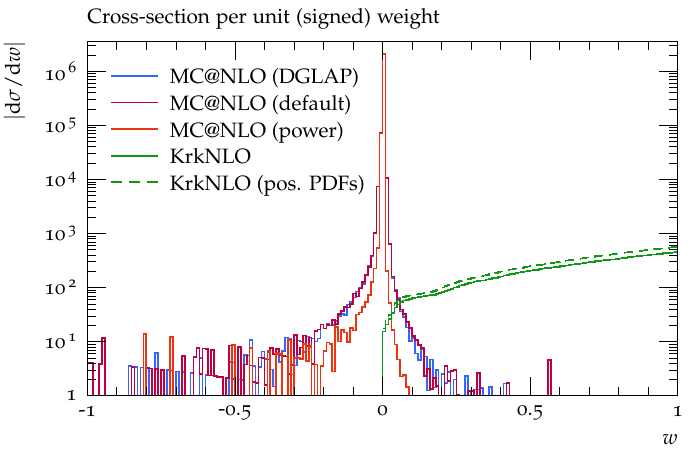}
		\caption{$W$ \label{fig:dxsdw_W}}
	\end{subfigure}
	\begin{subfigure}[t]{0.41\textwidth}
		\includegraphics[width=\textwidth]{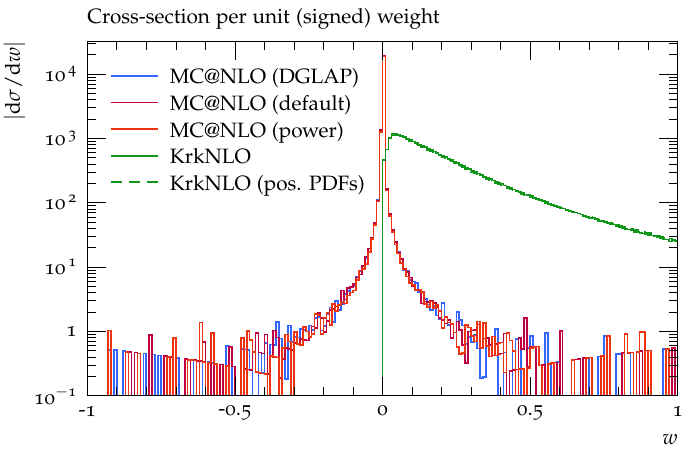}
		\caption{$Z\gamma$ \label{fig:dxsdw_Zgam}}
	\end{subfigure}
	\begin{subfigure}[t]{0.41\textwidth}
		\includegraphics[width=\textwidth]{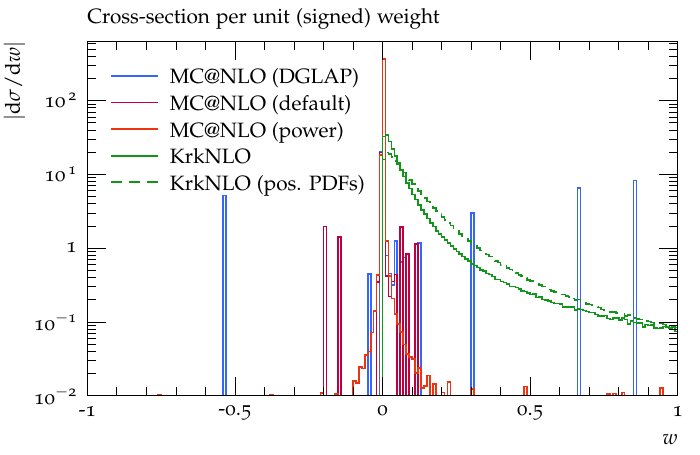}
		\caption{$WW$ \label{fig:dxsdw_WW}}
	\end{subfigure}
	\begin{subfigure}[t]{0.41\textwidth}
		\includegraphics[width=\textwidth]{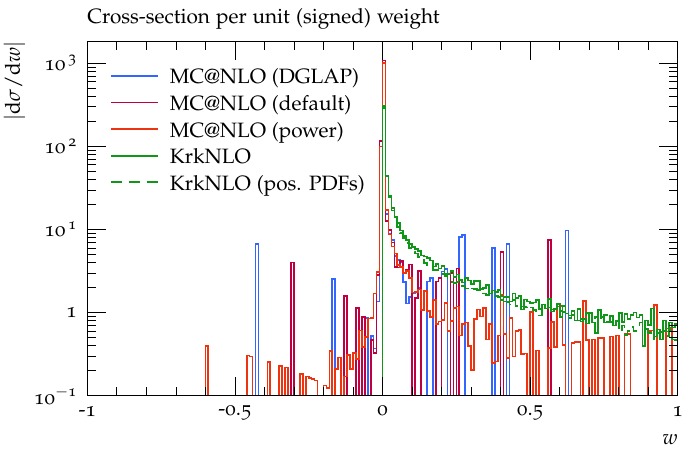}
		\caption{$ZZ$ \label{fig:dxsdw_ZZ}}
	\end{subfigure}        
	\caption{Magnitude of the differential cross-section with respect to event weight for the \krknlo and \mcatnlo methods,
		for the generator set-up used in \cref{sec:matching_full,sec:LHCpheno}.
		Note that only the generation cuts of \cref{tab:generator_cuts} have been applied, and no analysis cuts.
		\label{fig:dxsdw}}
\end{figure}

In \cref{fig:weight_distributions} we show the empirical weight-distributions obtained from the generator set-up used for \cref{sec:matching_full,sec:LHCpheno},
prior to the analysis cuts.
In \cref{fig:dxsdw} we show the differential cross-section with respect to
the event weight.
As can be seen from the weight distributions, within the \krknlo method the problem of negative weights
is almost entirely eliminated, with a small number visible
in the bin adjoining 0 in \cref{fig:weight_distributions}
(though not \cref{fig:dxsdw}, due to their vanishing contribution to the cross-section).
This is in contrast with the \mcatnlo predictions, which show the characteristic
weight-distributions of subtractive calculations.

We illustrate the origin of these negative weights
by also showing equivalent KrkNLO runs made using explicitly non-negative PDF sets.%
\footnote{In practice, we achieve this using the \texttt{ForcePositive} switch within the LHAPDF library for \texttt{CT18NLO} in the \krk scheme, and the positive-definite \texttt{NNPDF40MC} PDF set for the shower.}
These results confirm that weight-negativity observed in the KrkNLO method
for these processes is almost entirely attributable to
the negativity of the transformed \krk-scheme \texttt{CT18NLO} 
PDF set in certain corners of $(x,Q)$-space, as explored in \cite{Delorme:2025teo}.
For practical purposes both choices can be seen to have negligible
event-weight-negativity, for these processes,
and therefore to provide a solution to the negative-weight problem in event generation.

\section{Unabridged double-differential distributions}
\label{sec:app_ddplots}

\begin{figure}[p]
	\centering
    \vspace{-10mm}
        \includegraphics[width=.32\textwidth]{fig/pheno/Zgam/oneemission/double_differential/mlly_ptj1_1.pdf}
        \includegraphics[width=.32\textwidth]{fig/pheno/WW/oneemission/mc_vector_diboson/m_WW_fine_ptj1_in_1_to_20.pdf}
        \includegraphics[width=.32\textwidth]{fig/pheno/ZZ/oneemission/R_mc_vector_diboson_ZZ/m_ZZ_fine_ptj1_in_1_to_20.pdf}
        \includegraphics[width=.32\textwidth]{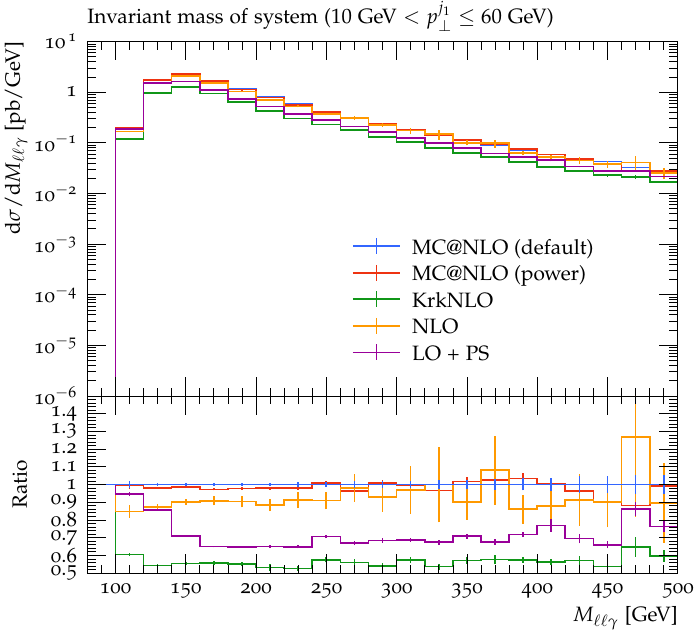}
        \includegraphics[width=.32\textwidth]{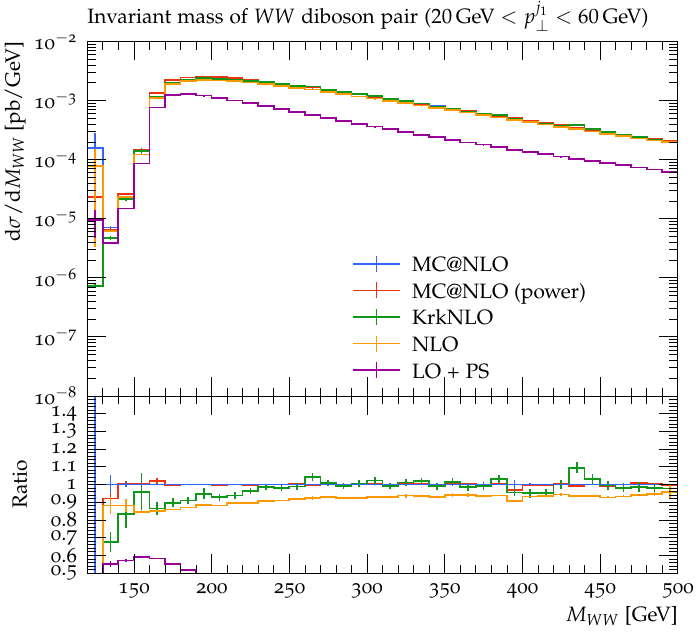}
        \includegraphics[width=.32\textwidth]{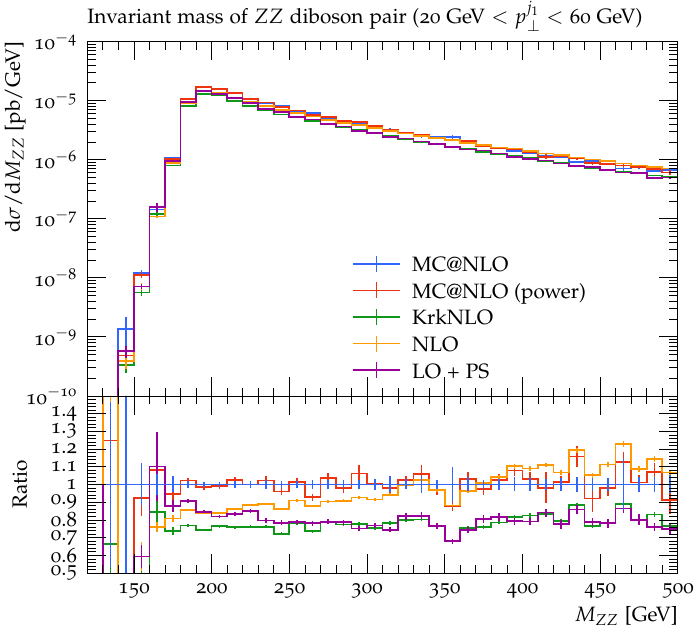}
        \includegraphics[width=.32\textwidth]{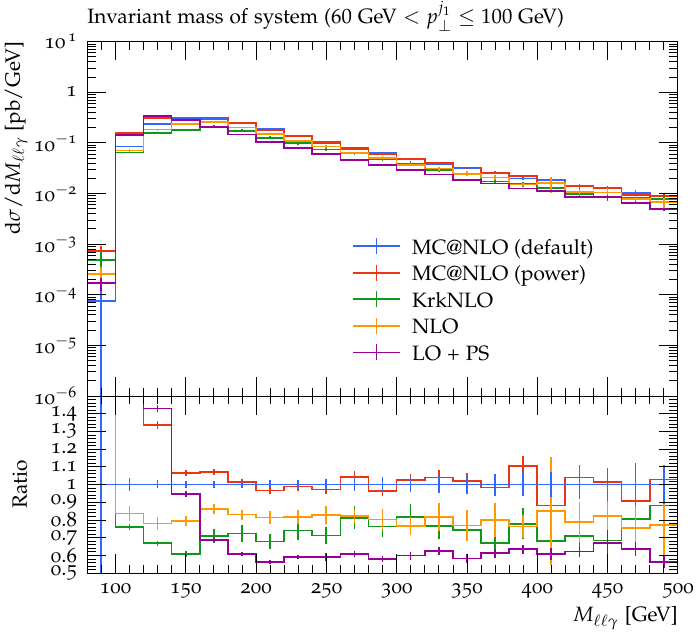}
        \includegraphics[width=.32\textwidth]{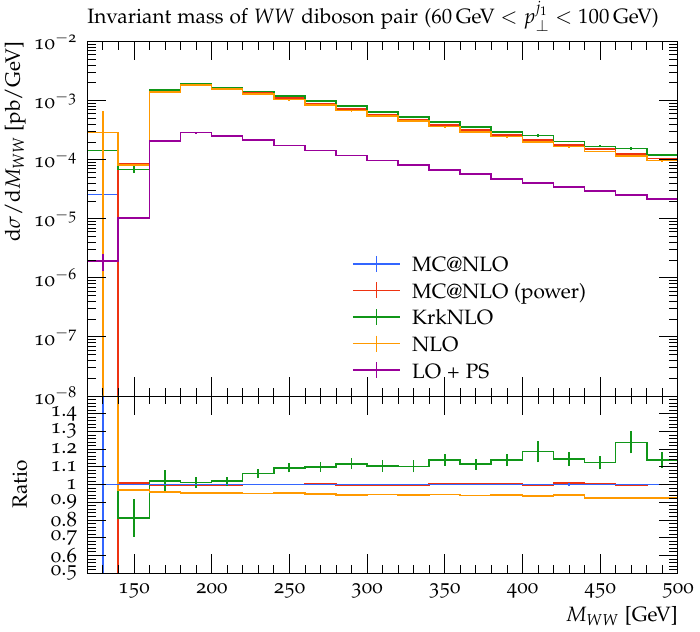}
        \includegraphics[width=.32\textwidth]{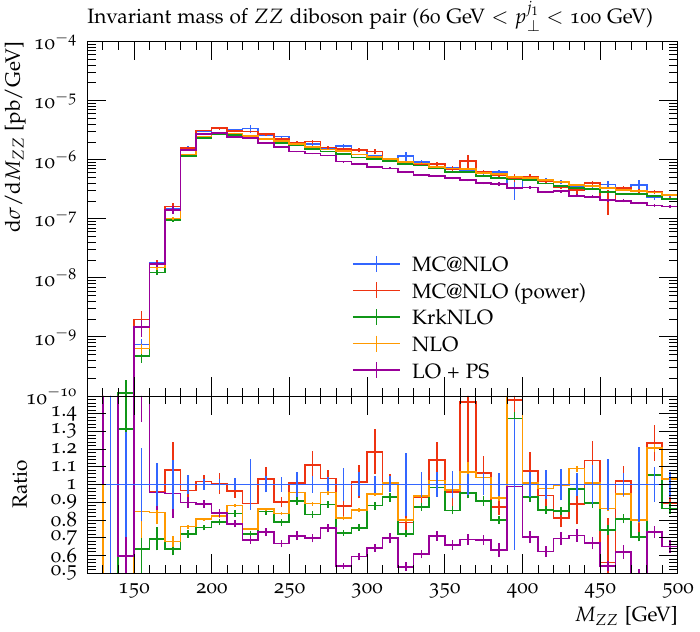}
        \includegraphics[width=.32\textwidth]{fig/pheno/Zgam/oneemission/double_differential/mlly_ptj1_4.pdf}
        \includegraphics[width=.32\textwidth]{fig/pheno/WW/oneemission/mc_vector_diboson/m_WW_fine_ptj1_in_100_to_500.pdf}
        \includegraphics[width=.32\textwidth]{fig/pheno/ZZ/oneemission/R_mc_vector_diboson_ZZ/m_ZZ_fine_ptj1_in_100_to_500.pdf}
        \begin{subfigure}[b]{.32\textwidth}
        \includegraphics[width=\textwidth]{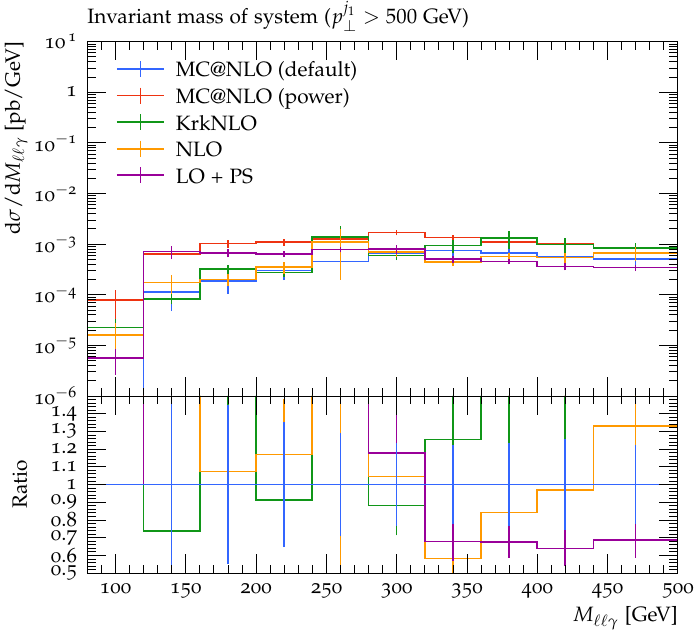}
        \caption{$Z \gamma$}
        \end{subfigure}
        \begin{subfigure}[b]{.32\textwidth}
        \includegraphics[width=\textwidth]{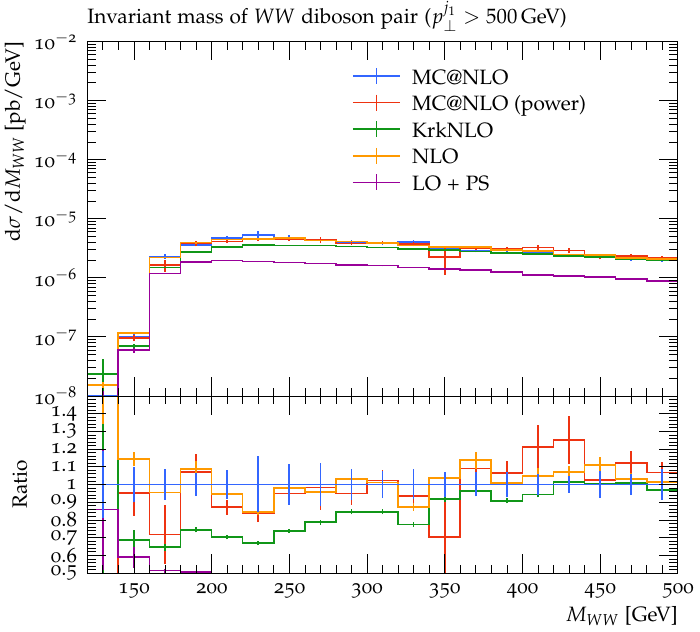}
        \caption{$WW$}
        \end{subfigure}
        \begin{subfigure}[b]{.32\textwidth}
        \includegraphics[width=\textwidth]{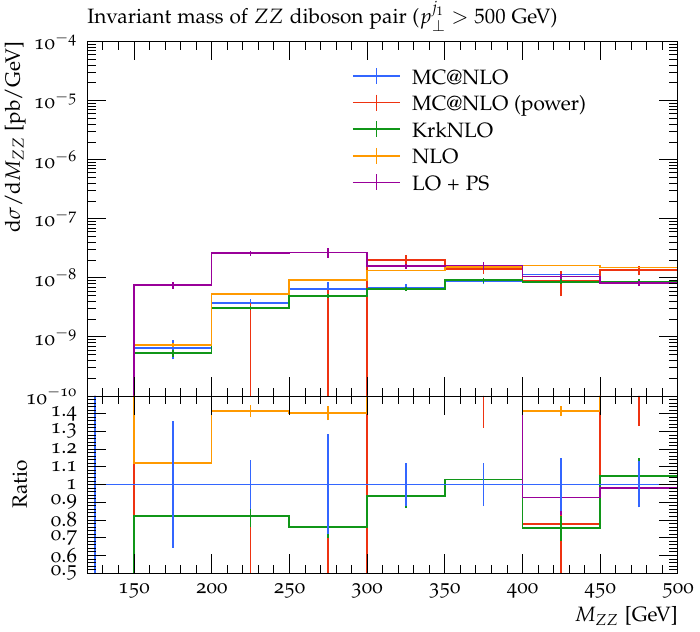}
        \caption{$ZZ$}
        \end{subfigure}
	\caption{`Parton-level' (first-emission) comparison of the invariant mass of the
		 colour-singlet system in slices of the transverse momentum
		 of the leading jet; the complete partition of which a subset is shown in \cref{fig:oneemission_dsigma_dM_dptj}.
    \label{fig:oneemission_dsigma_dM_dptj_full}}
    \vspace{-10mm}
\end{figure}

\begin{figure}[tp]
    \vspace{-10mm}
	\centering
        \includegraphics[width=.32\textwidth]{fig/pheno/Zgam/fullshower/double_differential/mlly_ptj1_1.pdf}
        \includegraphics[width=.32\textwidth]{fig/pheno/WW/fullshower/mc_vector_diboson/m_WW_fine_ptj1_in_1_to_20.pdf}
        \includegraphics[width=.32\textwidth]{fig/pheno/ZZ/fullshower/mc_vector_diboson_ZZ/m_ZZ_fine_ptj1_in_1_to_20.pdf}
        \includegraphics[width=.32\textwidth]{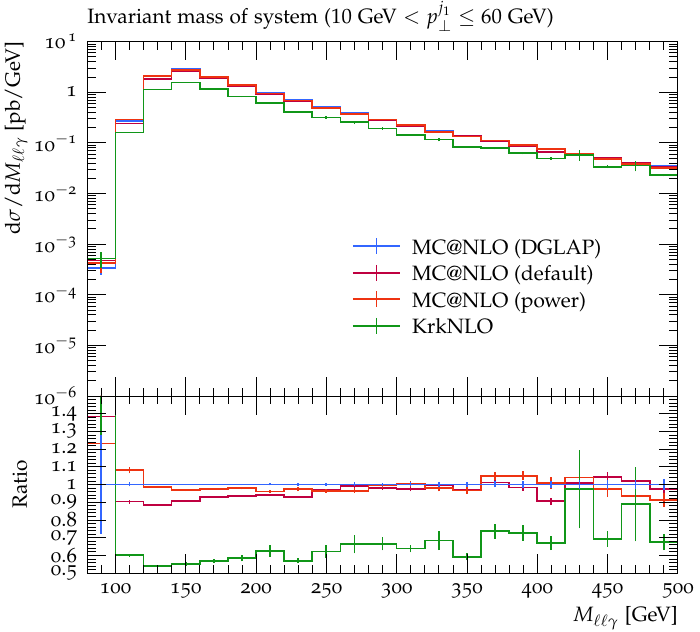}
        \includegraphics[width=.32\textwidth]{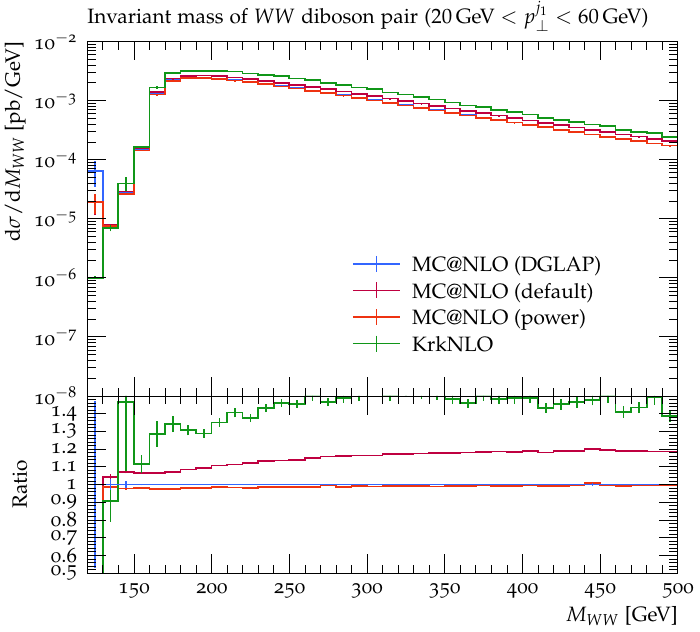}
        \includegraphics[width=.32\textwidth]{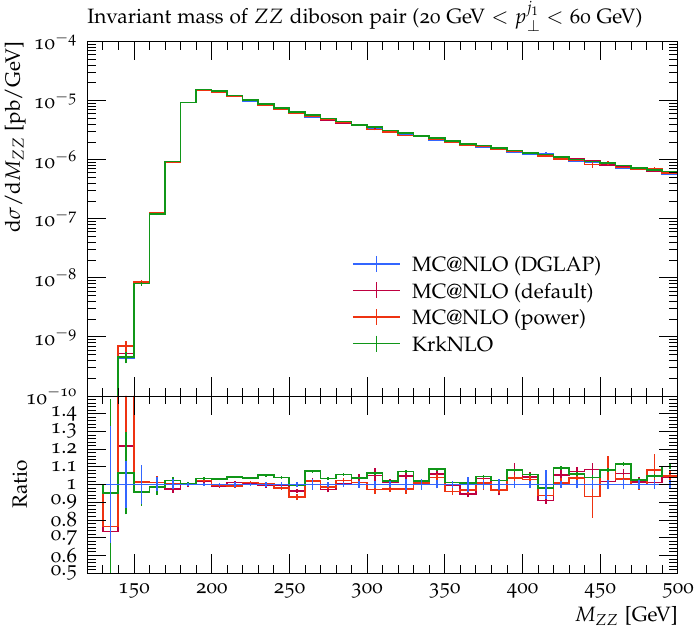}
        \includegraphics[width=.32\textwidth]{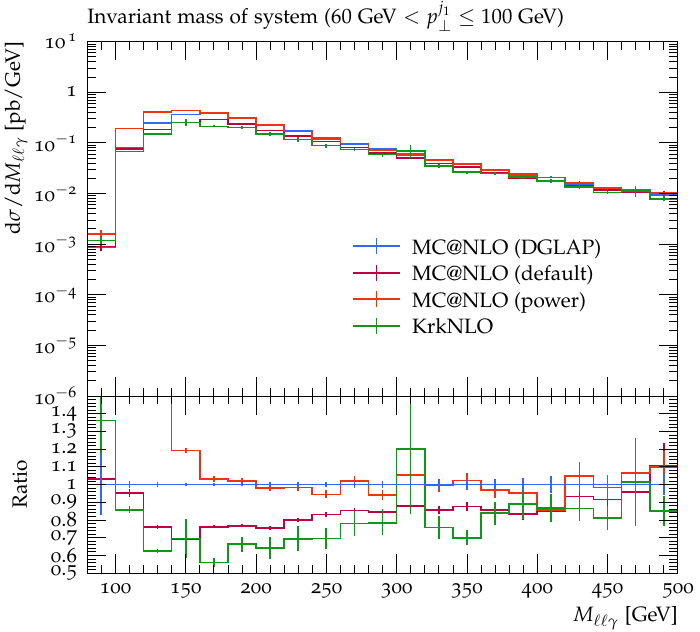}
        \includegraphics[width=.32\textwidth]{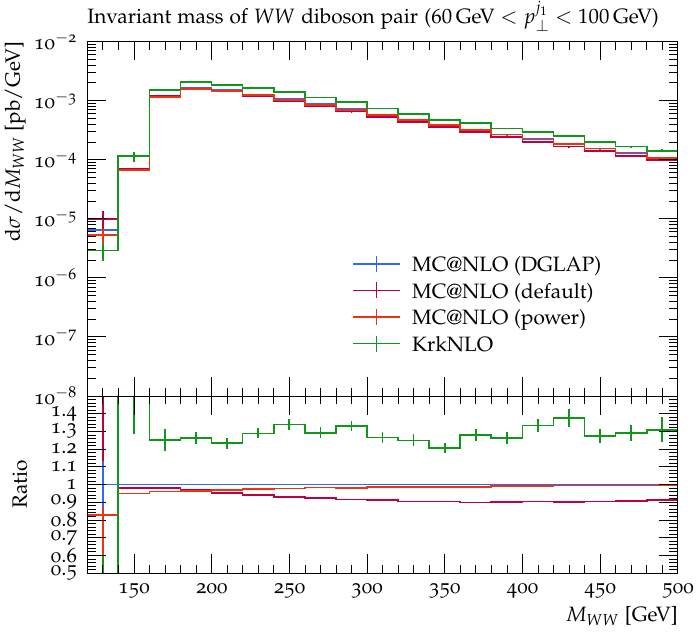}
        \includegraphics[width=.32\textwidth]{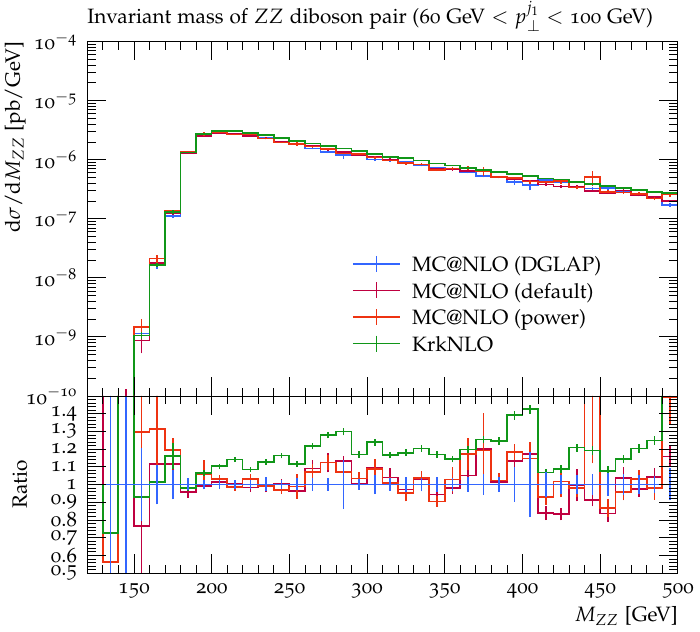}
        \includegraphics[width=.32\textwidth]{fig/pheno/Zgam/fullshower/double_differential/mlly_ptj1_4.pdf}
        \includegraphics[width=.32\textwidth]{fig/pheno/WW/fullshower/mc_vector_diboson/m_WW_fine_ptj1_in_100_to_500.pdf}
        \includegraphics[width=.32\textwidth]{fig/pheno/ZZ/fullshower/mc_vector_diboson_ZZ/m_ZZ_fine_ptj1_in_100_to_500.pdf}
        \begin{subfigure}[b]{.32\textwidth}
        \includegraphics[width=\textwidth]{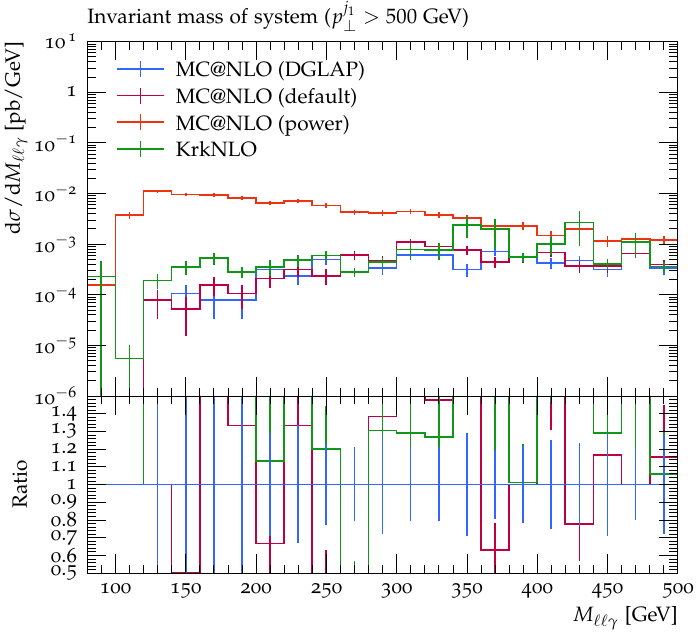}
        \caption{$Z \gamma$}
	    \end{subfigure}
        \begin{subfigure}[b]{.32\textwidth}
        \includegraphics[width=\textwidth]{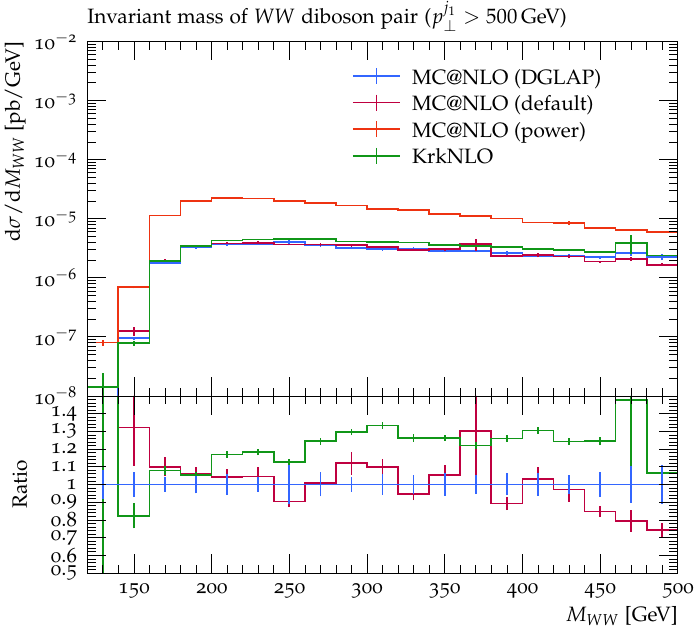}
        \caption{$WW$}
	    \end{subfigure}
        \begin{subfigure}[b]{.32\textwidth}
        \includegraphics[width=\textwidth]{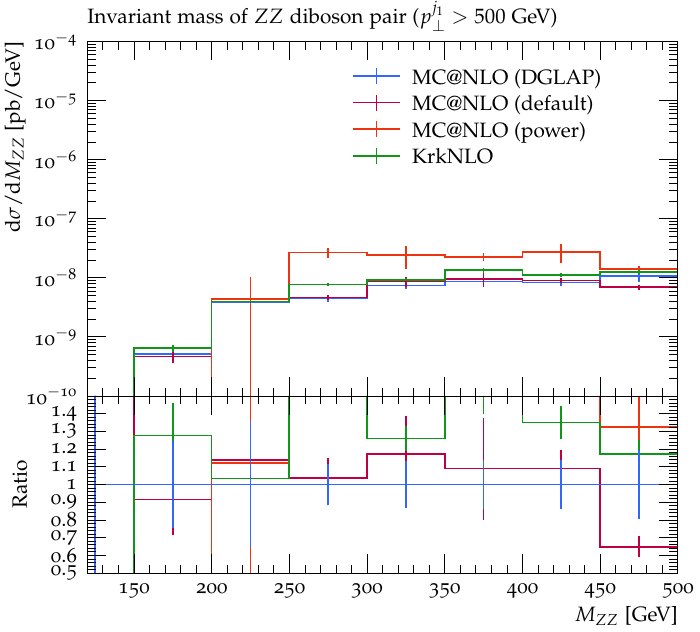}
        \caption{$ZZ$}
        \end{subfigure}
	\caption{Full-shower comparison of the invariant mass of the
		 colour-singlet system in slices of the transverse momentum
		 of the leading jet;
    the complete partition of which a subset is shown in \cref{fig:fullshower_dsigma_dM_dptj}.
    \label{fig:fullshower_dsigma_dM_dptj_full}}
    \vspace{-10mm}
\end{figure}

\begin{figure}[tp]
	\centering
        \vspace{-15mm}
        \makebox[\textwidth][c]{%
        \includegraphics[width=0.2\paperwidth]{fig/pheno/W/fullshower/double_differential/W_mass_fine_1.pdf}
        \includegraphics[width=0.2\paperwidth]{fig/pheno/Zgam/fullshower/double_differential/m_lly_fine_dphilly_1.pdf}
        \includegraphics[width=0.2\paperwidth]{fig/pheno/WW/fullshower/mc_vector_diboson/m_WW_fine_1.pdf}
        \includegraphics[width=0.2\paperwidth]{fig/pheno/ZZ/fullshower/mc_vector_diboson_ZZ/m_ZZ_fine_1.pdf}
        } \makebox[\textwidth][c]{%
        \includegraphics[width=0.2\paperwidth]{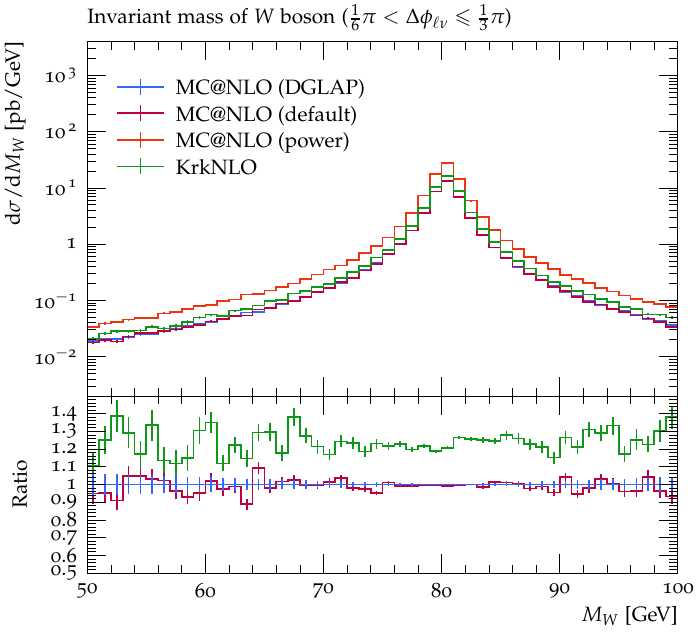}
        \includegraphics[width=0.2\paperwidth]{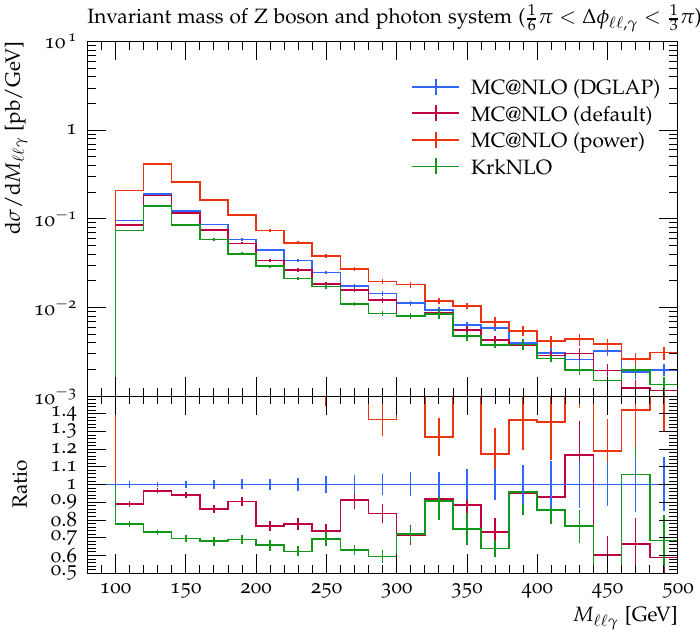}
        \includegraphics[width=0.2\paperwidth]{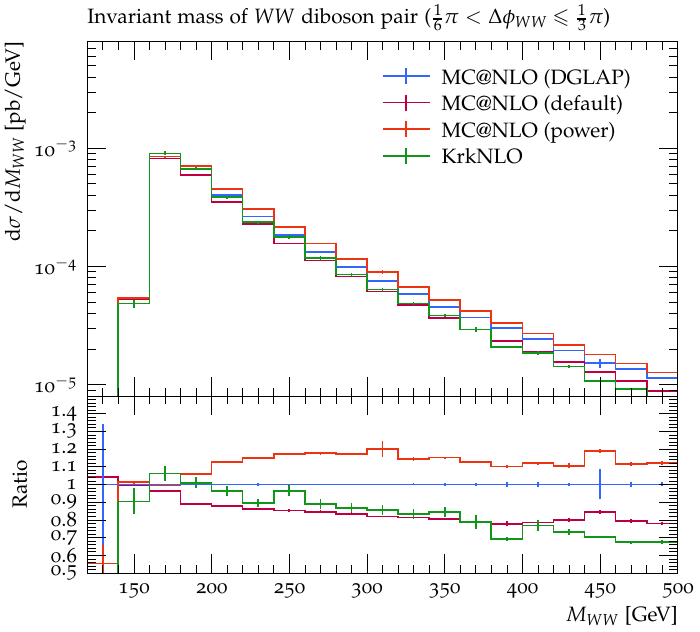}
        \includegraphics[width=0.2\paperwidth]{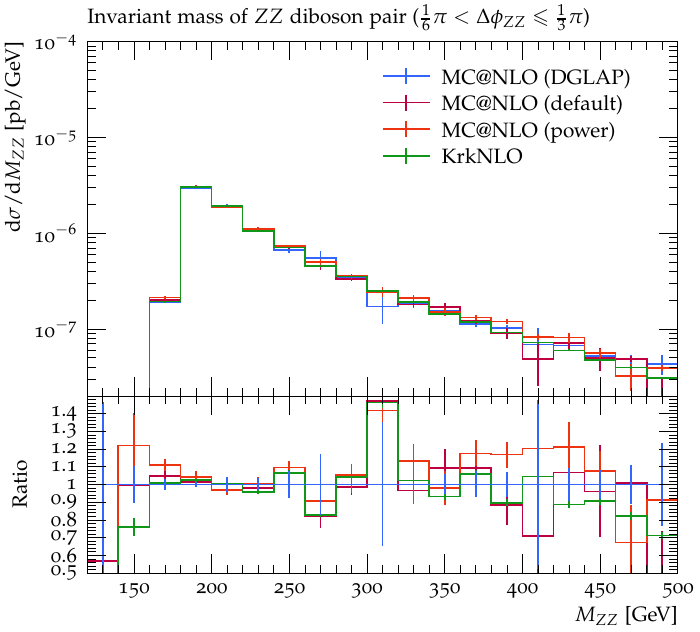}
        } \makebox[\textwidth][c]{%
        \includegraphics[width=0.2\paperwidth]{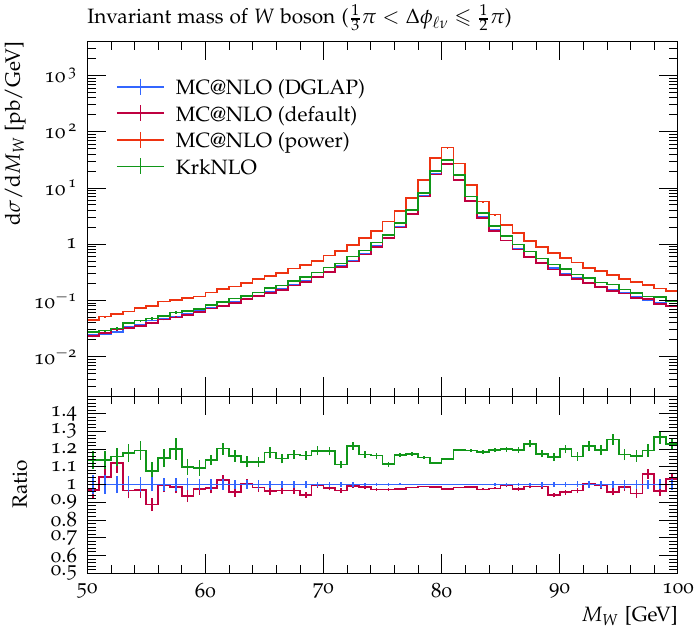}
        \includegraphics[width=0.2\paperwidth]{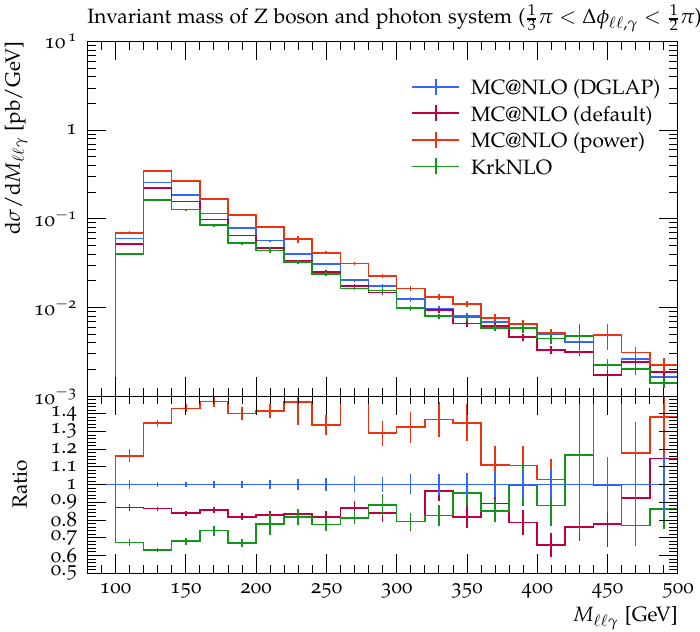}
        \includegraphics[width=0.2\paperwidth]{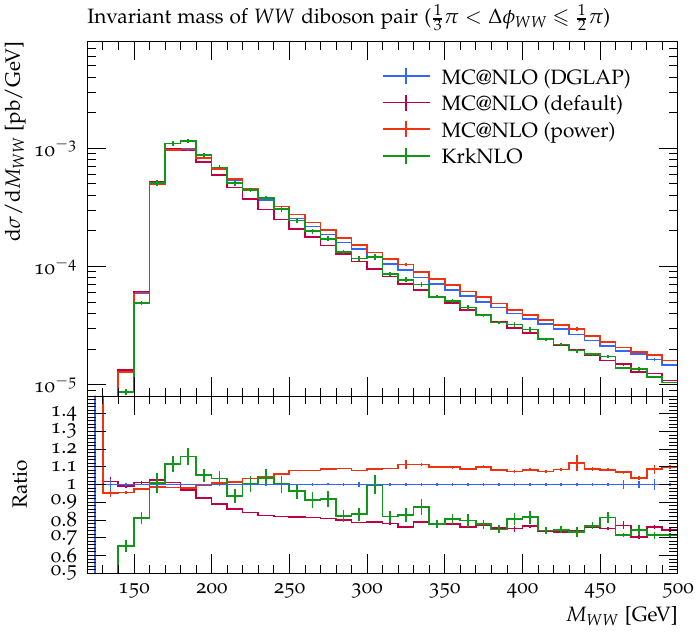}
        \includegraphics[width=0.2\paperwidth]{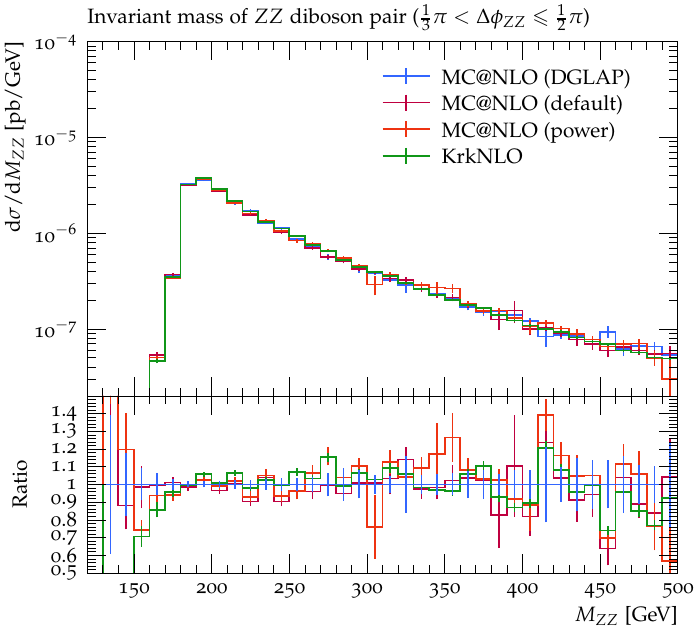}
        } \makebox[\textwidth][c]{%
        \includegraphics[width=0.2\paperwidth]{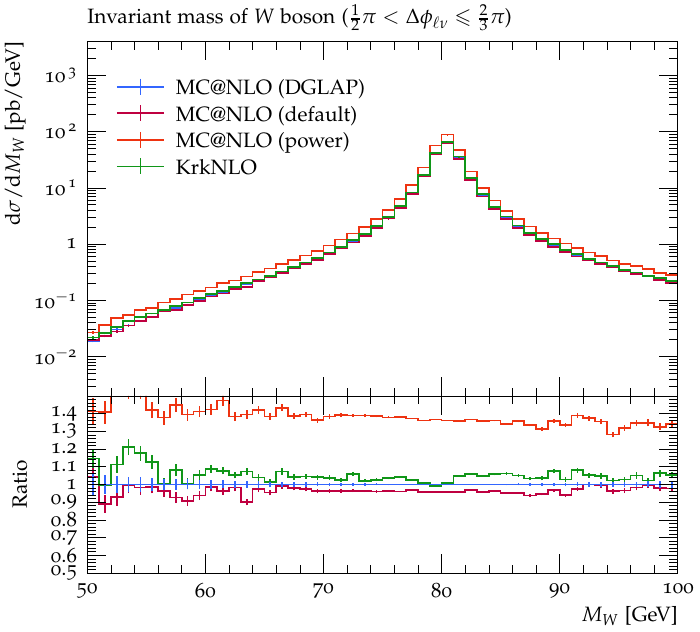}
        \includegraphics[width=0.2\paperwidth]{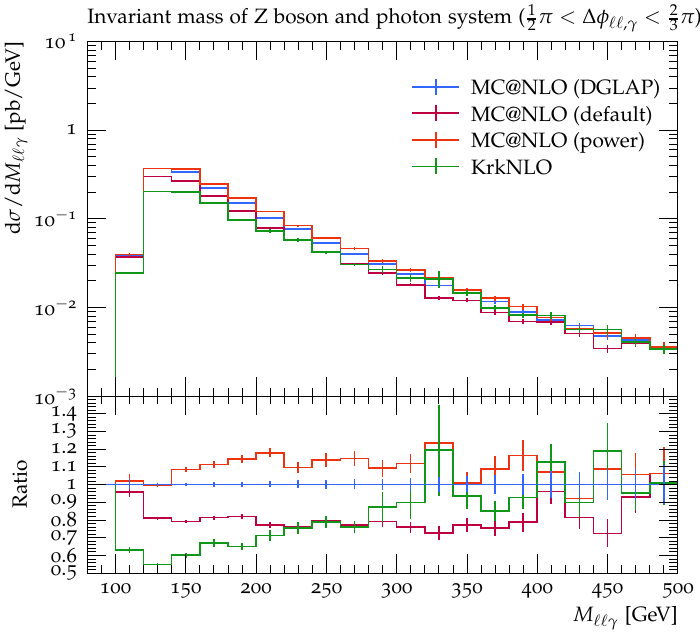}
        \includegraphics[width=0.2\paperwidth]{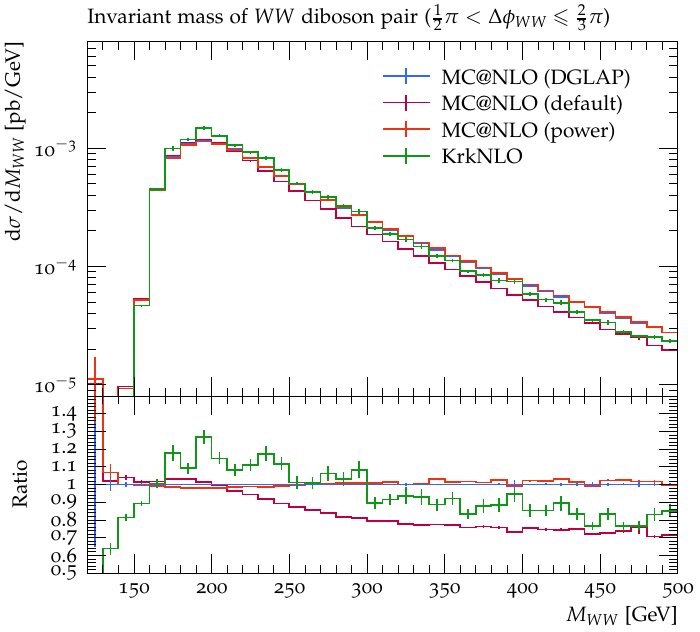}
        \includegraphics[width=0.2\paperwidth]{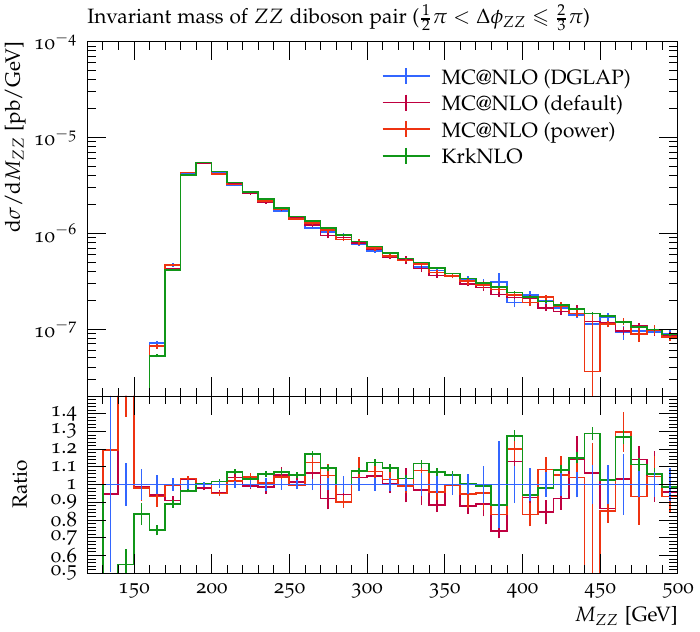}
        } \makebox[\textwidth][c]{%
        \includegraphics[width=0.2\paperwidth]{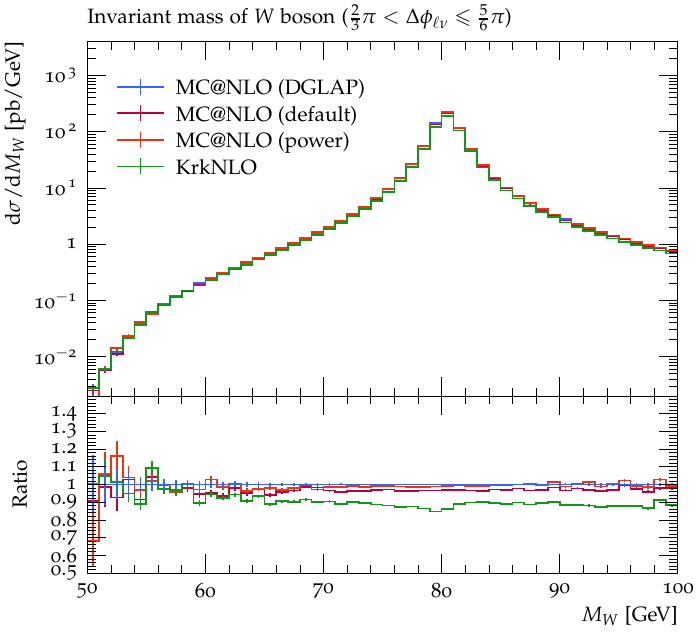}
        \includegraphics[width=0.2\paperwidth]{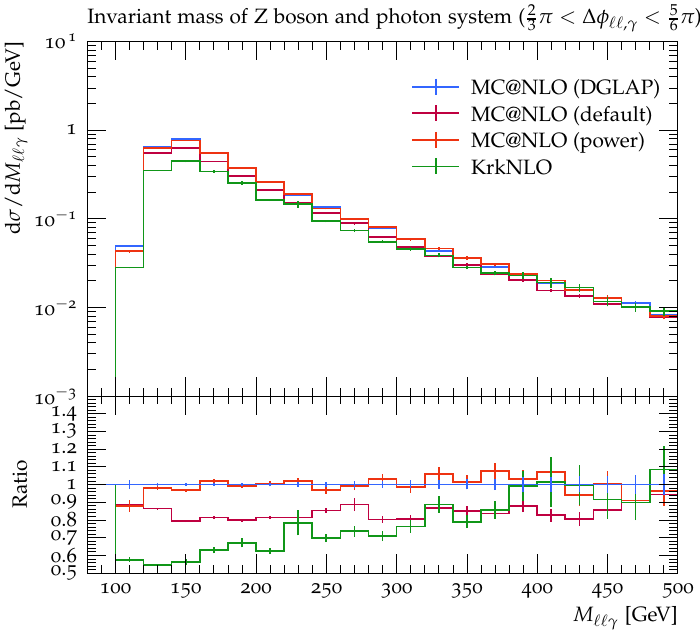}
        \includegraphics[width=0.2\paperwidth]{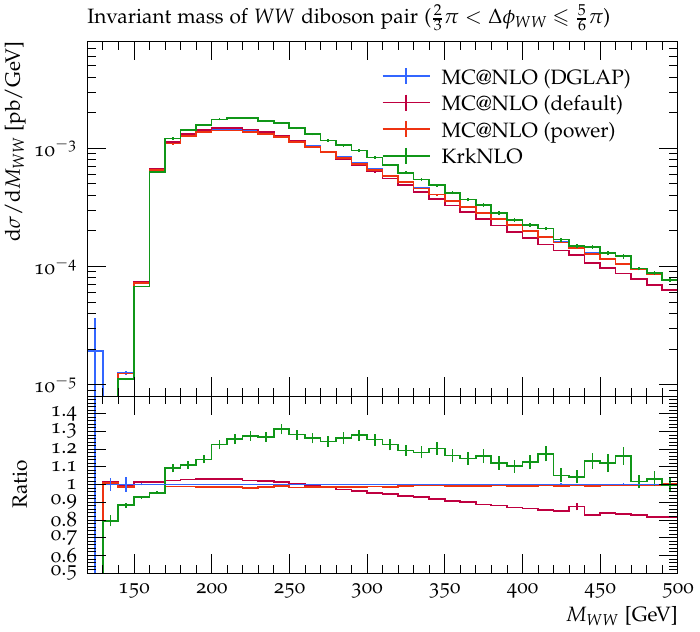}
        \includegraphics[width=0.2\paperwidth]{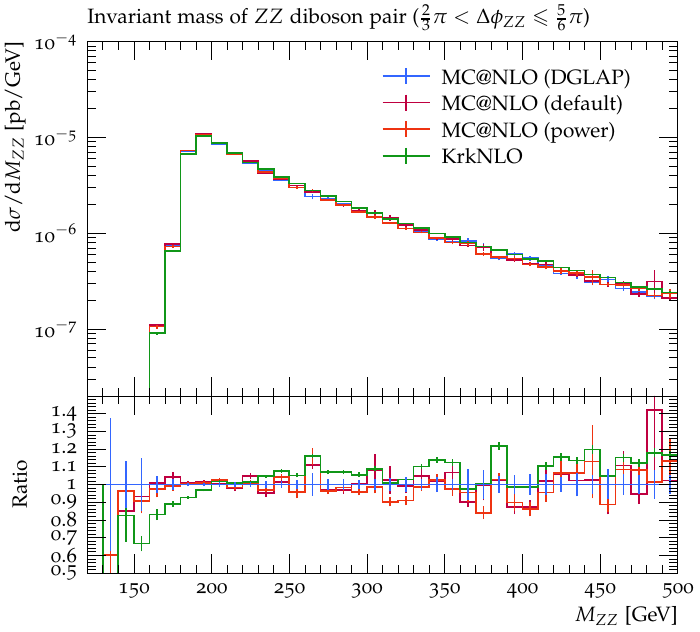}
        } \makebox[\textwidth][c]{%
        \begin{subfigure}[b]{0.2\paperwidth}
        \includegraphics[width=\textwidth]{fig/pheno/W/fullshower/double_differential/W_mass_fine_6.pdf}
        \caption{$W$}
        \end{subfigure}
        \begin{subfigure}[b]{0.2\paperwidth}
        \includegraphics[width=\textwidth]{fig/pheno/Zgam/fullshower/double_differential/m_lly_fine_dphilly_6.pdf}
        \caption{$Z\gamma$}
        \end{subfigure}
        \begin{subfigure}[b]{0.2\paperwidth}
        \includegraphics[width=\textwidth]{fig/pheno/WW/fullshower/mc_vector_diboson/m_WW_fine_6.pdf}
        \caption{$WW$}
        \end{subfigure}
        \begin{subfigure}[b]{0.2\paperwidth}
        \includegraphics[width=\textwidth]{fig/pheno/ZZ/fullshower/mc_vector_diboson_ZZ/m_ZZ_fine_6.pdf}
        \caption{$ZZ$}
        \end{subfigure}
        }
	\caption{Full-shower comparison of the invariant mass of the
		 colour-singlet system in slices of the azimuthal angle between the
         (reconstructed)
         two-particle constituents of the colour-singlet system;
         the complete partition of which a subset is shown in \cref{fig:fullshower_dsigma_dM_dphi}.
    \label{fig:fullshower_dsigma_dM_dphi_full}}
    \vspace{-15mm}
\end{figure}

\begin{figure}[htp]
	\centering
        \vspace{-15mm}
        \makebox[\textwidth][c]{%
        \includegraphics[width=0.2\paperwidth]{fig/pheno/W/fullshower/double_differential/ptj1_fine_1.pdf}
        \includegraphics[width=0.2\paperwidth]{fig/pheno/Zgam/fullshower/double_differential/pT_j1_fine_dphilly_1.pdf}
        \includegraphics[width=0.2\paperwidth]{fig/pheno/WW/fullshower/mc_vector_diboson/pT_j1_fine_1.pdf}
        \includegraphics[width=0.2\paperwidth]{fig/pheno/ZZ/fullshower/mc_vector_diboson_ZZ/pT_j1_fine_1.pdf}
        }
        \makebox[\textwidth][c]{
        \includegraphics[width=0.2\paperwidth]{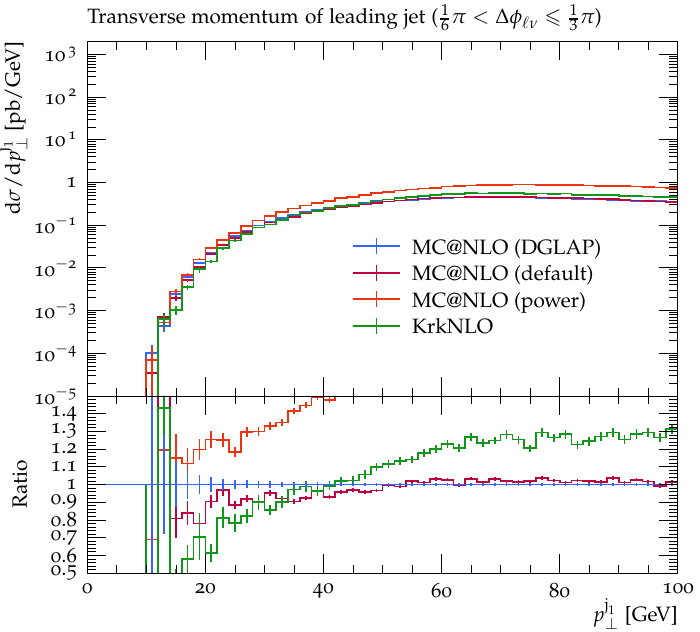}
        \includegraphics[width=0.2\paperwidth]{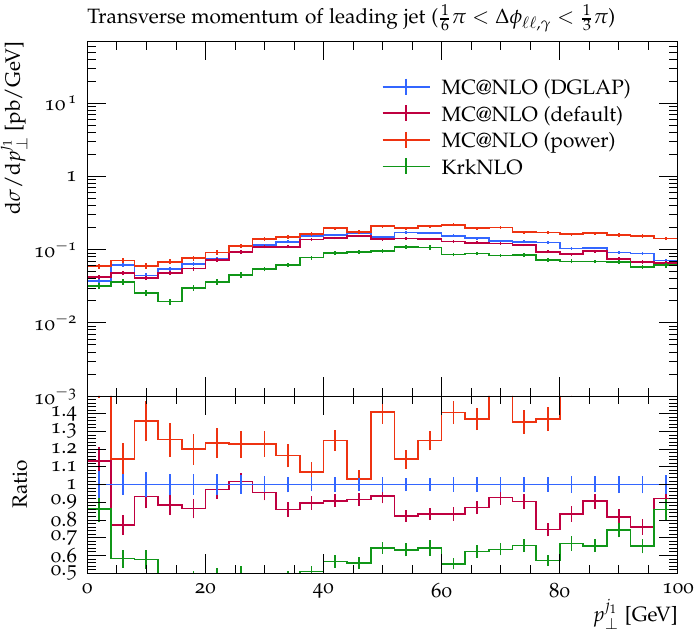}
        \includegraphics[width=0.2\paperwidth]{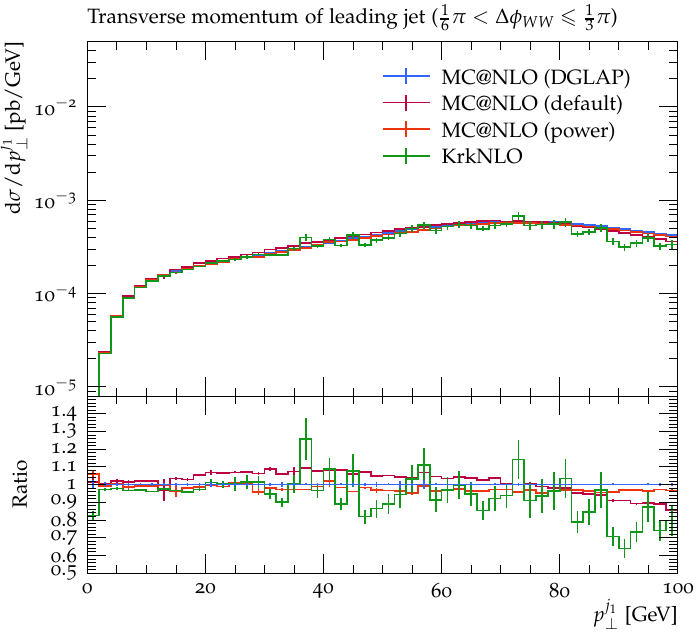}
        \includegraphics[width=0.2\paperwidth]{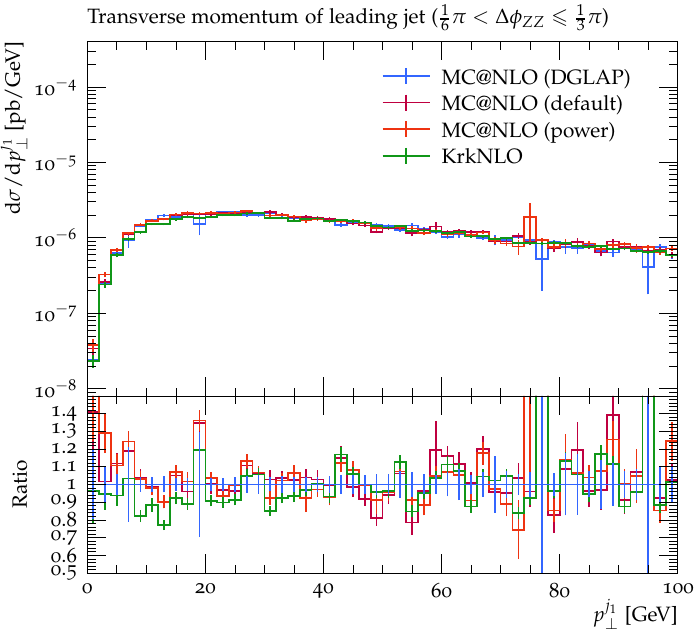}
        } \makebox[\textwidth][c]{
        \includegraphics[width=0.2\paperwidth]{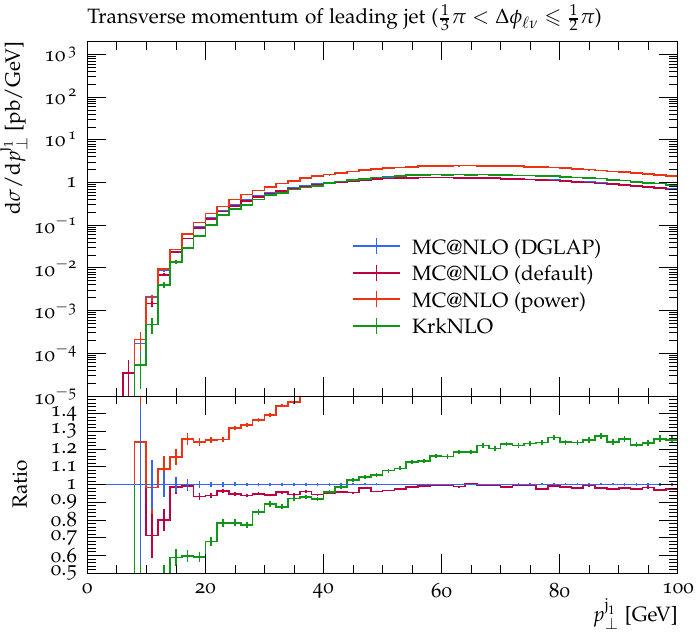}
        \includegraphics[width=0.2\paperwidth]{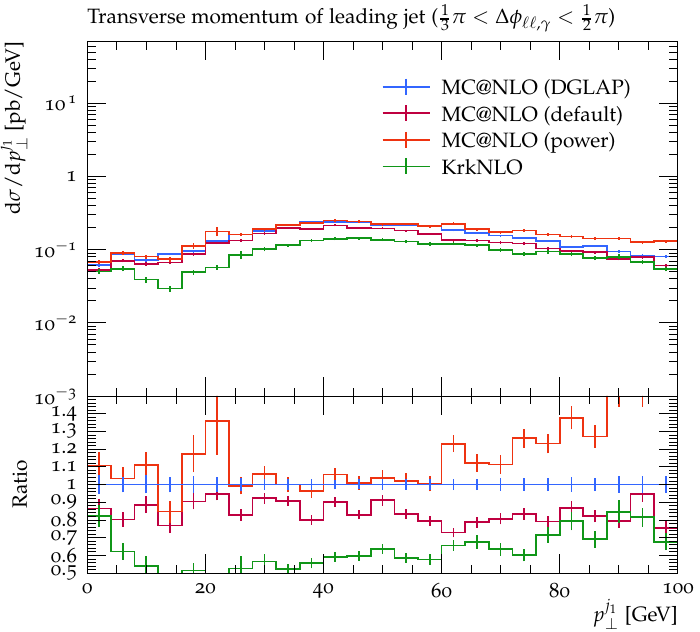}
        \includegraphics[width=0.2\paperwidth]{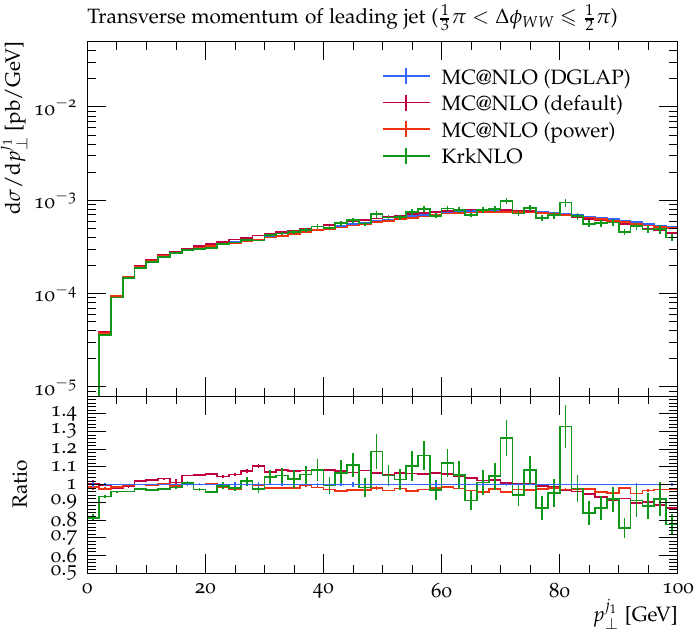}
        \includegraphics[width=0.2\paperwidth]{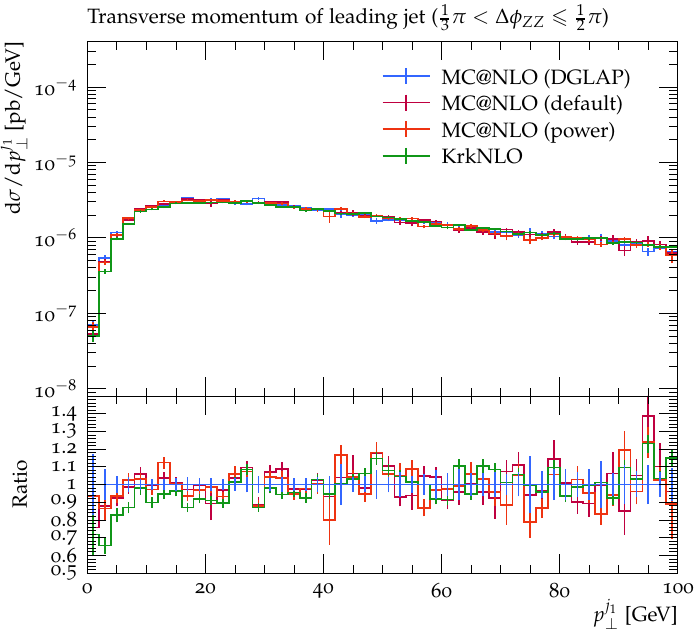}
        } \makebox[\textwidth][c]{
        \includegraphics[width=0.2\paperwidth]{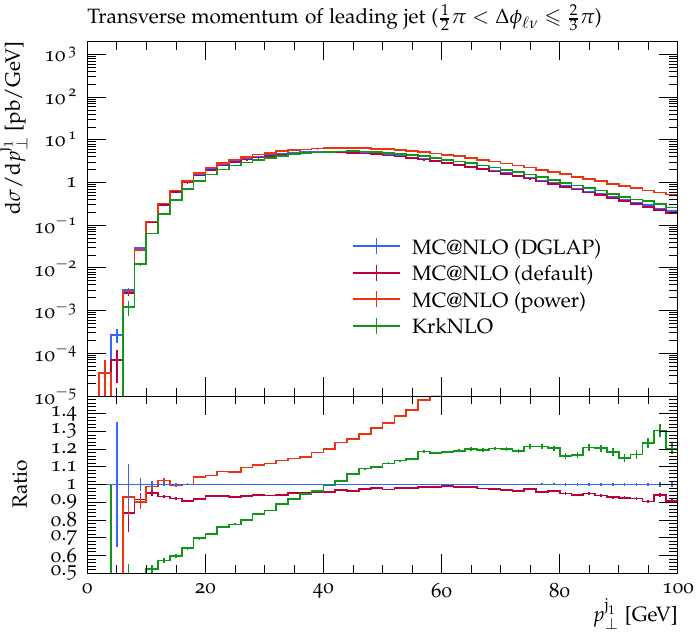}
        \includegraphics[width=0.2\paperwidth]{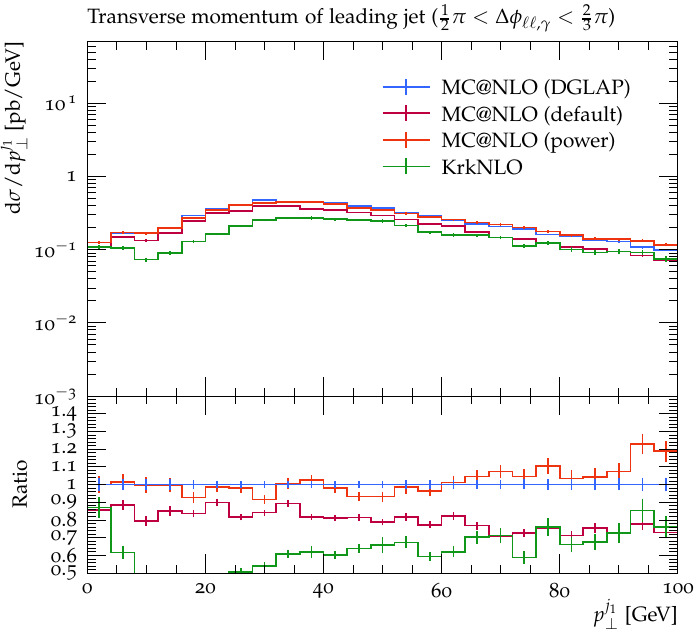}
        \includegraphics[width=0.2\paperwidth]{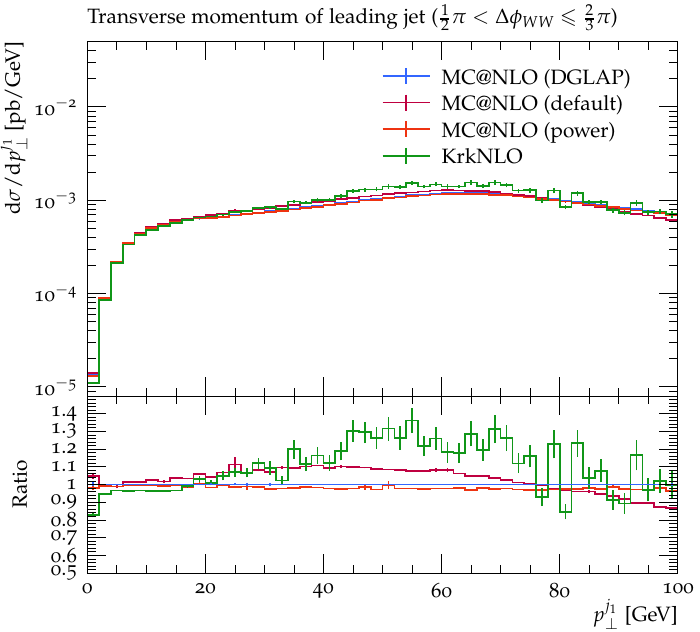}
        \includegraphics[width=0.2\paperwidth]{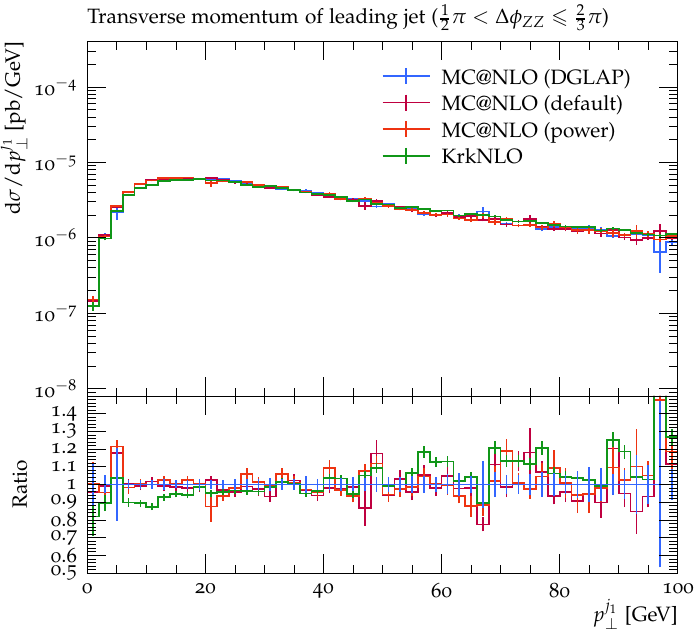}
        } \makebox[\textwidth][c]{
        \includegraphics[width=0.2\paperwidth]{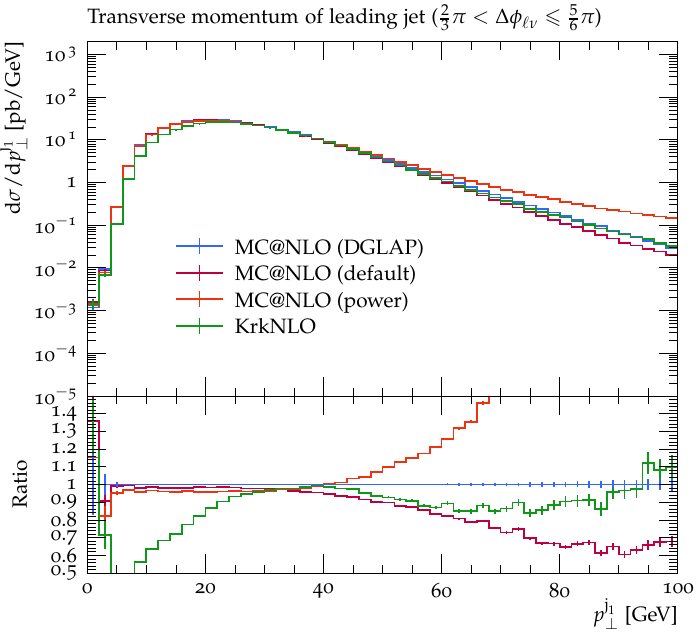}
        \includegraphics[width=0.2\paperwidth]{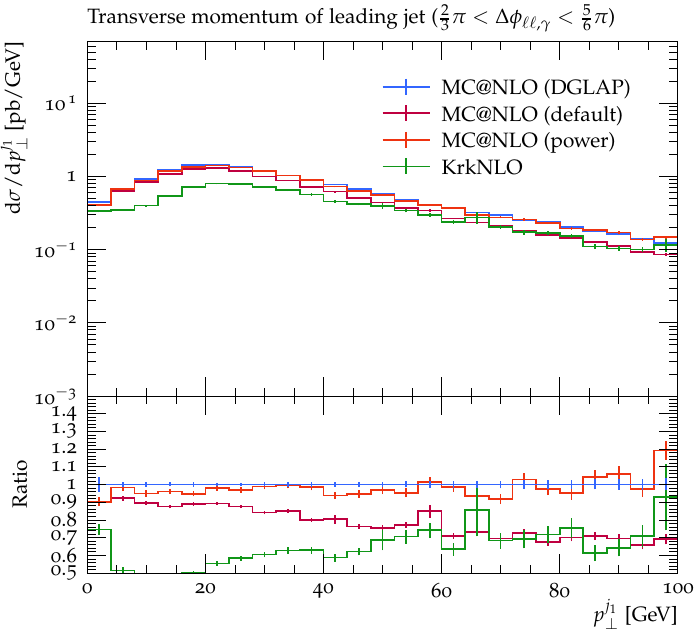}
        \includegraphics[width=0.2\paperwidth]{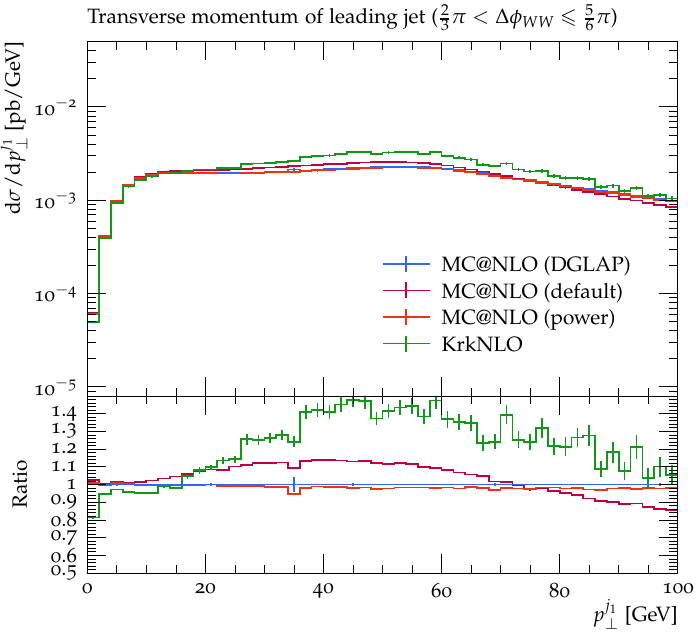}
        \includegraphics[width=0.2\paperwidth]{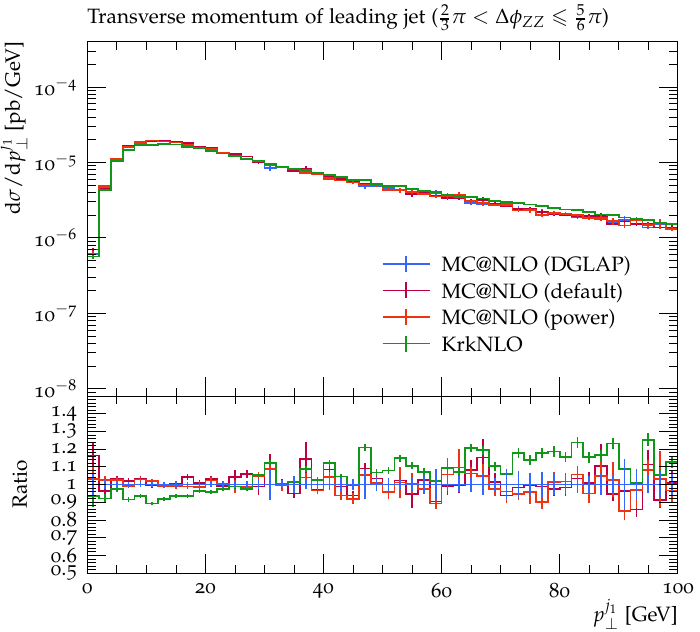}
        } \makebox[\textwidth][c]{
        \begin{subfigure}[b]{0.2\paperwidth}
        \includegraphics[width=\textwidth]{fig/pheno/W/fullshower/double_differential/ptj1_fine_6.pdf}
        \caption{$W$}
        \end{subfigure}
        \begin{subfigure}[b]{0.2\paperwidth}
        \includegraphics[width=\textwidth]{fig/pheno/Zgam/fullshower/double_differential/pT_j1_fine_dphilly_6.pdf}
        \caption{$Z \gamma$}
        \end{subfigure}
        \begin{subfigure}[b]{0.2\paperwidth}
        \includegraphics[width=\textwidth]{fig/pheno/WW/fullshower/mc_vector_diboson/pT_j1_fine_6.pdf}
        \caption{$WW$}
        \end{subfigure}
        \begin{subfigure}[b]{0.2\paperwidth}
        \includegraphics[width=\textwidth]{fig/pheno/ZZ/fullshower/mc_vector_diboson_ZZ/pT_j1_fine_6.pdf}
        \caption{$ZZ$}
        \end{subfigure}
        }
	\caption{Full-shower comparison of the transverse momentum of the leading jet
    in slices of the azimuthal angle between the (reconstructed) two-particle constituents of the colour-singlet system;
     the complete partition of which a subset is shown in \cref{fig:fullshower_dsigma_dptj_dphi}.
    }
    \label{fig:fullshower_dsigma_dptj_dphi_full}
    \vspace{-15mm}
\end{figure}

For completeness, we here reproduce the double-differential distributions presented in the main text
in their full context, with no omitted phase-space `slices'.
For each distribution, the processes are presented left-to-right in a consistent order, 
while adjacent slices in the second observable are presented top-to-bottom.
In \cref{fig:fullshower_dsigma_dM_dphi_full,fig:fullshower_dsigma_dptj_dphi_full},
for which the second observable is $\Delta \phi$, the slices partition the inclusive phase-space,
so that every event passing the fiducial cuts contributes to exactly one $\Delta\phi$ interval.
In \cref{fig:oneemission_dsigma_dM_dptj_full,fig:fullshower_dsigma_dM_dptj_full},
for which the second observable is $\ptj{1}$, the slices partition the radiative phase-space,
omitting only events for which the parton shower evolution generates no radiation.

\clearpage

\bibliographystyle{JHEP}
\bibliography{main.bib}

\end{document}